\documentclass[11pt,letterpaper]{article}

\usepackage{jheppub}
\usepackage{color}
\usepackage[dvipsnames]{xcolor}
\usepackage{colortbl}
\usepackage{graphicx}
\usepackage{wrapfig,enumerate,slashed}
\usepackage{stmaryrd}
\usepackage{subfigure}
\usepackage{multirow,enumitem}
\usepackage[noabbrev]{cleveref}
\usepackage[normalem]{ulem}
\usepackage{footmisc}
\usepackage{amsmath}
\usepackage{wasysym} 
\usepackage{graphicx}
\usepackage{color}
\usepackage{comment}
\usepackage{hyperref}
\usepackage{enumitem}

\newcommand{\mathcl}{\mathcal}

\newcommand{\be}{\begin{eqnarray}}
	\newcommand{\ee}{\end{eqnarray}}

\newcommand{\bee}{\begin{eqnarray}}
	\newcommand{\eee}{\end{eqnarray}}
\newcommand{\beeq}{\begin{equation}}
	\newcommand{\eeeq}{\end{equation}}

\definecolor{verde}{cmyk}{0.92,0,0.59,0.25}
\newcommand{\Dfb}{\mathord{\buildrel{\lower3pt\hbox{$\scriptscriptstyle{\leftrightarrow \tiny{ \ \ \ } }$}}\over {D^{\mu}}}}

\newcommand{\Dfbd}{\mathord{\buildrel{\lower3pt\hbox{$\scriptscriptstyle\leftrightarrow$}}\over {D}_{\mu}}}

\newcommand{\tr}{\text{Tr}}

\newcommand{\eff}{\text{eff}}

\usepackage{bbold}
\usepackage{comment}
\renewcommand{\L}{\mathcal{L}}

\begin{document}
\flushbottom
\allowdisplaybreaks
	
	\title{Electroweak Scalar Effects Beyond Dimension-6 in SMEFT}
	
	\author[a]{Nilabhra Adhikary,}
	\author[a]{Tisa Biswas,}
	\author[a]{Joydeep Chakrabortty,}
	\author[b]{Christoph Englert,}
	\author[c]{and Michael Spannowsky}
	
	\affiliation[a]{Indian Institute of Technology Kanpur, Kalyanpur, Kanpur 208016, Uttar Pradesh, India}
	\affiliation[b]{School of Physics \& Astronomy, University of Glasgow, Glasgow G12 8QQ, UK}
	\affiliation[c]{Institute for Particle Physics Phenomenology, Department of Physics, Durham University, Durham DH1 3LE, UK}

	\emailAdd{nilabhra23@iitk.ac.in, tisab@iitk.ac.in, joydeep@iitk.ac.in,  christoph.englert@glasgow.ac.uk, michael.spannowsky@durham.ac.uk}

	\abstract{
	The Standard Model Effective Field Theory (SMEFT) provides a robust framework for probing deviations in the couplings of Standard Model particles from their theoretical predictions. This framework relies on an expansion in higher-dimensional operators, often truncated at dimension-six. In this work, we compute the effective dimension-eight operators generated by integrating out heavy scalar fields at one-loop order in the Green's basis within two extended scalar sector models: the Two Higgs Doublet Model and the Complex Triplet Scalar Model. We also investigate the impact of heavy scalar fields on the fermion sector, deriving the fermionic effective operators up to dimension eight for these models, and detail how contributions can be mapped onto non-redundant bases. To assess the importance of higher-order contributions in the SMEFT expansion, we analyze the dimension-eight effects for electroweak precision observables at the next frontier of precision lepton machines such as GigaZ.}	
	\preprint{IPPP/25/04}
	
	\maketitle

%%%%%%%%%%%%%%%%%%%%%%%%%%%%%%%%%%%%%%%%%%%%%%%%%%%%%
% Introduction
%%%%%%%%%%%%%%%%%%%%%%%%%%%%%%%%%%%%%%%%%%%%%%%%%%%%%
\section{Introduction}
\label{sec:intro}
In the past few decades, high-energy physics has made tremendous strides in understanding our universe's fundamental forces and particles. Central to this understanding is the Standard Model (SM) of particle physics, a remarkably successful framework for describing the known particles and interactions, most notably through experimental verification at facilities such as the Large Hadron Collider (LHC). However, known limitations of the SM have led physicists to look for traces of physics beyond the Standard Model (BSM), which might manifest at higher energy scales that are not yet accessible through direct experiments.

Effective Field Theories (EFTs) have become a prevalent tool in this context. EFTs allow us to describe the effects of unknown high-energy physics at low energies~\cite{Weinberg:1978kz,Georgi:1991ch}, where direct measurements are possible, by expanding the known theory in terms of higher-dimensional operators. Typically, these expansions include terms that capture the influence of new physics, even if the fundamental theory is not directly observed. The most commonly used EFT for collider experiments like those at the LHC is the SM Effective Field Theory (SMEFT). This formalism expands the SM by adding higher-dimension operators that encapsulate the possible deviations from the SM predictions due to unknown high-energy phenomena. These operators consist of only SM fields and respect the gauge symmetries of the SM.

The dimension-six operators~\cite{BUCHMULLER1986621,Grzadkowski:2010es,Dedes:2019uzs} are often the leading terms considered in these expansions, as they offer the first corrections to the SM predictions. In practice, SMEFT analyses often focus on dimension-six operators up to linear order (e.g., ~\cite{Giudice:2007fh,Englert:2015hrx,Ellis:2018gqa,Banerjee:2019twi,Banerjee:2018bio}), corresponding to $\mathcal{O}(1/\Lambda^2)$ in the new physics scale $\Lambda$ (e.g.,~\cite{Pomarol:2013zra,Butter:2016cvz,Berthier:2015gja,Berthier:2015oma,deBlas:2019rxi}). Some studies extend to quadratic contributions at $\mathcal{O}(1/\Lambda^4)$ (e.g.,~\cite{Buckley:2015lku,Brivio:2019ius,Ellis:2021jhep,Durieux:2019rbz,Biswas:2021qaf,Biswas:2022fsr,Ethier:2021bye}), but a consistent treatment at this order generally requires accounting for terms linear in dimension-eight operators. In many cases, such as rare processes or regions of parameter space with limited data, these dimension-six terms might not capture all relevant effects, especially when the data lack precision~\cite{Hays:2018zze,Degrande:2013kka,Dawson:2021xei}. As experimental searches, such as those conducted at the high-luminosity LHC (HL-LHC), push the boundaries of sensitivity, it becomes increasingly important to explore higher-order effects. This is where dimension-eight operators, e.g., \cite{Dedes:2023zws,Bakshi:2024wzz,DasBakshi:2022mwk}, come into play~\cite{Dawson:2021xei,Li:2023pfw,Degrande:2023iob,Chen:2023bhu,Ellis:2023zim,Dawson:2024ozw}. These terms can provide a more refined description of possible BSM effects, offering corrections to processes where the dimension-six terms may be insufficient~\cite{Adhikary:2025pbb}. The impact of dimension-eight operators is crucial to validate the truncation of SMEFT at a fixed order in light of collider searches. This has been facilitated by the recent development of comprehensive and non-redundant bases for dimension-eight operators ~\cite{Li:2020gnx,Murphy:2020rsh}. Another useful approach involves matching the SMEFT framework with a particular complete UV model containing new particles on the high energy scale, \(\Lambda\). This process often results in a limited set of effective operators, whose impact can be analyzed in low energy and electroweak-scale phenomena~\cite{Brivio:2017vri,Isidori:2023pyp,Anisha:2021hgc,Gorbahn:2015gxa,delAguila:2016zcb}.

This work investigates the role of dimension-eight operators within the EFT framework, with the aim of enhancing the predictive accuracy of SMEFT and its ability to constrain or identify BSM physics. Specifically, we focus on two well-motivated BSM scenarios that extend the SM with additional heavy scalar fields, one involving a complex scalar triplet and the other involving a complex scalar doublet. These models are of interest because such additional scalar fields are common in various extensions of the SM, including models related to neutrino mass generation and grand unification theories~\cite{Branco:2011iw,Gunion:1989we,delAguila:2008ks,Akeroyd:2005gt,Akeroyd:2009hb,Han:2015sca,Akeroyd:2009nu} and can leave characteristic signatures at different collider experiments~\cite{Dey:2018uvu,Cao:2009as,Bhattacharyya:2015nca,Bambhaniya:2013yca,Cai:2017mow,BhupalDev:2018tox,Fuks:2019clu,Antusch:2018svb,delAguila:2013mia,CMS:2022cbe}.  Although, the matching of dimension-six operators to UV models with a single heavy particle is well-established at both tree level and one-loop~\cite{Henning:2014wua,Bakshi:2018ics,Henning:2016lyp,vonGersdorff:2022kwj,Fuentes-Martin:2022jrf,Cohen:2020qvb,Carmona:2021xtq,Criado:2017khh,Celis:2017hod,Haisch:2020ahr}, the inclusion of dimension-eight operators in more complex models remains relatively unexplored. Previous studies have addressed only a limited number of cases~\cite{Jiang:2018pbd,Gherardi:2020det,Dawson:2021xei,Du:2022vso}. In this study, we extend the framework by integrating out heavy scalars up to one-loop, focusing on the complex scalar triplet and doublet scenarios.

We use the model-independent dimension-eight effective action, derived in \cite{Banerjee:2023iiv} to integrate out heavy scalar fields at the one-loop level for specific UV models: electroweak complex triplet and doublet.
 We obtain the effective operators up to  dimension-eight and the associated coefficients in Sec.~\ref{sec:models}. We further capture the imprint of heavy field interactions with the SM fermions in Sec.~\ref{fermionic_int_heavyfd}. In general, these results serve a dual purpose. First, they allow us to understand the impact of extending the analysis to dimension-eight but not at the cost of increasing the number of free parameters, as they are functions of the same UV model parameters. We expect these additional effects may allow us to adjust the model parameter space more precisely and enlarge the parameter space where EFT truncation is still validated. We impose constraints on the EFT parameter space for both models by utilizing electroweak precision data in Sec.~\ref{sec:obsd8}. Specific focus is put on the analysis of future precision lepton machines. Finally, we conclude in Sec.~\ref{sec:conclusions}. We put all the details of the computations in the Appendix~\ref{matching} for the complex triplet and doublet scalar models. In Appendix~\ref{eom_fdrefinition}, we tabulate the contributions for dimension-eight structures after removing redundancies at dimension-eight.
\bigskip
%%%%%%%%%%%%%%%%%%%%%%%%%%%%%%%%%%%%%%%%%%%%%%%%%%%%%
\section{A model-independent approach to integrating out a heavy scalar}\label{sec:models}
%%%%%%%%%%%%%%%%%%%%%%%%%%%%%%%%%%%%%%%%%%%%%%%%%%%%%
\bigskip
Consider a UV-complete action $S$, which depends on heavy ($\Phi$) and light ($\phi$) fields.  The effective action at one-loop after integrating out the fluctuations\footnote{We assume $\Phi = \Phi_c+\eta$, where $\eta$ are the quantum fluctuations.} around its classical solution $(\Phi_c)$ is given by
\begin{align}\label{eq:effective-action}
        e^{i\,S_{\text{eff}}[\phi]} = \int [\mathcal{D}\Phi]e^{i\,S[\Phi,\phi]}
        &= \int [\mathcal{D}\eta]\, \exp\left[i\,\left(S[\Phi_c,\phi]\,+\,\frac{1}{2}\frac{\delta^2 S}{\delta \Phi^2}\bigg\vert_{\Phi=\Phi_c}\,\eta^2\,+\,\mathcal{O}(\eta^3)\right)\right],\nonumber\\
        S_{\text{eff}} &\approx S[\Phi_c]\,+\,\frac{i}{2}  \ln\left(\text{Det}\,\frac{\delta^2 S}{\delta \Phi^2}\bigg\vert_{\Phi=\Phi_c}\right).
\end{align}
After integrating out heavy scalars at one-loop, the effective action consisting of operators up to mass dimension-eight is computed for the first time in Ref.~\cite{Banerjee:2023iiv} employing the Heat-Kernel method. The universal one-loop effective action\footnote{In Ref.~\cite{Chakrabortty:2023yke}, the same has been derived for fermions and noted its equivalence with the scalar case.} up to dimension-eight is given as  \cite{Banerjee:2023iiv}
\begin{align}\label{eq:finite}
		%\begin{split}
		\mathcal{L}_{\eff}^{d \leq 8}    &=  \cfrac{c_s}{(4\pi)^{2}}   M^4\ \left[-\frac{1}{2}\,\left(\ln\left[\frac{M^2}{\mu^2}\right]-\frac{3}{2}\right)\right ]  + \cfrac{c_s}{(4\pi)^{2}}  \tr   \,\Bigg\{ M^2 \ \Bigg[-\left(\ln\left[\frac{M^2}{\mu^2}\right] { \color{verde} - 1}\right) \, { \color{verde} U}\Bigg] \nonumber \\
		& \hspace{1cm}+ M^0\ \frac{1}{2}  \Bigg[- \ln\left[\frac{M^2}{\mu^2}\right] \, U^2 -\frac{1}{6} \ln\left[\frac{M^2}{\mu^2}\right] \, (G_{\mu\nu})^2\Bigg] \nonumber\\
		& \hspace{1cm}+ \frac{1}{M^2}  {\color{magenta} \frac{1}{6}  \,\Bigg[ -U^3 - \frac{1}{2} (P_\mu U)^2-\frac{1}{2}U\,(G_{\mu\nu})^2  - \frac{1}{10}(J_\nu)^2   + \frac{1}{15}\,G^{\mu\nu}\,G_{\nu\rho}\,G^{\rho}_{~\mu} \Bigg] } \nonumber\\
		& \hspace{1cm}+  \frac{1}{M^4} {\color{blue} \frac{1}{24}   \, \Bigg[U^4 - U^2 (P^2 U) + \frac{4}{5}U^2 (G_{\mu\nu})^2 + \frac{1}{5} (U\,G_{\mu\nu})^2 +  \frac{1}{5} (P^2 U)^2}  \nonumber \\
		& \hspace{2.8cm} {\color{blue} - \frac{2}{5} U\, (P_\mu U)\,J^{\mu} + \frac{2}{5} U(J_\mu)^2 - \frac{2}{15} (P^2 U) (G_{\rho\sigma})^2 +\frac{1}{35}(P_\nu J_{\mu})^2 } \nonumber\\ 
		& \hspace{2.8cm} {\color{blue} - \frac{4}{15} U\,G^{\mu\nu}\,G_{\nu\rho}\,G^{\rho}_{~\mu}  - \frac{8}{15} (P_\mu P^\nu U)\, G^{\rho\mu} G_{\rho\nu} + \frac{16}{105}G^{\mu\nu}J_{\mu}J_{\nu} } \nonumber \\
		& \hspace{2.8cm} {\color{blue} +\frac{16}{105} (P^\mu J_{\nu}) G^{\nu\sigma}G_{\sigma\mu}  + \frac{1}{420} (G_{\mu\nu}G_{\rho\sigma})^2 } \nonumber\\
		& \hspace{2.8cm} {\color{blue} +\frac{17}{210}(G_{\mu\nu})^2(G_{\rho\sigma})^2 +\frac{2}{35}(G_{\mu\nu}G_{\nu\rho})^2  + \frac{1}{105} G^{\mu\nu}G_{\nu\rho}G^{\rho\sigma}G_{\sigma\mu}\Bigg]  },
\end{align}
where $c_s = {1}/{2} $ for real scalar fields while $c_s = 1$ for complex scalar fields. 

The stress tensor is given, as usual, $G_{\mu\nu}=[P_\mu, P_\nu]$; $J_\mu = D_\rho G_{\rho \mu}=D_\rho[D_\rho, D_\mu]$ denotes to the gauge current. The trace in the above expression accounts for all symmetry indices, including gauge, Lorentz (spinor and vector), flavour indices etc. Additionally, Hermitian conjugates have been incorporated to ensure the self-Hermitian nature of the effective Lagrangian.

Let us consider a toy Lagrangian with a heavy scalar field $\Phi$ of mass $M_s$:
	\begin{equation}
		\mathcal{L}^\Phi = \Phi^\dagger (P^2 - M_s^2) \Phi + \mathcal{L}_{\text{int}}.
		\label{total_lagr}
	\end{equation}
Here, $\mathcal{L}_{\text{int}}$ represents the interaction part of the Lagrangian that involves both heavy and light fields ($\phi$). We consider, without loss of generality, the interactions as
	\begin{equation}
		\mathcal{L}_{\text{int}} \supset \lambda\,(\phi^{\dagger}\phi)^2 + \lambda_{1}\,(\Phi^{\dagger}\Phi)^2  + \eta\,(\phi^{\dagger}\phi)(\Phi^{\dagger}\Phi) +\big(\kappa_1 \, (\phi^{\dagger}\phi) \Phi + \kappa_2\,\phi^{\dagger}(\Phi^{\dagger}\Phi)+\text{h.c.}\big).
	\end{equation}		
The parameters $\kappa_1$ and $\kappa_2$ have dimensions of mass and we assume $\kappa_1, \kappa_2 \ll M$.
The effective action at the tree level arises from the term linear in the heavy field in the UV model. The Lagrangian in Eqn.~\ref{total_lagr} takes the form :
	\begin{equation}
		\label{eq:the-toy-Lag}
		\mathcal{L}_{\text{int}} \ni \Big\{ \Phi^{\dagger} B_s + \mathrm{h.c.} \Big\} + \Phi^{\dagger} \Big(P^2 - M_s^2 - U_s \Big) \Phi + \mathcal{O}(\Phi^3),
	\end{equation}	
such that $B_s$ and $U_s$ contain interactions that are linear and quadratic in $\Phi$, respectively; they are functions of the light fields only. For our example, we have $B_s = \kappa_1^\ast (\phi^{\dagger}\phi)$ and $U_s = \eta\,(\phi^{\dagger}\phi) + \lambda_1 (\Phi^\dagger \Phi)+ \kappa_2 \phi + \kappa_2^\ast \phi^\dagger $. 
We solve the equation of motion for the heavy field $\Phi$ and substitute it into the action. To leading approximation, the classical solution of $\Phi$ reads as:
\begin{equation}
\Phi_c = -\frac{1}{P^2 - M_s^2 - U_s} B_s.
\label{eqn:trip_Ph_c}
\end{equation}
 Employing the covariant derivative expansion, we find\footnote{Here, we assume the background field varies very slowly compared to the UV mass scale $M_s$ and local expansion of the UV-propagator is validated in the domain of the energy of our interest.}
\begin{align}
\Phi_c &= \left[1 - \frac{1}{M_s^2}\big(P^2 - U_s\big)\right]^{-1}\frac{1}{M_s^2}B_s \nonumber \\
&= \frac{1}{M_s^2} B_s ~\rule[-10pt]{1pt}{30pt}_{\rm {~dim=2}} + \frac{1}{M_s^2}\big(P^2 - U_s\big)\frac{1}{M_s^2}B_s ~\rule[-10pt]{1pt}{30pt}_{\rm {~dim=4}} \nonumber\\ &\hspace{3.5cm}+ \frac{1}{M_s^2}\big(P^2 - U_s\big)\frac{1}{M_s^2}\big(P^2 - U_s\big)\frac{1}{M_s^2}B_s~\rule[-10pt]{1pt}{30pt}_{\rm {~dim=6}} + \dots~ \nonumber \\
 &\equiv \Phi_c^{(0)} + \dfrac{\Phi_c^{(1)}}{M_s^2}  + \dfrac{\Phi_c^{(2)}}{M_s^4} + \dfrac{\Phi_c^{(2)}}{{M_s}^6} + \dots \hspace{2mm}.
\label{phic_expand}
\end{align}
In the above equation, we highlight the dimension for each term in the $\mathcal{O}(1/M^2)$ series expansion of $\Phi_c$, assuming the contribution of light fields only to $U_s$.\footnote{A more detailed discussion for the dimension of $U$ and $\Phi_c$ is presented when considering specific UV models in the subsequent sections.} Thus, $\Phi_c$ can be determined by inserting $B_s$ and $U_s$ in Eqn.~\ref{phic_expand} and solving order by order in $M_s^{-2}$.

After integrating out $\Phi$, a series of operators is obtained that can be systematically organized based on their canonical mass dimension. The effective Lagrangian can then be obtained up to the desired order by substituting $\Phi$ into Eqn.~\ref{eq:the-toy-Lag} via $\Phi_c$. Now, if the effective Lagrangian is to be truncated at order $M^{-4}$, we do not need to consider $\Phi_c^{(2)}$, which is the coefficient $\sim M_s^{-4}$ in the solution of $\Phi_c$ (Eqn.~\ref{phic_expand}), or higher order terms. To see this, we note that at order $M_s^0$ and $M_s^{-2}$, we have $\Phi_c^{(0)} =0$ and $\Phi_c^{(1)} =B_s$, respectively. Using these results and substituting the solution from Eqn.~\ref{phic_expand} (up to the $\Phi_c^{(2)}$ term) in Eqn.~\ref{eq:the-toy-Lag}, terms up to order ${1}/{M_s^4}$ can be obtained as :
\begin{eqnarray}
		\mathcal{L}_{\rm int} &\ni&
		\Bigg\{ \Bigg[\dfrac{B}{M_s^2} +  \dfrac{\Phi_c^{(2)}}{M_s^4} \Bigg]^\dagger B \Bigg\} + \Bigg[\dfrac{B}{M_s^2} + \dfrac{\Phi_c^{(2)}}{M_s^4}\Bigg]^{\dagger} \Big(P^2 - M_s^2 - U_s \Big) \Bigg[\dfrac{B}{M_s^2} + \dfrac{\Phi_c^{(2)}}{M_s^4}\Bigg], \nonumber\\[3mm]
		&=& \dfrac{|B|^2}{M_s^2}  + \dfrac{B^{\dagger} (P^2 - U_s) B}{{M_s}^4} + \mathcal{O}\left(\dfrac{1}{M_s^6}\right),
		\label{cde}
\end{eqnarray}
so that the $\Phi_c^{(2)}$ contribution is canceled.

\bigskip

We are ready to apply this general procedure to the particular cases of two BSM-relevant UV models, namely, the complex scalar triplet model and the 2HDM, to enumerate the possible operators up to dimension-eight. We compute the corresponding operator Wilson coefficients in terms of the parameters of the UV theory.  An implicit assumption throughout is that the heavy-mass scales are much greater than the weak scale. This condition is essential for the SMEFT to be valid for the UV models by ensuring the decoupling of the heavy degrees of freedom. In the following subsections, we present the leading contributions from loop-generated operators with integrated-out contributions at dimension-eight.
%%%%%%%%%%%%%%%%%%%%%%%%%%%%%
\subsection{Case I: Electroweak Complex Triplet}
\label{subsec:triplet}
%%%%%%%%%%%%%%%%%%%%%%%%%%%%%
We consider an electroweak $SU(2)_L$ scalar triplet $\Delta = \Delta^i T^i$ with hypercharge $Y=1$. The generators of $SU(2)_L$ in the fundamental representation are given by $T^i = \sigma^i / 2$ with the Pauli matrices $\sigma^i$ with $i=1,2,3$. 
The relevant part of the Lagrangian that involves a complex scalar triplet~\cite{Barrie:2022cub} is the following. 
\begin{align}
		\label{lbsmCTS}
		\mathcal{L}_{BSM} &=  \mathcal{L}_{SM} \ + \left(\tr [ (\mathcal{D}_{\mu} \Delta)^{\dagger} (\mathcal{D}^{\mu} \Delta ) ] - M_{\Delta}^{2} \tr [ \Delta^{\dagger} \Delta ] + \mathcal{L}_{Y} - V(H,\Delta) \right), 
\end{align}
where the scalar potential is 
\begin{equation}
		\begin{split}
			V(H,\Delta) &=  \lambda_{1} \left(H^{\dagger} H\right) \text{Tr}[ \Delta^{\dagger} \Delta ] + \lambda_{2} \left(\text{Tr}[ \Delta^{\dagger} \Delta ]\right)^2 + \lambda_{3}\text{Tr}\left[ \left(\Delta^{\dagger} \Delta\right)^2 \right]   \\
			& \hspace{1cm}+ \lambda_{4}\left( H^\dagger \Delta {\Delta}^\dagger H \right) + \left[ \mu_{\Delta}\left( H^{T} i \sigma^{2} \Delta^{\dagger} H \right) + \text{h.c.}\right].
			\label{triplet_potential}
		\end{split}
\end{equation}
The covariant derivative has the usual form as $\mathcal{D}_\mu \equiv \partial_\mu - i g' Y B_\mu - i g T^i W^i_\mu$. Here, $Y$ and $T^i$ represent the generators of $U(1)_Y$ and $SU(2)_L$, respectively, with $i=1,2,3$. In the adjoint representation $(T^i)_{jk}=i \epsilon^{ijk}$ with $i,j,k=1,2,3$.
 The Yukawa term in the Lagrangian is
\begin{align}
		{\cal L}_{Y} = Y_{\Delta_{ab}} L^T_a C i\sigma^2 \Delta L_b + \text{h.c.},
		\label{triplet_fermi}
\end{align}
where $C$ represents the charge conjugation matrix, while $a,b = e, \mu, \tau$ denote the flavor indices. This triplet model allows the generation of the lepton number violating Majorana masses for the SM neutrinos, and that feature is also reflected through the effective operators. 

The functional $B \equiv \mathcl{L}/mod(\Delta)$ is derived from the interaction where the heavy field $\Delta$ couples with the lighter fields linearly.  For the scalar Lagrangian Eqn.~\ref{triplet_potential}, it is given as: 
	\begin{equation}
		B^i = \mu_\Delta H^T i \sigma_2 \tau^i H = - \mu_\Delta \tilde{H}^\dagger \tau^i H,
	\end{equation}
where $\tilde{H} = i \sigma_2 H^*$ is the charge-conjugated Higgs field. Note that $\mu_\Delta$ is a mass-dimension one coupling.
	
The scalar potential for the heavy triplet field (using Eqn.~\ref{triplet_decomposition}) can be written in the following basis of heavy fields:
	\begin{equation}
		\mathcal{L_{\rm int}} \supset  \dfrac{1}{2} \begin{pmatrix}
			\Delta^\ast_i & \Delta_i 
		\end{pmatrix} \, 
		\begin{pmatrix}
			\left(U_{11}\right)_{ij} & \left(U_{12}\right)_{ij} \\
			\left(U_{21}\right)_{ij} & \left(U_{22}\right)_{ij}
		\end{pmatrix} \,
		\begin{pmatrix}
			\Delta_j \\ \Delta^\ast_j
		\end{pmatrix},\label{Lint}
	\end{equation}
where the indices $i,j$ range from 1 to 3, corresponding to the three component fields. Thus, $U$ is a $6\times6$ hermitian matrix, where each $3\times3$ block (Eqn.~\ref{Lint}) in the $(\Delta_i, \Delta_j^*)$ basis is given by:
	\begin{equation}
		\left(U_{11}\right)_{ij} = \dfrac{\delta^2 V}{\delta \Delta_i^\ast \delta \Delta_j},\quad
		\left(U_{12}\right)_{ij} = \dfrac{\delta^2 V}{\delta \Delta_i \delta \Delta_j}, \nonumber\\
		\left(U_{21}\right)_{ij} = \dfrac{\delta^2 V}{\delta \Delta_i^\ast \delta \Delta_j^\ast}, \quad \left(U_{22}\right)_{ij} = \dfrac{\delta^2 V}{\delta \Delta_i \delta \Delta_j^\ast}. \nonumber
	\end{equation}
	Hence, we obtain these matrix elements as : 
	\begin{eqnarray}
		(U_{{11}})_{ij} &=& \dfrac{1}{4} (2 \lambda_1 + \lambda_4) (H^\dag H) \delta_{ij} + \dfrac{1}{2}\lambda_2   \Big(\, \Delta_a^* \Delta_a~\delta_{ij} + \Delta_i \Delta^*_j \, \Big) \nonumber \\
		& &+ \dfrac{1}{2} \lambda_3 \Big(\, \Delta_a^* \Delta_a~\delta_{ij} + \Delta_i \Delta^*_j - \Delta^* _i \Delta_j \, \Big) - \frac{i}{2} \lambda_4 \epsilon_{ijk} ( H^\dagger \tau^k H) \nonumber,\label{matrix_u11}
		 \\   
		(U_{{12}})_{ij} & =& \dfrac{1}{2} \lambda_2  (\Delta^\ast_i \Delta^\ast_j) + \dfrac{1}{4}\lambda_3  \left(2 \Delta^\ast_i \Delta^\ast_j -\Delta^\ast_a \Delta^\ast_a \delta_{ij}\right) ,   \label{matrix_u12}\\ 
		(U_{{21}})_{ij} & =& \dfrac{1}{2} \lambda_2  ( \Delta_i \Delta_j) + \dfrac{1}{4}\lambda_3 \left(2 \Delta_i \Delta_j -\Delta_a \Delta_a \delta_{ij}\right),  \label{matrix_u21} \\
		(U_{{22}})_{ij}  
		& =& \dfrac{1}{4} (2 \lambda_1 + \lambda_4) (H^\dag H) \delta_{ij} + \dfrac{1}{2}\lambda_2 \Big(\, \Delta_a^* \Delta_a~\delta_{ij} + \Delta_i \Delta^*_j \, \Big) \nonumber \\
		& &+ \dfrac{1}{2} \lambda_3 \Big(\, \Delta_a^* \Delta_a~\delta_{ij} + \Delta^*_i \Delta_j - \Delta_i \Delta^*_j \, \Big)
		+  \frac{i}{2} \lambda_4 \epsilon_{ijk} ( H^\dagger \tau^k H) . 
		\label{matrix_u22}
	\end{eqnarray}
	These matrix elements need to be evaluated for the classical heavy field solution, which is given by:
	\begin{equation}
		\left(\Delta_c\right)_i = \frac{B_i}{M_\Delta^2} + \frac{(P^2 - U)_{ij}}{M_\Delta^4} B_j + \frac{(P^2 - U)_{ij}^2}{M_\Delta^6} B_j + \cdots.
		\label{tripletsol1}
	\end{equation}
Using the matrix $U$ mentioned above and Eqn.~\ref{tripletsol1}, we derive the effective Lagrangian, see Eqn.~\ref{eq:finite}, up to dimension-eight. The WCs are functions of the following BSM parameters: $\lambda_1, \lambda_2, \lambda_3, \lambda_4, \mu_\Delta$ along with the SM ones.   
The dimension-eight operators in the Green's basis~\cite{Chala:2021cgt}, generated by integrating out the heavy-scalar triplet at the one-loop level within the above model framework, are given in Tables~\ref{tab:matching_triplet_complex_scalar_1} and ~\ref{tab:matching_triplet_complex_scalar_2}. The details of the computation and the corresponding WCs derived from the various operator structures are presented in Appendix~\ref{matching_d8_ctm}. 
%%%matching results
\begin{table}[htb]
	\center
	\resizebox{\linewidth}{!}{
		\renewcommand\arraystretch{2.3}
		\begin{tabular}{| c | l |}
			\hline
			{\bf EFT operators} & \multicolumn{1}{|c|}{{\bf Matching results at scale $M_\Delta (\gg v ) $}}
			\\
			\hline \hline
			\multirow{1}{*}{$\mathcal{O}_{\phi^8} = (\phi^\dag \phi)^4$}
			& $\mathbb{C}_{\phi^8}= {\textcolor{verde}{\frac{1}{16 \pi^2} c_{\phi^8}^{\llbracket U \rrbracket}}} - \frac{1}{32 \pi^2} {\rm ln} \left( \frac{M_\Delta^2}{\mu^2} \right)c_{\phi^8}^{\llbracket U^2 \rrbracket} - {\textcolor{magenta}{\frac{1}{96 \pi^2} c_{\phi^8}^{\llbracket U^3 \rrbracket}}} + {\textcolor{blue}{\frac{1}{384 \pi^2} c^{\llbracket U^{4} \rrbracket}_{\phi^8}}}  $
			\\
			\hline
			\multirow{1}{*}{$\mathcal{O}_{\phi^6}^{(1)} = (\phi^{\dag} \phi)^2 (D_{\mu} \phi^{\dag} D^{\mu} \phi)$}
			& $\mathbb{C}_{\phi^6}^{(1)}= {\textcolor{verde}{\frac{1}{16 \pi^2} c_{\phi^6}^{(1), \llbracket U \rrbracket}}} - {\textcolor{blue}{\frac{1}{384 \pi^2} c_{\phi^6}^{(1), {\llbracket U^{2} (P^2 U) \rrbracket}}}}$
			\\
			\hline
			\multirow{1}{*}{$\mathcal{O}_{\phi^6}^{(2)} = (\phi^{\dag} \phi) (\phi^{\dag} \sigma^I \phi) (D_{\mu} \phi^{\dag} \sigma^I D^{\mu} \phi)$}
			& $\mathbb{C}_{\phi^6}^{(2)}= - {\textcolor{blue}{\frac{1}{384 \pi^2} c_{\phi^6}^{(2), {\llbracket U^{2} (P^2 U) \rrbracket}}}}$
			\\
			\hline
			\multirow{1}{*}{$\mathcal{O}_{\phi^6}^{(3)} = (\phi^\dagger\phi)^2 (\phi^\dagger D^2\phi + \text{h.c.}) $}
			& $\mathbb{C}_{\phi^6}^{(3)}=   {\textcolor{verde}{\frac{1}{16 \pi^2} c_{\phi^6}^{(3), \llbracket U \rrbracket}}} - {\textcolor{magenta}{\frac{1}{192 \pi^2} c_{\phi^6}^{(3), \llbracket (P_\mu U)^2 \rrbracket}}} - {\textcolor{blue}{\frac{1}{384 \pi^2} c_{\phi^6}^{(3), {\llbracket U^{2} (P^2 U) \rrbracket}}}}$
			\\
			\hline
			\multirow{1}{*}{$\mathcal{O}^{(1)}_{W^2 \phi^4}=(\phi^\dag \phi)^2 W^I_{\mu\nu} W^{I\mu\nu}$} 
			& $\mathbb{C}^{(1)}_{W^2 \phi^4}= {\textcolor{magenta}{- \frac{1}{192 \pi^2} c^{(1) , \llbracket U (G_{\mu \nu})^2 \rrbracket}_{W^2 \phi^4}}} + {\textcolor{blue}{\frac{1}{480 \pi^2} c^{(1) , \llbracket U^2 (G_{\mu \nu})^2 \rrbracket}_{W^2 \phi^4}}} + {\textcolor{blue}{\frac{1}{1920 \pi^2} c^{(1) , \llbracket (U G_{\mu \nu})^2 \rrbracket}_{W^2 \phi^4}}} $
			\\
			\hline
			\multirow{1}{*}{$\mathcal{O}^{(3)}_{W^2 \phi^4}=(\phi^\dag \sigma^I \phi) (\phi^\dag \sigma^J \phi) W^I_{\mu\nu} W^{J\mu\nu}$} 
			& $\mathbb{C}^{(3)}_{W^2 \phi^4}= {\textcolor{magenta}{- \frac{1}{192 \pi^2} c^{(3) , \llbracket U (G_{\mu \nu})^2 \rrbracket}_{W^2 \phi^4}}} + {\textcolor{blue}{\frac{1}{480 \pi^2} c^{(3) , \llbracket U^2 (G_{\mu \nu})^2 \rrbracket}_{W^2 \phi^4}}} + {\textcolor{blue}{\frac{1}{1920 \pi^2} c^{(3) , \llbracket (U G_{\mu \nu})^2 \rrbracket}_{W^2 \phi^4}}} $
			\\
			\hline
			\multirow{1}{*}{$\mathcal{O}^{(1)}_{B^2  \phi^4} = (\phi^\dag \phi)^2 B_{\mu\nu} B^{\mu\nu}$} 
			& $\mathbb{C}^{(1)}_{B^2  \phi^4}  =  {\textcolor{magenta}{- \frac{1}{192 \pi^2} c^{(1) , \llbracket U (G_{\mu \nu})^2 \rrbracket}_{B^2 \phi^4}}} + {\textcolor{blue}{\frac{1}{480 \pi^2} c^{(1) , \llbracket U^2 (G_{\mu \nu})^2 \rrbracket}_{B^2 \phi^4}}} + {\textcolor{blue}{\frac{1}{1920 \pi^2} c^{(1) , \llbracket (U G_{\mu \nu})^2 \rrbracket}_{B^2 \phi^4}}} $
			\\
			\hline
			\multirow{1}{*}{$\mathcal{O}^{(1)}_{W B \phi^4} = (\phi^\dag \phi) (\phi^\dag \sigma^I \phi) W^I_{\mu\nu} B^{\mu\nu} $} & 
			$\mathbb{C}^{(1)}_{W B \phi^4} = {\textcolor{blue}{\frac{1}{480 \pi^2} c^{(1) , \llbracket U^2 (G_{\mu \nu})^2 \rrbracket}_{W B \phi^4}}} + {\textcolor{blue}{\frac{1}{1920 \pi^2} c^{(1) , \llbracket (U G_{\mu \nu})^2 \rrbracket}_{W B \phi^4}}} $
			\\
			\hline
			\multirow{1}{*}{$\mathcal{O}^{(3)}_{\phi^4} = (D^{\mu} \phi^{\dag} D_{\mu} \phi) (D^{\nu} \phi^{\dag} D_{\nu} \phi) $} & 
			$\mathbb{C}^{(3)}_{\phi^4} = {\textcolor{verde}{\frac{1}{16 \pi^2} c_{\phi^6}^{(3), \llbracket U \rrbracket}}}+ {\textcolor{blue}{\frac{1}{1920 \pi^2} c^{(3) , \llbracket (P^2 U)^2 \rrbracket}_{\phi^4}}}$
			\\
			\hline
			\multirow{1}{*}{$\mathcal{O}^{(4)}_{\phi^4}= (D_\mu\phi^\dagger D^\mu\phi)(\phi^\dagger D^2\phi + \text{h.c.}) $} & 
			$\mathbb{C}^{(4)}_{\phi^4} = {\textcolor{verde}{\frac{1}{16 \pi^2} c_{\phi^4}^{(4), \llbracket U \rrbracket}}} + {\textcolor{blue}{\frac{1}{1920 \pi^2} c^{(4) , \llbracket (P^2 U)^2 \rrbracket}_{\phi^4}}}$
			\\
			\hline
			\multirow{1}{*}{$\mathcal{O}^{(8)}_{\phi^4} = ((D^2\phi^\dagger\phi) (D^2\phi^\dagger\phi)+\text{h.c.})$} & 
			$\mathbb{C}^{(8)}_{\phi^4} = {\textcolor{verde}{\frac{1}{16 \pi^2} c_{\phi^4}^{(8), \llbracket U \rrbracket}}} + {\textcolor{blue}{\frac{1}{1920 \pi^2} c^{(8) , \llbracket (P^2 U)^2 \rrbracket}_{\phi^4}}}$
			\\
			\hline
		\end{tabular}
	}
	\caption{The matching results for relevant dimension-eight operators in the SMEFT for the complex triplet model with the heavy triplet integrated up to one-loop. The terms in blue, magenta, black and green denote the contributions in Eqn.~\ref{eq:finite} with terms proportional to : ${M_\Delta^{-4}},{M_\Delta^{-2}}, M_\Delta^0$, and $M_\Delta^2$, respectively.}
	\label{tab:matching_triplet_complex_scalar_1}
\end{table}
%%%%%%%%%%%%%%%%%%%%

%%%%%%%%%%%%%%%%%%%%
\begin{table}[htb]
	\center
	\resizebox{\linewidth}{!}{
		\renewcommand\arraystretch{1.5}
		\begin{tabular}{| c | l |}
			\hline
			{\bf EFT operators} & \multicolumn{1}{|c|}{{\bf Matching results at scale $M_\Delta (\gg v ) $}}
			\\
			\hline \hline
			\multirow{1}{*}{$\mathcal{O}^{(10)}_{\phi^4} = (D^2\phi^\dagger D^2\phi) (\phi^\dagger\phi)$} & 
			$\mathbb{C}^{(10)}_{\phi^4} = {\textcolor{verde}{\frac{1}{16 \pi^2} c_{\phi^4}^{(10), \llbracket U \rrbracket}}} +  {\textcolor{blue}{\frac{1}{1920 \pi^2} c^{(10) , \llbracket (P^2 U)^2 \rrbracket}_{\phi^4}}}$
			\\
			\hline
			\multirow{1}{*}{$\mathcal{O}^{(11)}_{\phi^4} = (\phi^\dagger D^2\phi) (D^2\phi^\dagger\phi)$} & 
			$\mathbb{C}^{(11)}_{\phi^4} = {\textcolor{verde}{\frac{1}{16 \pi^2} c_{\phi^4}^{(11), \llbracket U \rrbracket}}} + {\textcolor{blue}{\frac{1}{1920 \pi^2} c^{(11) , \llbracket (P^2 U)^2 \rrbracket}_{\phi^4}}}$
			\\
			\hline
			\multirow{1}{*}{$\mathcal{O}_{W\phi^4 D^2}^{(6)} = (\phi^\dag \phi) D_\nu W^{I \mu\nu}(D_\mu\phi^\dagger \text{i} \sigma^I \phi + \text{h.c.})$} & 
			$\mathbb{C}_{W\phi^4 D^2}^{(6)} = - {\textcolor{blue}{\frac{1}{960 \pi^2} c^{(6) , \llbracket U (P_\mu U) J_\mu \rrbracket}_{W \phi^4 D^2}}}$
			\\
			\hline
			\multirow{1}{*}{$\mathcal{O}_{W\phi^4 D^2}^{(7)} = \epsilon^{IJK} (D_\mu \phi^\dag \sigma^I \phi) (\phi^\dag \sigma^J D_\nu \phi) W^{K \mu\nu} $} & 
			$\mathbb{C}_{W\phi^4 D^2}^{(7)} = - {\textcolor{blue}{\frac{1}{960 \pi^2} c^{(7) , \llbracket U (P_\mu U) J_\mu \rrbracket}_{W \phi^4 D^2}}}$
			\\
			\hline
			\multirow{1}{*}{$\mathcal{O}_{B\phi^4 D^2}^{(1)} = \text{i}(\phi^{\dag} \phi) (D^{\mu} \phi^{\dag} D^{\nu} \phi) B_{\mu\nu} $} & 
			$\mathbb{C}_{B\phi^4 D^2}^{(1)} =  - {\textcolor{blue}{\frac{1}{960 \pi^2} c^{(1) , \llbracket U (P_\mu U) J_\mu \rrbracket}_{B \phi^4 D^2}}} $
			\\
			\hline
			\multirow{1}{*}{$\mathcal{O}^{(4)}_{W^2 \phi^2 D^2} = i \epsilon^{IJK} (D^{\mu} \phi^{\dag} \sigma^I D^{\nu} \phi) W_{\mu\rho}^J W_{\nu}^{K \rho} $} & 
			$\mathbb{C}^{(4)}_{W^2 \phi^2 D^2} = - {\textcolor{blue}{\frac{1}{720 \pi^2} c^{(4) , \llbracket (P_\mu P_\nu U) G_{\rho \mu} G_{\rho \nu} \rrbracket}_{W^2 \phi^2 D^2}}}  $
			\\
			\hline
			\multirow{1}{*}{$\mathcal{O}^{(11)}_{W^2 \phi^2 D^2} = (\phi^\dagger D_\nu\phi+D_\nu\phi^\dagger \phi) D_\mu W^{I\mu\rho} W^{I\nu}_{\,\,\,\,\rho} $} & 
			$\mathbb{C}^{(11)}_{W^2 \phi^2 D^2} =  - {\textcolor{blue}{\frac{1}{720 \pi^2} c^{(11) , \llbracket (P_\mu P_\nu U) G_{\rho \mu} G_{\rho \nu} \rrbracket}_{W^2 \phi^2 D^2}}}  $
			\\
			\hline
			\multirow{1}{*}{$\mathcal{O}^{(13)}_{W^2 \phi^2 D^2} = \phi^\dagger\phi D_\mu W^{I\mu\rho} D_\nu W^{I\nu}_{\,\,\,\,\rho} $} & 
			$\mathbb{C}^{(13)}_{W^2 \phi^2 D^2} =  {\textcolor{blue}{\frac{1}{960 \pi^2} c^{(13) , \llbracket  U (J_{\mu})^2  \rrbracket}_{W^2 \phi^2 D^2}}} $
			\\
			\hline
			\multirow{1}{*}{$\mathcal{O}^{(14)}_{W^2 \phi^2 D^2} = (D_\mu\phi^\dagger\phi+\phi^\dagger D_\mu\phi) W^{I\nu\rho} D^\mu W^{I}_{\nu\rho} $} & 
			$\mathbb{C}^{(14)}_{W^2 \phi^2 D^2} = - {\textcolor{blue}{\frac{1}{2880 \pi^2} c^{(14) , \llbracket (P^2 U) (G_{\rho \sigma})^2 \rrbracket}_{W^2 \phi^2 D^2}}}  $
			\\
			\hline
			\multirow{1}{*}{$\mathcal{O}^{(2)}_{B^2 \phi^2 D^2} = (D^{\mu} \phi^{\dag} D_{\mu} \phi) B_{\nu\rho} B^{\nu\rho} $} & 
			$\mathbb{C}^{(2)}_{B^2 \phi^2 D^2} = - {\textcolor{blue}{\frac{1}{2880 \pi^2} c^{(14) , \llbracket (P^2 U) (G_{\rho \sigma})^2 \rrbracket}_{B^2 \phi^2 D^2}}} $
			\\
			\hline
			\multirow{1}{*}{$\mathcal{O}^{(4)}_{B^2 \phi^2 D^2} = (D_\mu\phi^\dagger \phi+\phi^\dagger D_\mu\phi) D_\nu B^{\mu\rho} B^\nu_{\,\,\rho} $} & 
			$\mathbb{C}^{(4)}_{B^2 \phi^2 D^2} = - {\textcolor{blue}{\frac{1}{720 \pi^2} c^{(4) , \llbracket (P_\mu P_\nu U) G_{\rho \mu} G_{\rho \nu} \rrbracket}_{B^2 \phi^2 D^2}}} $
			\\ \hline
			\multirow{1}{*}{$\mathcal{O}^{(6)}_{B^2 \phi^2 D^2} = \phi^\dagger\phi D_\mu D_\nu B^{\mu\rho} B^{\nu}_{\,\,\rho}  $} & 
			$\mathbb{C}^{(6)}_{B^2 \phi^2 D^2} =  {\textcolor{blue}{\frac{1}{960 \pi^2} c^{(6) , \llbracket  U (J_{\mu})^2  \rrbracket}_{B^2 \phi^2 D^2}}} $
			\\
			\hline
			\multirow{1}{*}{$\mathcal{O}^{(8)}_{B^2 \phi^2 D^2} = (\phi^\dagger D_\nu\phi+D_\nu\phi^\dagger\phi) D_\mu B^{\mu\rho} B^\nu_{\,\,\rho} $} & 
			$\mathbb{C}^{(8)}_{B^2 \phi^2 D^2}  =  {\textcolor{blue}{\frac{1}{960 \pi^2} c^{(8) , \llbracket  U (J_{\mu})^2  \rrbracket}_{B^2 \phi^2 D^2}}}  $
			\\
			\hline
			\multirow{1}{*}{$\mathcal{O}^{(1)}_{W B \phi^2 D^2} = (D^{\mu} \phi^{\dag} \sigma^I D_{\mu} \phi) B_{\nu\rho} W^{I \nu\rho} $} & 
			$\mathbb{C}^{(1)}_{W B \phi^2 D^2} =  - {\textcolor{blue}{\frac{1}{2880 \pi^2} c^{(1) , \llbracket (P^2 U) (G_{\rho \sigma})^2 \rrbracket}_{W B \phi^2 D^2}}}  $
			\\
			\hline
			\multirow{1}{*}{$\mathcal{O}^{(8)}_{W B \phi^2 D^2} = (\phi^\dagger\sigma^I D^\nu\phi+D^\nu\phi^\dagger\sigma^I\phi) D_\mu B^{\mu\rho} W^{I}_{\nu\rho} $} & 
			$\mathbb{C}^{(8)}_{W B \phi^2 D^2} =  - {\textcolor{blue}{\frac{1}{720 \pi^2} c^{(8) , \llbracket (P_\mu P_\nu U) G_{\rho \mu} G_{\rho \nu} \rrbracket}_{W B \phi^2 D^2}}} $
			\\
			\hline
			\multirow{1}{*}{$\mathcal{O}^{(10)}_{W B \phi^2 D^2} = (\phi^\dagger\sigma^I\phi) D^\mu B_{\mu\rho} D_\nu W^{I\nu\rho} $} & 
			$\mathbb{C}^{(10)}_{W B \phi^2 D^2} =    {\textcolor{blue}{\frac{1}{960 \pi^2} c^{(10) , \llbracket  U (J_{\mu})^2  \rrbracket}_{W B \phi^2 D^2}}}  $
			\\
			\hline
			\multirow{1}{*}{$\mathcal{O}^{(11)}_{W B \phi^2 D^2} = (D_\nu\phi^\dagger\sigma^I\phi+\phi^\dagger\sigma^I D_\nu\phi) B_{\mu\rho} D^\mu W^{I\nu\rho} $} & 
			$\mathbb{C}^{(11)}_{W B \phi^2 D^2} = - {\textcolor{blue}{\frac{1}{720 \pi^2} c^{(11) , \llbracket (P_\mu P_\nu U) G_{\rho \mu} G_{\rho \nu} \rrbracket}_{W B \phi^2 D^2}}}  $
			\\
			\hline
			\multirow{1}{*}{$\mathcal{O}^{(1)}_{W^3 \phi^2 } = \epsilon^{IJK} (\phi^\dag \phi) W_{\mu}^{I\nu} W_{\nu}^{J\rho} W_{\rho}^{K\mu} $} & 
			$\mathbb{C}^{(1)}_{W^3 \phi^2 } = - {\textcolor{blue}{\frac{1}{1440 \pi^2}  c^{(1), \llbracket U G_{\mu \nu} G_{\nu \rho} G_{\rho \mu} \rrbracket}_{W^3 \phi^2}}}$
			\\
			\hline
			\multirow{1}{*}{$\mathcal{O}_{W^2 D^4} = (D_\sigma D_\mu W^{I\mu\nu}) (D^\sigma D^\rho W^I_{\rho\nu}) $}
			& $\mathbb{C}_{W^2 D^4}= {\textcolor{blue}{\frac{1}{13440 \pi^2}  c^{ \llbracket (P_\nu J_{\nu})^2 \rrbracket}_{W^2 D^4}}} $
			\\
			\hline
			\multirow{1}{*}{$\mathcal{O}_{B^2 D^4} = (D_\sigma D_\mu B^{\mu\nu}) (D^\sigma D^\rho B_{\rho\nu}) $}
			& $\mathbb{C}_{B^2 D^4}=  {\textcolor{blue}{\frac{1}{13440 \pi^2}  c^{ \llbracket (P_\nu J_{\nu})^2 \rrbracket}_{B^2 D^4}}} $
			\\
			\hline
			\multirow{1}{*}{$\mathcal{Q}^{(3)}_{W^4} = (W_{\mu\nu}^I W^{J\mu\nu}) (W_{\rho\sigma}^I W^{J\rho\sigma})$}
			& $\mathbb{C}^{(3)}_{W^4}=  {\textcolor{blue}{\frac{1}{161280 \pi^2}  c^{(1),  \llbracket (G_{\mu \nu} G_{\rho \sigma})^2 \rrbracket}_{W^2 D^4}}} + {\textcolor{blue}{\frac{17}{161280 \pi^2}  c^{(3), \llbracket (G_{\mu \nu})^2 (G_{\rho \sigma})^2 \rrbracket}_{W^4}}} $
			\\
			\hline
			\multirow{1}{*}{$\mathcal{Q}^{(1)}_{B^4} = (B_{\mu\nu} B^{\mu\nu}) (B_{\rho\sigma} B^{\rho\sigma}) $}
			& $\mathbb{C}^{(1)}_{B^4}= {\textcolor{blue}{\frac{1}{161280 \pi^2}  c^{(1),  \llbracket (G_{\mu \nu} G_{\rho \sigma})^2 \rrbracket}_{B^2 D^4}}}  + {\textcolor{blue}{\frac{17}{161280 \pi^2}  c^{(1), \llbracket (G_{\mu \nu})^2 (G_{\rho \sigma})^2 \rrbracket}_{B^4}}} $
			\\
			\hline
			\multirow{1}{*}{$\mathcal{Q}^{(1)}_{W^2 B^2}=(B_{\mu\nu} B^{\mu\nu}) (W_{\rho\sigma}^I W^{I\rho\sigma})$} 
			& $\mathbb{C}^{(1)}_{W^2 B^2}=  {\textcolor{blue}{\frac{1}{161280 \pi^2}  c^{(1), \llbracket (G_{\mu \nu} G_{\rho \sigma})^2 \rrbracket}_{W^2 B^2}}}  + {\textcolor{blue}{\frac{17}{161280 \pi^2}  c^{(1), \llbracket (G_{\mu \nu})^2 (G_{\rho \sigma})^2 \rrbracket}_{W^2 B^2}}}$
			\\
			\hline
			\multirow{1}{*}{$\mathcal{Q}^{(3)}_{W^2 B^2}=(B_{\mu\nu} W^{I\mu\nu}) (B_{\rho\sigma} W^{I\rho\sigma})$} 
			& $\mathbb{C}^{(3)}_{W^2 B^2}=  {\textcolor{blue}{\frac{1}{161280 \pi^2}  c^{(3), \llbracket (G_{\mu \nu} G_{\rho \sigma})^2 \rrbracket}_{W^2 B^2}}} + {\textcolor{blue}{\frac{17}{161280 \pi^2}  c^{(3), \llbracket (G_{\mu \nu})^2 (G_{\rho \sigma})^2 \rrbracket}_{W^2 B^2}}}$
			\\
			\hline
		\end{tabular} 
	}
	\caption{Summary of matching results for relevant dimension-eight operators in the SMEFT for the complex triplet model with the heavy triplet integrated up to one-loop. The terms in blue and green denote the contributions from $M_\Delta^{-4}$ and $M_\Delta^{2}$ from proportional terms, respectively.}
	\label{tab:matching_triplet_complex_scalar_2}
\end{table}
%%%%%%%%%%%%%%%%%%%%

We also present the results for dimension-six operators arising from this procedure on a SILH~\cite{Giudice:2007fh} basis that are in agreement with \cite{DasBakshi:2018vni}. In addition, in Sec.~\ref{fermionic_int_heavyfd}, we capture the impact of fermionic interactions of the heavy triplet scalar within the effective operators, which are of great phenomenological interest~\cite{Li:2018jns,Bolton:2022lrg,Primulando:2019evb}.
%%%%%%%%%%%%%%%%%%%%

\subsection{Case II: Electroweak Complex Doublet}
\label{subsec:doublet}
%%%%%%%%%%%%%%%%%%%%%%%%%%%%%
In this section, we consider the electroweak complex  $SU(2)_L$ doublet with hypercharge $Y=-\frac{1}{2}$, also known as 2HDM.
The part of the Lagrangian of our interest is ~\cite{Henning:2014wua,Bakshi:2018ics}: 
\begin{align}
\mathcal{L}_{\mathcal{H}_2} &= ~\mathcal{L}_{\text{SM}} + |D_\mu \mathcal{H}_2|^2 - m_{\mathcal{H}_2}^2|\mathcal{H}_2|^2+V_{\text{scalar}} + \L_{\text{Yuk}}~,
\label{2hdm_lagrangian}
\end{align}
where $V_{\text{scalar}}$ refers to the scalar potential term: 
\begin{align}
V_{\text{scalar}} &= \frac{\lambda_{\mathcal{H}_2}}{4}|\mathcal{H}_2|^4 - \left( \eta_H |\tilde{H}|^2 + \eta_{\mathcal{H}_2} |\mathcal{H}_2|^2 \right) \left( \tilde{H}^\dagger \mathcal{H}_2 + \mathcal{H}_2^\dagger \tilde{H} \right) \nonumber \\
& \hspace{1cm}+ \lambda_{\mathcal{H}_2,1}|\tilde{H}|^2|\mathcal{H}_2|^2 + \lambda_{\mathcal{H}_2,2}|\tilde{H}^\dagger \mathcal{H}_2|^2  + \lambda_{\mathcal{H}_2,3}\left[ (\tilde{H}^\dagger \mathcal{H}_2)^2 + (\mathcal{H}_2^\dagger \tilde{H})^2 \right]~,  
\label{2hdm_potential}
\end{align}
and  $\L_{\text{Yuk}}$ is the Yukawa interactions of the SM fermions with the heavy scalar: 
\begin{align}
\L_{\text{Yuk}} & = - \left\{ Y_{\mathcal{H}_2}^{(e)} \bar{\ell}_L \mathcal{H}_2 e_R + Y_{\mathcal{H}_2}^{(u)} \bar{q}_L \tilde{\mathcal{H}}_2 u_R + Y_{\mathcal{H}_2}^{(d)} \bar{q}_L \mathcal{H}_2 d_R + \text{h.c.} \right\},
\label{2hdm_yukawa}
\end{align}
where $\widetilde{\mathcal{H}}_2 = i \sigma_2 \mathcal{H}_2^\ast$. We suppress the family indices of the fermions. 
Here, we discuss the emergence of effective operators after integrating out heavy scalar doublet $\mathcal{H}_2$.

The potential $V_{\text{scalar}}$ has distinct interactions such as $( \tilde{H}^\dagger \mathcal{H}_2)^2$ and $(\mathcal{H}_2^\dagger \tilde{H} )^2$, so it is convenient to treat $\mathcal{H}_2$ and $\mathcal{H}_2^*$ as different variables. $U$ is a $4\times4$ hermitian matrix, where each $2\times2$ block in $(\mathcal{H}_2, \mathcal{H}_2^*)$ basis is given by:
\begin{eqnarray}
	\left(U_{11}^{\mathcal{H}_2}\right)_{ij} = \dfrac{\delta^2 V_{\text{scalar}}}{\delta {\mathcal{H}_2}_i^\ast \delta {\mathcal{H}_2}_j},\quad
	\left(U_{12}^{\mathcal{H}_2}\right)_{ij} = \dfrac{\delta^2 V_{\text{scalar}}}{\delta {\mathcal{H}_2}_i \delta {\mathcal{H}_2}_j}, \nonumber\\
	\left(U_{21}^{\mathcal{H}_2}\right)_{ij} = \dfrac{\delta^2 V_{\text{scalar}}}{\delta {\mathcal{H}_2}_i^\ast \delta {\mathcal{H}_2}_j^\ast}, \quad 
	\left(U_{22}^{\mathcal{H}_2}\right)_{ij} = \dfrac{\delta^2 V_{\text{scalar}}}{\delta {\mathcal{H}_2}_i \delta {\mathcal{H}_2}_j^\ast}. \nonumber
\end{eqnarray}
Here, indices $ i,j$ run from 1 to 2 and refer to the two-component fields of each complex scalar doublet. We, thus, obtain these matrix elements as follows: 
\begin{eqnarray}
	\left(U_{11}^{\mathcal{H}_2}\right)_{i j}  &=& \frac{\lambda_{\mathcal{H}_2}}{2} \left[\delta_{ij} \left({\mathcal{H}_2}_k^* {\mathcal{H}_2}_k \right) + {\mathcal{H}_2}_j^* {\mathcal{H}_2}_i \right]  + \lambda_{\mathcal{H}_2,1} \delta_{ij} \left( H_k^* H_k \right) + \lambda_{\mathcal{H}_2,2} \left(\tilde{H}_j \tilde{H}_i^* \right) \nonumber \\
	&&- \eta_{\mathcal{H}_2} \left[ \delta_{ij} \left( \tilde{H}_k^* {\mathcal{H}_2}_k + {\mathcal{H}_2}_k^* \tilde{H}_k \right) + {\mathcal{H}_2}_i \tilde{H}_j^* + {\mathcal{H}_2}_j^* \tilde{H}_i \right] , \\
	\left(U_{12}^{\mathcal{H}_2}\right)_{i j} &= &\frac{\lambda_{\mathcal{H}_2}}{4} \left[{\mathcal{H}_2}_i^* {\mathcal{H}_2}_j^* + {\mathcal{H}_2}_j^* {\mathcal{H}_2}_i^*\right] 
	- \eta_{\mathcal{H}_2} \left[{\mathcal{H}_2}_i^* \tilde{H}_j^* + {\mathcal{H}_2}_j^* \tilde{H}_i^*\right] \nonumber \\
	&& + \lambda_{\mathcal{H}_2,3} \left[\tilde{H}_i^* \tilde{H}_j^* + \tilde{H}_j^* \tilde{H}_i^*\right], \\
	\left(U_{21}^{\mathcal{H}_2}\right)_{i j}  &=& \frac{\lambda_{\mathcal{H}_2}}{4} \left[{\mathcal{H}_2}_i {\mathcal{H}_2}_j + {\mathcal{H}_2}_j {\mathcal{H}_2}_i\right] 
	- \eta_{\mathcal{H}_2} \left[{\mathcal{H}_2}_i \tilde{H}_j + {\mathcal{H}_2}_j \tilde{H}_i\right] \nonumber \\
	&& + \lambda_{\mathcal{H}_2,3} \left[ \tilde{H}_i \tilde{H}_j + \tilde{H}_j \tilde{H}_i\right] ,\\
	\left(U_{22}^{\mathcal{H}_2}\right)_{i j} &=& \frac{\lambda_{\mathcal{H}_2}}{2} \left[\delta_{ij} \left({\mathcal{H}_2}_k^* {\mathcal{H}_2}_k \right) + {\mathcal{H}_2}_j {\mathcal{H}_2}_i^* \right] + \lambda_{\mathcal{H}_2,1} \delta_{ij} \left(H_k^* H_k \right) + \lambda_{\mathcal{H}_2,2} \left( \tilde{H}_j^* \tilde{H}_i \right) \nonumber \\
	&&- \eta_{\mathcal{H}_2} \left[ \delta_{ij} \left( \tilde{H}_k^* {\mathcal{H}_2}_k + {\mathcal{H}_2}_k^* H_k \right) + {\mathcal{H}_2}_i^* \tilde{H}_j + {\mathcal{H}_2}_j H_i^* \right] .
	\label{2hdmmatrix_u22}
\end{eqnarray}
These matrix elements must be evaluated at the classical heavy field solution for $\mathcal{H}_2$. 

We obtain the functional $B$  from the linear term in the heavy field $\mathcal{H}_2$ of the scalar Lagrangian (Eqn.~\ref{2hdm_potential}) as $B_i = \eta_H (H^\dag H) \tilde{H_i}$ (and $B^\dag_i = \eta_H (H^\dag H) \tilde{H}^\dag_i$). The mass dimension of $B$ is three, and $\eta_H$ is dimensionless coupling.

The equation of motion for the heavy doublet field $\mathcal{H}_2$ is given by:
\begin{equation}
	\left(\mathcal{H}_{\rm 2,c}\right)_i = \frac{B_i}{M_{\mathcal{H}_{2}}^2} + \frac{(P^2 - U)_{ij}}{M_{\mathcal{H}_{2}}^4} B_j + \frac{(P^2 - U)_{ij}^2}{M_{\mathcal{H}_{2}}^6} B_j + \cdots .
	\label{doubletsol1}
\end{equation}

Using Eqn.~\ref{eq:finite}, we obtain the dimension-eight operators arising by integrating out the heavy doublet scalar at the one-loop level and list them in the Green's basis in Tables~\ref{tab:matching_doublet_complex_scalar_1}- \ref{tab:matching_doublet_complex_scalar_2}.  The details of the derivation for the associated Wilson coefficients up to $\mathcal{O}(M_{\mathcal{H}_2}^{-4})$ are discussed in Appendix~\ref{app:2hdm}.

%%%%%%%%%%%%%%%%%%%%
\begin{table}[pht]
	\center
	\resizebox{\linewidth}{!}{
		\renewcommand\arraystretch{2.3}
		\begin{tabular}{| c | l |}
			\hline
			{\bf EFT operators} & \multicolumn{1}{|c|}{{\bf Matching results at scale $M_{\mathcal{H}_2} (\gg v ) $}}
			\\
			\hline \hline
			\multirow{1}{*}{$\mathcal{O}_{\phi^8} = (\phi^\dag \phi)^4$}
			& $\mathbb{C}_{\phi^8}= {\textcolor{verde}{\frac{1}{16 \pi^2} c_{\phi^8}^{\llbracket U \rrbracket}}} - \frac{1}{32 \pi^2} {\rm ln} \left( \frac{M_{\mathcal{H}_2}^2}{\mu^2} \right)c_{\phi^8}^{\llbracket U^2 \rrbracket} - {\textcolor{magenta}{\frac{1}{96 \pi^2} c_{\phi^8}^{\llbracket U^3 \rrbracket}}} + {\textcolor{blue}{\frac{1}{384 \pi^2} c^{\llbracket U^{4} \rrbracket}_{\phi^8}}}$
			\\
			\hline
			\multirow{1}{*}{$\mathcal{O}_{\phi^6}^{(1)} = (\phi^{\dag} \phi)^2 (D_{\mu} \phi^{\dag} D^{\mu} \phi)$}
			& $\mathbb{C}_{\phi^6}^{(1)}=  - {\textcolor{magenta}{\frac{1}{192 \pi^2} c_{\phi^6}^{(1), \llbracket (P_\mu U)^2 \rrbracket}}} $
			\\
			\hline
			\multirow{1}{*}{$\mathcal{O}_{\phi^6}^{(3)} = (\phi^\dagger\phi)^2 (\phi^\dagger D^2\phi + \text{h.c.}) $}
			& $\mathbb{C}_{\phi^6}^{(3)}=   {\textcolor{verde}{\frac{1}{16 \pi^2} c_{\phi^6}^{(3), \llbracket U \rrbracket}}} - \frac{1}{32 \pi^2} {\rm ln} \left( \frac{M_{\mathcal{H}_2}^2}{\mu^2} \right)c_{\phi^6}^{(3), \llbracket U^2 \rrbracket} - {\textcolor{magenta}{\frac{1}{192 \pi^2} c_{\phi^6}^{(3), \llbracket (P_\mu U)^2 \rrbracket}}} - {\textcolor{blue}{\frac{1}{384 \pi^2} c_{\phi^6}^{(3), {\llbracket U^{2} (P^2 U) \rrbracket}}}}$
			\\
			\hline
			\multirow{1}{*}{$\mathcal{O}^{(1)}_{W^2 \phi^4}=(\phi^\dag \phi)^2 W^I_{\mu\nu} W^{I\mu\nu}$} 
			& $\mathbb{C}^{(1)}_{W^2 \phi^4}= {\textcolor{magenta}{- \frac{1}{192 \pi^2} c^{(1) , \llbracket U (G_{\mu \nu})^2 \rrbracket}_{W^2 \phi^4}}} + {\textcolor{blue}{\frac{1}{480 \pi^2} c^{(1) , \llbracket U^2 (G_{\mu \nu})^2 \rrbracket}_{W^2 \phi^4}}} + {\textcolor{blue}{\frac{1}{1920 \pi^2} c^{(1) , \llbracket (U G_{\mu \nu})^2 \rrbracket}_{W^2 \phi^4}}} $
			\\
			\hline
			\multirow{1}{*}{$\mathcal{O}^{(1)}_{B^2  \phi^4} = (\phi^\dag \phi)^2 B_{\mu\nu} B^{\mu\nu}$} 
			& $\mathbb{C}^{(1)}_{B^2  \phi^4}  =  {\textcolor{magenta}{- \frac{1}{192 \pi^2} c^{(1) , \llbracket U (G_{\mu \nu})^2 \rrbracket}_{B^2 \phi^4}}} + {\textcolor{blue}{\frac{1}{480 \pi^2} c^{(1) , \llbracket U^2 (G_{\mu \nu})^2 \rrbracket}_{B^2 \phi^4}}} + {\textcolor{blue}{\frac{1}{1920 \pi^2} c^{(1) , \llbracket (U G_{\mu \nu})^2 \rrbracket}_{B^2 \phi^4}}}$
			\\
			\hline
			\multirow{1}{*}{$\mathcal{O}^{(1)}_{W B \phi^4} = (\phi^\dag \phi) (\phi^\dag \sigma^I \phi) W^I_{\mu\nu} B^{\mu\nu} $} & 
			$\mathbb{C}^{(1)}_{W B \phi^4} = {\textcolor{magenta}{- \frac{1}{192 \pi^2} c^{(1) , \llbracket U (G_{\mu \nu})^2 \rrbracket}_{W B \phi^4}}} + {\textcolor{blue}{\frac{1}{480 \pi^2} c^{(1) , \llbracket U^2 (G_{\mu \nu})^2 \rrbracket}_{W B \phi^4}}} + {\textcolor{blue}{\frac{1}{1920 \pi^2} c^{(1) , \llbracket (U G_{\mu \nu})^2 \rrbracket}_{W B \phi^4}}} $
			\\
			\hline
			\multirow{1}{*}{$\mathcal{O}^{(3)}_{\phi^4} = (D^{\mu} \phi^{\dag} D_{\mu} \phi) (D^{\nu} \phi^{\dag} D_{\nu} \phi) $} & 
			$\mathbb{C}^{(3)}_{\phi^4} = {\textcolor{verde}{\frac{1}{16 \pi^2} c_{\phi^6}^{(3), \llbracket U \rrbracket}}}$
			\\
			\hline
			\multirow{1}{*}{$\mathcal{O}^{(4)}_{\phi^4}= (D_\mu\phi^\dagger D^\mu\phi)(\phi^\dagger D^2\phi + \text{h.c.}) $} & 
			$\mathbb{C}^{(4)}_{\phi^4} = {\textcolor{verde}{\frac{1}{16 \pi^2} c_{\phi^4}^{(4), \llbracket U \rrbracket}}} + {\textcolor{blue}{\frac{1}{1920 \pi^2} c^{(4) , \llbracket (P^2 U)^2 \rrbracket}_{\phi^4}}}$
			\\
			\hline
			\multirow{1}{*}{$\mathcal{O}^{(8)}_{\phi^4} = ((D^2\phi^\dagger\phi) (D^2\phi^\dagger\phi)+\text{h.c.})$} & 
			$\mathbb{C}^{(8)}_{\phi^4} = {\textcolor{verde}{\frac{1}{16 \pi^2} c_{\phi^4}^{(8), \llbracket U \rrbracket}}} + {\textcolor{blue}{\frac{1}{1920 \pi^2} c^{(8) , \llbracket (P^2 U)^2 \rrbracket}_{\phi^4}}}$
			\\
			\hline
		\end{tabular} 
	}
	\caption{The matching results for relevant dimension-eight operators in the SMEFT for the complex doublet model with the heavy triplet integrated up to one-loop.	The terms in blue, magenta, black and green denote the contributions  in Eqn.~\ref{eq:finite} with terms proportional to : ${M_{\mathcal{H}_2}^{-4}},{M_{\mathcal{H}_2}^{-2}}, M_{\mathcal{H}_2}^0$ and $M_{\mathcal{H}_2}^2$, respectively.}
	\label{tab:matching_doublet_complex_scalar_1}
\end{table}
%%%%%%%%%%%%%%%%%%%%

%%%%%%%%%%%%%%%%%%%%
\begin{table}[htb]
	\center
	\resizebox{\linewidth}{!}{
		\renewcommand\arraystretch{1.5}
		\begin{tabular}{| c | l |}
			\hline
			{\bf EFT operators} & \multicolumn{1}{|c|}{{\bf Matching results at scale $M_{\mathcal{H}_2} (\gg v ) $}}
			\\
			\hline \hline
			\multirow{1}{*}{$\mathcal{O}^{(10)}_{\phi^4} = (D^2\phi^\dagger D^2\phi) (\phi^\dagger\phi)$} & 
			$\mathbb{C}^{(10)}_{\phi^4} = {\textcolor{verde}{\frac{1}{16 \pi^2} c_{\phi^4}^{(10), \llbracket U \rrbracket}}}$
			\\
			\hline
			\multirow{1}{*}{$\mathcal{O}^{(11)}_{\phi^4} = (\phi^\dagger D^2\phi) (D^2\phi^\dagger\phi)$} & 
			$\mathbb{C}^{(11)}_{\phi^4} = {\textcolor{verde}{\frac{1}{16 \pi^2} c_{\phi^4}^{(11), \llbracket U \rrbracket}}} + {\textcolor{blue}{\frac{1}{1920 \pi^2} c^{(11) , \llbracket (P^2 U)^2 \rrbracket}_{\phi^4}}}$
			\\
			\hline
			\multirow{1}{*}{$\mathcal{O}^{(12)}_{\phi^4} = (D_\mu\phi^\dagger \phi)(D^\mu\phi^\dagger D^2\phi) + \text{h.c.}$} & 
			$\mathbb{C}^{(12)}_{\phi^4} = {\textcolor{blue}{\frac{1}{1920 \pi^2} c^{(12) , \llbracket (P^2 U)^2 \rrbracket}_{\phi^4}}}$
			\\
			\hline
			\multirow{1}{*}{$\mathcal{O}_{B\phi^4 D^2}^{(3)} =  (\phi^{\dag} \phi) D_{\nu} B^{\mu\nu} (D_\mu \phi^\dagger \text{i} \phi + \text{h.c.})$} & 
			$\mathbb{C}_{B\phi^4 D^2}^{(3)} =  - {\textcolor{blue}{\frac{1}{960 \pi^2} c^{(1) , \llbracket U (P_\mu U) J_\mu \rrbracket}_{B \phi^4 D^2}}} $
			\\
			\hline
			\multirow{1}{*}{$\mathcal{O}^{(2)}_{W^2 \phi^2 D^2} =  (D^\mu\phi^\dagger D_\mu\phi) W_{\nu\rho}^I W^{I\nu\rho}$} & 
			$\mathbb{C}^{(2)}_{W^2 \phi^2 D^2} = - {\textcolor{blue}{\frac{1}{2880 \pi^2} c^{(2) , \llbracket (P^2 U) (G_{\rho \sigma})^2 \rrbracket}_{W^2 \phi^2 D^2}}}  $
			\\
			\hline
			\multirow{1}{*}{$\mathcal{O}^{(4)}_{W^2 \phi^2 D^2} = i \epsilon^{IJK} (D^{\mu} \phi^{\dag} \sigma^I D^{\nu} \phi) W_{\mu\rho}^J W_{\nu}^{K \rho} $} & 
			$\mathbb{C}^{(4)}_{W^2 \phi^2 D^2} = - {\textcolor{blue}{\frac{1}{720 \pi^2} c^{(4) , \llbracket (P_\mu P_\nu U) G_{\rho \mu} G_{\rho \nu} \rrbracket}_{W^2 \phi^2 D^2}}}  - {\textcolor{blue}{\frac{1}{2880 \pi^2} c^{(4) , \llbracket (P^2 U) (G_{\rho \sigma})^2 \rrbracket}_{W^2 \phi^2 D^2}}}  $
			\\
			\hline
			\multirow{1}{*}{$\mathcal{O}^{(13)}_{W^2 \phi^2 D^2} = \phi^\dagger\phi D_\mu W^{I\mu\rho} D_\nu W^{I\nu}_{\,\,\,\,\rho} $} & 
			$\mathbb{C}^{(13)}_{W^2 \phi^2 D^2} =  {\textcolor{blue}{\frac{1}{960 \pi^2} c^{(13) , \llbracket  U (J_{\mu})^2  \rrbracket}_{W^2 \phi^2 D^2}}} $
			\\
			\hline
			\multirow{1}{*}{$\mathcal{O}^{(14)}_{W^2 \phi^2 D^2} = (D_\mu\phi^\dagger\phi+\phi^\dagger D_\mu\phi) W^{I\nu\rho} D^\mu W^{I}_{\nu\rho} $} & 
			$\mathbb{C}^{(14)}_{W^2 \phi^2 D^2} = - {\textcolor{blue}{\frac{1}{720 \pi^2} c^{(14) , \llbracket (P_\mu P_\nu U) G_{\rho \mu} G_{\rho \nu} \rrbracket}_{W^2 \phi^2 D^2}}}  $
			\\
			\hline
			\multirow{1}{*}{$\mathcal{O}^{(2)}_{B^2 \phi^2 D^2} = (D^{\mu} \phi^{\dag} D_{\mu} \phi) B_{\nu\rho} B^{\nu\rho} $} & 
			$\mathbb{C}^{(2)}_{B^2 \phi^2 D^2} = - {\textcolor{blue}{\frac{1}{2880 \pi^2} c^{(14) , \llbracket (P^2 U) (G_{\rho \sigma})^2 \rrbracket}_{B^2 \phi^2 D^2}}} $
			\\
			\hline
			\multirow{1}{*}{$\mathcal{O}^{(11)}_{W^2 \phi^2 D^2} = (\phi^\dagger D_\nu\phi+D_\nu\phi^\dagger \phi) D_\mu W^{I\mu\rho} W^{I\nu}_{\,\,\,\,\rho} $} & 
			$\mathbb{C}^{(11)}_{W^2 \phi^2 D^2} =  - {\textcolor{blue}{\frac{1}{720 \pi^2} c^{(11) , \llbracket (P_\mu P_\nu U) G_{\rho \mu} G_{\rho \nu} \rrbracket}_{W^2 \phi^2 D^2}}}  $
			\\
			\hline
			\multirow{1}{*}{$\mathcal{O}^{(4)}_{B^2 \phi^2 D^2} = (D_\mu\phi^\dagger \phi+\phi^\dagger D_\mu\phi) D_\nu B^{\mu\rho} B^\nu_{\,\,\rho} $} & 
			$\mathbb{C}^{(4)}_{B^2 \phi^2 D^2} = - {\textcolor{blue}{\frac{1}{720 \pi^2} c^{(4) , \llbracket (P_\mu P_\nu U) G_{\rho \mu} G_{\rho \nu} \rrbracket}_{B^2 \phi^2 D^2}}} $
			\\
			\hline
			\multirow{1}{*}{$\mathcal{O}^{(6)}_{B^2 \phi^2 D^2} = \phi^\dagger\phi D_\mu D_\nu B^{\mu\rho} B^{\nu}_{\,\,\rho}  $} & 
			$\mathbb{C}^{(6)}_{B^2 \phi^2 D^2} =  {\textcolor{blue}{\frac{1}{960 \pi^2} c^{(6) , \llbracket  U (J_{\mu})^2  \rrbracket}_{B^2 \phi^2 D^2}}} $
			\\
			\hline
			\multirow{1}{*}{$\mathcal{O}^{(8)}_{B^2 \phi^2 D^2} = (\phi^\dagger D_\nu\phi+D_\nu\phi^\dagger\phi) D_\mu B^{\mu\rho} B^\nu_{\,\,\rho} $} & 
			$\mathbb{C}^{(8)}_{B^2 \phi^2 D^2}  =  {\textcolor{blue}{\frac{1}{960 \pi^2} c^{(8) , \llbracket  U (J_{\mu})^2  \rrbracket}_{B^2 \phi^2 D^2}}} - {\textcolor{blue}{\frac{1}{720 \pi^2} c^{(8) , \llbracket (P_\mu P_\nu U) G_{\rho \mu} G_{\rho \nu} \rrbracket}_{B^2 \phi^2 D^2}}}  $
			\\
			\hline
			\multirow{1}{*}{$\mathcal{O}^{(1)}_{W B \phi^2 D^2} = (D^{\mu} \phi^{\dag} \sigma^I D_{\mu} \phi) B_{\nu\rho} W^{I \nu\rho} $} & 
			$\mathbb{C}^{(1)}_{W B \phi^2 D^2} =  - {\textcolor{blue}{\frac{1}{2880 \pi^2} c^{(1) , \llbracket (P^2 U) (G_{\rho \sigma})^2 \rrbracket}_{W B \phi^2 D^2}}}  $
			\\
			\hline
			\multirow{1}{*}{$\mathcal{O}^{(8)}_{W B \phi^2 D^2} = (\phi^\dagger\sigma^I D^\nu\phi+D^\nu\phi^\dagger\sigma^I\phi) D_\mu B^{\mu\rho} W^{I}_{\nu\rho} $} & 
			$\mathbb{C}^{(8)}_{W B \phi^2 D^2} =  - {\textcolor{blue}{\frac{1}{720 \pi^2} c^{(8) , \llbracket (P_\mu P_\nu U) G_{\rho \mu} G_{\rho \nu} \rrbracket}_{W B \phi^2 D^2}}} $
			\\
			\hline
			\multirow{1}{*}{$\mathcal{O}^{(10)}_{W B \phi^2 D^2} = (\phi^\dagger\sigma^I\phi) D^\mu B_{\mu\rho} D_\nu W^{I\nu\rho} $} & 
			$\mathbb{C}^{(10)}_{W B \phi^2 D^2} =    {\textcolor{blue}{\frac{1}{960 \pi^2} c^{(10) , \llbracket  U (J_{\mu})^2  \rrbracket}_{W B \phi^2 D^2}}}  - {\textcolor{blue}{\frac{1}{720 \pi^2} c^{(10) , \llbracket (P_\mu P_\nu U) G_{\rho \mu} G_{\rho \nu} \rrbracket}_{W B \phi^2 D^2}}}  $
			\\
			\hline
			\multirow{1}{*}{$\mathcal{O}^{(11)}_{W B \phi^2 D^2} = (D_\nu\phi^\dagger\sigma^I\phi+\phi^\dagger\sigma^I D_\nu\phi) B_{\mu\rho} D^\mu W^{I\nu\rho} $} & 
			$\mathbb{C}^{(11)}_{W B \phi^2 D^2} = - {\textcolor{blue}{\frac{1}{720 \pi^2} c^{(11) , \llbracket (P_\mu P_\nu U) G_{\rho \mu} G_{\rho \nu} \rrbracket}_{W B \phi^2 D^2}}}  $
			\\
			\hline
			\multirow{1}{*}{$\mathcal{O}^{(13)}_{W B \phi^2 D^2} = (\phi^\dagger\sigma^I\phi) B_{\mu\rho}  D_\nu D^\mu W^{I\nu\rho} $} & 
			$\mathbb{C}^{(13)}_{W B \phi^2 D^2} =    {\textcolor{blue}{\frac{1}{960 \pi^2} c^{(10) , \llbracket  U (J_{\mu})^2  \rrbracket}_{W B \phi^2 D^2}}}  - {\textcolor{blue}{\frac{1}{720 \pi^2} c^{(13) , \llbracket (P_\mu P_\nu U) G_{\rho \mu} G_{\rho \nu} \rrbracket}_{W B \phi^2 D^2}}}  $
			\\
			\hline
			\multirow{1}{*}{$\mathcal{O}^{(1)}_{W^3 \phi^2 } = \epsilon^{IJK} (\phi^\dag \phi) W_{\mu}^{I\nu} W_{\nu}^{J\rho} W_{\rho}^{K\mu} $} & 
			$\mathbb{C}^{(1)}_{W^3 \phi^2 } = - {\textcolor{blue}{\frac{1}{1440 \pi^2}  c^{(1), \llbracket U G_{\mu \nu} G_{\nu \rho} G_{\rho \mu} \rrbracket}_{W^3 \phi^2}}}$
			\\
			\hline
		\end{tabular} 
	}
	\caption{Summary of matching results for relevant dimension-eight operators in the SMEFT for the complex doublet model with the heavy triplet integrated up to one-loop. The terms in blue, and green denote the contributions from terms proportional to  ${M_{\mathcal{H}_2}^{-4}}$ and $M_{\mathcal{H}_2}^2$ respectively.}
	\label{tab:matching_doublet_complex_scalar_2}
\end{table}
%%%%%%%%%%%%%%%%%%%%

%%%%%%%%%%%%%%%%%%%%%%%%%%%%%%%%%%%%%%%%%%%%%%%%%%%%%%%%%%%%%%%%%%%%%%%%%%%
\section{Impact of fermionic interactions with the heavy scalar field}
\label{fermionic_int_heavyfd}
%%%%%%%%%%%%%%%%%%%%%%%%%%%%%%%%%%%%%%%%%%%%%%%%%%%%%%%%%%%%%%%%%%%%%%%%%%%
In this section, we analyze the impact of introducing fermionic interactions with the heavy scalar field for the two models under consideration: the complex scalar triplet model and the complex scalar doublet model. In the BSM Lagrangian of both models, there is a linear term in the heavy field that describes its interaction with fermions (see Eqns.~\ref{triplet_fermi} and~\ref{2hdm_yukawa}). Including this interaction in the functional $B$ will play a crucial role in not only giving intriguing collider signatures for each of the two models but will also be useful in distinguishing distinctly between the two UV models from an effective theory framework. 
%%%%%%%%%%%%%%%%%%%%%%%%%%%%%%%%%%%%%%%%%%%%%%%%%%%%%%%%%%%%%%%%%%%%%%%%%%%
\subsection{Fermionic interactions mediated by the heavy triplet scalar field}
\label{fermionic_int_tripletfd}
%%%%%%%%%%%%%%%%%%%%%%%%%%%%%%%%%%%%%%%%%%%%%%%%%%%%%%%%%%%%%%%%%%%%%%%%%%%
Until now, we derived the functional $B$ from the linear term in the heavy field $\Delta$ of the scalar Lagrangian of complex triplet model (Eqn.~\ref{triplet_potential}). We now focus on the interactions between fermions and the heavy field and derive the EFT operators up to dimension-eight for the fermionic interactions after the heavy triplet scalar field has been integrated out at one-loop. The Yukawa term in the Lagrangian for the interaction of the SM leptons with the heavy field $\Delta$ is expressed as:
	\begin{align}
		{\cal L}_{Y} = Y_{\Delta_{\alpha \beta}} \ell^T_\alpha C i\sigma^2 \Delta \ell_\beta + \text{h.c.},
		\label{triplet_fermi_2}
	\end{align}
	
Considering the total Lagrangian of Eqn.~\ref{lbsmCTS}, one obtains the functional  $\widehat{B}$ as :
	\begin{eqnarray}
		\widehat{B}_i = - \mu_\Delta \tilde{H}^\dagger \tau^i H + Y_{\Delta_{\alpha \beta}}^\ast \ell_{L_\alpha}^{r \dag} \tau_i i \sigma_2 C \ell_{L_\beta}^{s \ast}  = - \mu_\Delta \tilde{H}^\dagger \tau^i H  + Y_{\Delta_{\alpha \beta}}^\ast \bar{\ell}_{L_\alpha}^{r} \tau_i \tilde{\ell}_{L_\beta}^{s} ,
	\end{eqnarray}
where indices $p,q,r,s$ denote the components of the lepton doublet $\ell_L$, $Y_\Delta$ represents a symmetric matrix in the generation space and $\tilde{\ell}_L = i \sigma_2 (\ell_L)^c$ (i.e. $\bar{\tilde{\ell}}_L = -\ell_L^T C i \sigma_2$). 
    
Like the previous section, we derive the resulting dimension-eight, seven, and six operators, emerging from the total effective Lagrangian~\ref{eq:finite}. The details are given in Appendix~\ref{fermi_int_matching_ctm}. We find that the Weinberg operator of dimension-five, the four-Fermi operator of the dimension-six Warsaw basis~\cite{Grzadkowski:2010es} from $\widehat{B}_i^\dagger \widehat{B}_i$, the dimension-seven, and -eight operators are generated and the results are depicted in Table~\ref{tab:matching_triplet_complex_scalar_3}.

	\begin{table}[htb]
		\center
		\resizebox{0.8\linewidth}{!}{
			\renewcommand\arraystretch{1.5}
			\begin{tabular}{| c | l |}
				\hline
				{\bf Dim-8 EFT operators} & \multicolumn{1}{|c|}{{\bf Matching results at scale $M_\Delta (\gg v ) $}}
				\\
				\hline \hline
				\multirow{1}{*}{$\mathcal{Q}_{l^4 \phi^2}^{(1)} = \epsilon_{ij} \epsilon_{mn} (\ell^i C \ell^m )  \phi^j \phi^n (\phi^\dagger \phi)$}
				& $\mathbb{C}_{l^4 \phi^2}^{(1)} = -\frac{1}{32 \pi^2} {\rm ln} \left( \frac{M_\Delta^2}{\mu^2} \right) c^{(1),\llbracket U^2 \rrbracket}_{l^4 \phi^2} + {\textcolor{verde}{\frac{1}{16 \pi^2} c^{(1),\llbracket U \rrbracket}_{l^4 \phi^2}}} $
				\\
				\hline
				\multirow{1}{*}{$\mathcal{Q}_{l^4 \phi^2}^{(2)} = \epsilon_{ij} \epsilon_{mn} (\ell^i C \ell^m )  \phi^j \phi^n (\phi^\dagger \phi)$}
				& $\mathbb{C}_{l^4 \phi^2}^{(2)}= -\frac{1}{32 \pi^2} {\rm ln} \left( \frac{M_\Delta^2}{\mu^2} \right) c^{(2),\llbracket U^2 \rrbracket}_{l^4 \phi^2} + {\textcolor{verde}{\frac{1}{16 \pi^2} c^{(2),\llbracket U \rrbracket}_{l^4 \phi^2}}} $
				\\
				\hline
				\multirow{1}{*}{$\mathcal{Q}_{l^4 D^2}^{(1)} = D^\nu (\bar{\ell}_p \gamma^\mu \ell_r)D_\nu (\bar{\ell}_s \gamma_\mu \ell_t)$}
				& $\mathbb{C}_{l^4 D^2}^{(1)}= {\textcolor{verde}{\frac{1}{16 \pi^2} c^{(1), \llbracket U \rrbracket}_{l^4 D^2}}} $
				\\
				\hline
				\hline
				{\bf Dim-7 EFT operators} & \multicolumn{1}{|c|}{{\bf Matching results at scale $M_\Delta (\gg v ) $}}
				\\
				\hline \hline
				\multirow{1}{*}{$\mathcal{Q}^{(1)}_{L \phi D} = \epsilon_{ij} \epsilon_{mn} \ell^i C (D^\mu \ell^j) \phi^m (D_\mu \phi^n)$}
				& $\mathbb{C}^{(1)}_{L \phi D}= {\textcolor{verde}{\frac{1}{16 \pi^2} c^{(1),\llbracket U \rrbracket}_{L \phi D}}} $
				\\
				\hline
				\multirow{1}{*}{$\mathcal{Q}^{(2)}_{L \phi D} = \epsilon_{im} \epsilon_{jn} \ell^i C (D^\mu \ell^j) \phi^m (D_\mu \phi^n) $}
				& $\mathbb{C}^{(2)}_{L \phi D}= {\textcolor{verde}{\frac{1}{16 \pi^2} c^{(2),\llbracket U \rrbracket}_{L \phi D}}} $
				\\
				\hline
				\multirow{1}{*}{$\mathcal{Q}_{L \phi} = \epsilon_{ij} \epsilon_{mn} (\ell^i C \ell^m )  \phi^j \phi^n (\phi^\dagger \phi)$}
				& $\mathbb{C}_{L \phi}= -\frac{1}{32 \pi^2} {\rm ln} \left( \frac{M_\Delta^2}{\mu^2} \right) c^{\llbracket U^2 \rrbracket}_{L \phi}  $
				\\
				\hline
				\hline
				{\bf Dim-6 EFT operators} & \multicolumn{1}{|c|}{{\bf Matching results at scale $M_\Delta (\gg v ) $}}
				\\
				\hline \hline
				\multirow{1}{*}{$\mathcal{Q}_{ll} = \left( \bar{\ell}_{L_\alpha}^{r} \gamma_\mu { \ell}_{L_\gamma}^{p} \right)_i \left( \bar{\ell}_{L_\beta}^{s} \gamma^\mu  \ell_{L_\delta}^{q}  \right)$}
				& $\mathbb{C}_{ll}= {\textcolor{verde}{\frac{1}{16 \pi^2} c^{\llbracket U \rrbracket}_{ll}}} $
				\\
				\hline
				\hline
				{\bf Dim-5 EFT operators} & \multicolumn{1}{|c|}{{\bf Matching results at scale $M_\Delta (\gg v ) $}}
				\\
				\hline \hline
				\multirow{1}{*}{$\mathcal{Q}_{\phi^2 L^2} = \epsilon_{ij} \epsilon_{mn} \phi^i \phi^m (\ell_p^j)^T C \ell_r^n$}
				& $\mathbb{C}_{\phi^2 L^2}= {\textcolor{verde}{\frac{1}{16 \pi^2} c^{\llbracket U \rrbracket}_{\phi^2 L^2}}}$
				\\
				\hline
				\hline	
			\end{tabular} 
		}
		\caption{The matching results for relevant fermionic dimension-eight, -seven, -six, -five operators in the SMEFT for the complex triplet model with the heavy triplet integrated up to one-loop. The terms in black and green denote the contributions from lower dimensional terms of total effective Lagrangian (Eqn.~\ref{eq:finite}), proportional to : $M_\Delta^0$ and $M_\Delta^2$, respectively.}
		\label{tab:matching_triplet_complex_scalar_3}
	\end{table}
	%%%%%%%%%%%%%%%%%%%%

	\begin{table}[htb]
		\center
		\resizebox{0.9\linewidth}{!}{
			\renewcommand\arraystretch{1.6}
			\begin{tabular}{| c | l |}
				\hline
				{\bf Dim-8 EFT operators} & \multicolumn{1}{|c|}{{\bf Matching results at scale $M_{\mathcal{H}_2} (\gg v ) $}}
				\\
				\hline \hline
				\multirow{1}{*}{$\mathcal{Q}_{le\phi^5} = (\phi^\dag \phi)^2 (\bar{l}_p e_r \phi)$}
				& $\mathbb{C}_{le\phi^5}= {\textcolor{magenta}{\frac{-1}{96 \pi^2} c^{\llbracket U^3 \rrbracket}_{le\phi^5}}} -\frac{1}{32 \pi^2} {\rm ln} \left( \frac{M_{\mathcal{H}_2}^2}{\mu^2} \right) c^{\llbracket U^2 \rrbracket}_{le\phi^5} + {\textcolor{verde}{\frac{1}{16 \pi^2} c^{\llbracket U \rrbracket}_{le\phi^5}}}$
				\\
				\hline
				\multirow{1}{*}{$\mathcal{Q}^{(1)}_{le \phi^3 D^2} = (D_\mu \phi^\dagger D^\mu \phi) (\bar{l}_p e_r \phi) $}
				& $\mathbb{C}^{(1)}_{le \phi^3 D^2}= {\textcolor{magenta}{\frac{-1}{96 \pi^2} c^{(1),\llbracket U^3 \rrbracket}_{le \phi^3 D^2}}} -\frac{1}{32 \pi^2} {\rm ln} \left( \frac{M_{\mathcal{H}_2}^2}{\mu^2} \right) c^{(1),\llbracket U^2 \rrbracket}_{le \phi^3 D^2} + {\textcolor{verde}{\frac{1}{16 \pi^2} c^{(1),\llbracket U \rrbracket}_{le \phi^3 D^2}}} $
				\\
				\hline
				\multirow{1}{*}{$\mathcal{Q}^{(5)}_{le\phi^3 D^2} = (\phi^\dagger D_\mu \phi) (\bar{l}_p e_r D^\mu \phi)$}
				& $\mathbb{C}^{(5)}_{le\phi^3 D^2}= {\textcolor{magenta}{\frac{-1}{96 \pi^2} c^{(5), \llbracket U^3 \rrbracket}_{le\phi^3 D^2}}} -\frac{1}{32 \pi^2} {\rm ln} \left( \frac{M_{\mathcal{H}_2}^2}{\mu^2} \right) c^{(5),\llbracket U^2 \rrbracket}_{le\phi^3 D^2} + {\textcolor{verde}{\frac{1}{16 \pi^2} c^{(5),\llbracket U \rrbracket}_{le\phi^3 D^2}}}$
				\\
				\hline
				\multirow{1}{*}{$\mathcal{Q}^{(1)}_{le W^2 \phi} =  (\bar{l}_p e_r) \phi W^I_{\mu\nu} W^{I\mu\nu}$}
				& $\mathbb{C}^{(1)}_{le W^2 \phi} =  {\textcolor{magenta}{\frac{-1}{96 \pi^2} c^{(1),\llbracket U^3 \rrbracket}_{le W^2 \phi}}} $
				\\
				\hline
				\multirow{1}{*}{$\mathcal{Q}^{(1)}_{le B^2 \phi}  = (\bar{l}_p e_r)  \phi B_{\mu\nu} B^{\mu\nu}$}
				& $\mathbb{C}^{(1)}_{le B^2 \phi} = {\textcolor{magenta}{\frac{-1}{96 \pi^2} c^{(1),\llbracket U^3 \rrbracket}_{le B^2 \phi}}}$
				\\
				\hline
				\multirow{1}{*}{$\mathcal{Q}^{(1)}_{le W B \phi} = (\bar{l}_p e_r)\tau^I \phi W^I_{\mu\nu} B^{\mu\nu}$}
				& $\mathbb{C}^{(1)}_{le W B \phi} =  {\textcolor{magenta}{\frac{-1}{96 \pi^2} c^{(1),\llbracket U^3 \rrbracket}_{le W B \phi} }} $
				\\
				\hline
				\multirow{1}{*}{$\mathcal{Q}_{l^2 e^2 \phi^2}^{(1)} = (\bar{l}_p e_r \phi) (\bar{l}_s e_t \phi) $}
				& $\mathbb{C}_{l^2 e^2 \phi^2}^{(1)} = -\frac{1}{32 \pi^2} {\rm ln} \left( \frac{M_{\mathcal{H}_2}^2}{\mu^2} \right) c^{(1),\llbracket U^2 \rrbracket}_{l^2 e^2 \phi^2} + {\textcolor{verde}{\frac{1}{16 \pi^2} c^{(1), \llbracket U \rrbracket}_{l^2 e^2 \phi^2}}}$
				\\
				\hline
				\multirow{1}{*}{$\mathcal{Q}_{l^2 e^2 \phi^2}^{(3)} =  (\bar{l}_p \gamma_\mu l_r) (\bar{e}_s \gamma^\mu e_r) \phi^\dagger \phi $}
				& $\mathbb{C}_{l^2 e^2 \phi^2}^{(3)} = {\textcolor{verde}{\frac{1}{16 \pi^2} c^{(3), \llbracket U \rrbracket}_{l^2 e^2 \phi^2}}} $
				\\
				\hline
				\multirow{1}{*}{$\mathcal{Q}_{l e q u \phi^2}^{(1)} = (\bar{l}_p^j e_r)\epsilon_{jk} (\bar{q}_s^t u_t) \phi^\dagger \phi$}
				& $\mathbb{C}_{l e q u \phi^2}^{(1)}= -\frac{1}{32 \pi^2} {\rm ln} \left( \frac{M_{\mathcal{H}_2}^2}{\mu^2} \right) c^{\llbracket U^2 \rrbracket}_{l e q u \phi^2} + {\textcolor{verde}{\frac{1}{16 \pi^2} c^{\llbracket U \rrbracket}_{l e q u \phi^2}}}$
				\\
				\hline	
				\multirow{1}{*}{$\mathcal{Q}_{l e q d \phi^2}^{(3)} =  (\bar{l}_p e_r \phi) (\bar{q}_s d_t \phi) $}
				& $\mathbb{C}_{l e q d \phi^2}^{(3)} = -\frac{1}{32 \pi^2} {\rm ln} \left( \frac{M_{\mathcal{H}_2}^2}{\mu^2} \right) c^{(3),\llbracket U^2 \rrbracket}_{l e q d \phi^2} + {\textcolor{verde}{\frac{1}{16 \pi^2} c^{(3), \llbracket U \rrbracket}_{l e q d \phi^2}}}$
				\\
				\hline
				\multirow{1}{*}{$\mathcal{Q}_{l^2 e^2 D^2}  =  (\bar{l}_p e_r \phi) (\bar{q}_s d_t \phi) $}
				& $\mathbb{C}_{l^2 e^2 D^2}  = {\textcolor{verde}{\frac{1}{16 \pi^2} c^{\llbracket U \rrbracket}_{l^2 e^2 D^2}}}$
				\\
				\hline \hline
				{\bf Dim-6 EFT operators} & \multicolumn{1}{|c|}{{\bf Matching results at scale $M_{\mathcal{H}_2} (\gg v ) $}}
				\\
				\hline \hline
				\multirow{1}{*}{$\mathcal{Q}_{e\phi} = (\phi^\dag\phi) (\bar{l}_p e_r \phi)$}
				& $\mathbb{C}_{e\phi}= -\frac{1}{32 \pi^2} {\rm ln} \left( \frac{M_{\mathcal{H}_2}^2}{\mu^2} \right) c^{\llbracket U^2 \rrbracket}_{e\phi} (1+ m_H^2) + {\textcolor{verde}{\frac{1}{16 \pi^2} c^{\llbracket U \rrbracket}_{e\phi}}}(1+m_H^2)$
				\\
				\hline
				\multirow{1}{*}{$\mathcal{Q}_{le} = \left(\bar{l}_p \gamma_\mu l_r\right)\left(\bar{e}_s \gamma^\mu e_t\right) $}
				& $\mathbb{C}_{le}= {\textcolor{verde}{\frac{1}{16 \pi^2} c^{\llbracket U \rrbracket}_{le}}} $
				\\
				\hline
				\multirow{1}{*}{$\mathcal{Q}_{ledq} = (\bar{l}_p^j e_r) (\bar{d}_s q_t^j) $}
				& $\mathbb{C}_{ledq}= {\textcolor{verde}{\frac{1}{16 \pi^2} c^{(1),\llbracket U \rrbracket}_{ledq}}} $
				\\
				\hline
				\multirow{1}{*}{$\mathcal{Q}_{lequ}^{(1)} =(\bar{l}_p^j e_r) \epsilon_{jk} (\bar{q}_s^k u_t)$}
				& $\mathbb{C}^{(1)}_{lequ}= {\textcolor{verde}{\frac{1}{16 \pi^2} c^{(1),\llbracket U \rrbracket}_{lequ}}}$
				\\
				\hline
				\hline
			\end{tabular} 
		}
		\caption{The matching results for relevant fermionic dimension-eight and dimension-six operators in the SMEFT for the complex doublet model with the heavy triplet integrated up to one-loop. The terms in magenta, black and green denote the contributions from lower dimensional terms of total effective Lagrangian (Eqn.~\ref{eq:finite}), proportional to : ${M_{\mathcal{H}_2}}, M_{\mathcal{H}_2}^0$ and $M_{\mathcal{H}_2}^2$, respectively.}
		\label{tab:matching_doublet_complex_scalar_3}
	\end{table}
	%%%%%%%%%%%%%%%%%%%%

%%%%%%%%%%%%%%%%%%%%%%%%%%%%%%%%%%%%%%%%%%%%%%%%%%%%%%%%%%%%%%%%%%%%%%%%%%%
\subsection{Fermionic interactions mediated by the heavy doublet scalar field}
\label{fermionic_int_doubletfd}
%%%%%%%%%%%%%%%%%%%%%%%%%%%%%%%%%%%%%%%%%%%%%%%%%%%%%%%%%%%%%%%%%%%%%%%%%%%
We now summarize the description of fermionic interactions with the heavy complex Higgs doublet in the EFT framework by truncating the EFT expansion up to dimension-eight operators. The Yukawa Lagrangian in the 2HDM model is given by :
	\begin{align}
		\L_{\text{Yuk}} &=  - Y_{\mathcal{H}_2}^{(e)} \bar{\ell}_L \tilde{\mathcal{H}}_2 e_R - Y_{\mathcal{H}_2}^{(u)} \bar{q}_L \mathcal{H}_2 u_R - Y_{\mathcal{H}_2}^{(d)} \bar{q}_L \tilde{\mathcal{H}}_2 d_R + \text{h.c.} \nonumber \\
		&=  - Y_{\mathcal{H}_2}^{(e)} \mathcal{H}_2^\dagger \widehat{\ell}_L e_R - Y_{\mathcal{H}_2}^{(u)} \bar{q}_L \mathcal{H}_2 u_R - Y_{\mathcal{H}_2}^{(d)} \mathcal{H}_2^\dagger \widehat{q}_L d_R + \text{h.c.},
		\label{2hdm_yukawa_1}
	\end{align}
where we define $\widehat{\ell}_L \equiv - i \sigma_2 (\bar{\ell}_L)^T$\footnote{Explicitly writing the components of the $SU(2)$ doublet for the 1st generation lepton doublet gives
		$
		\widehat{\ell}_L
		=
		%\begin{pmatrix}
		(	-\bar{e}_L, %\\
			{\bar{\nu}}_{e_L})^T
		%\end{pmatrix}
		$.
This implies the relation  $\bar{e}_R {\widehat{\ell}_L}^{\dagger} H_j = \bar{e}_R {\tilde{H}_j}^{\dagger} \ell_L$, along with its Hermitian conjugate, $H_j^{\dagger} \widehat{\ell}_L e_R = \bar{\ell}_L \tilde{H}_j e_R$.} and $\widehat{q}_L \equiv - i \sigma_2 (\bar{q}_L)^T$.
We derive the functional $\widehat{B}$ from the linear term in the heavy field $\mathcal{H}_2$ of the total Lagrangian (Eqn.~\ref{2hdm_lagrangian}), which  is given by 
	\begin{equation}
		\widehat{B}_i = \eta_H (H^\dag H) \tilde{H_i} - \Big(\left( Y_{\mathcal{H}_2}^{(e)}\right) \widehat{\ell}_L e_R + \left( Y_{\mathcal{H}_2}^{(u)}\right) \bar{u}_R q_L + \left( Y_{\mathcal{H}_2}^{(d)}\right) \widehat{q}_L d_R \Big)_i .
	\end{equation}
Note that, $\hat{B}$ has mass dimension three and $\left\{ Y_{\mathcal{H}_2}^{(e)}, Y_{\mathcal{H}_2}^{(u)},Y_{\mathcal{H}_2}^{(d)} \right\}$ are dimensionless couplings. The corresponding classical solution for the heavy doublet field $\mathcal{H}_2$ is given by :
\begin{equation}
	\left(\mathcal{H}_{\rm 2,c}\right)_i = \frac{\widehat{B}_i}{M_{\mathcal{H}_{2}}^2} + \frac{(P^2 - U)_{ij}}{M_{\mathcal{H}_{2}}^4} \widehat{B}_j + \frac{(P^2 - U)_{ij}^2}{M_{\mathcal{H}_{2}}^6} \widehat{B}_j + \cdots 
	\label{doubletsol3}
\end{equation}
Substituting this solution back into action, we derive the possible higher dimensional fermionic operators at dimension-eight (in Murphy basis~\cite{Murphy:2020rsh}) and at dimension-six (in Warsaw basis~\cite{Grzadkowski:2010es}). The results are summarized in Table~\ref{tab:matching_doublet_complex_scalar_3}, and details of the computation are given in Appendix~\ref{fermi_2hdm}. Although we present the operators considering the leptonic contribution only in $\widehat{B}$, in a similar way, operators for the quark involving interactions with the heavy doublet can be obtained. The non-redundant operator bases can be derived using the relations provided in Appendix~\ref{eom_fdrefinition}.

%%%%%%%%%%%%%%%%%%%%%%%%%%%%%%%%%%%%%%%%%%%%%%%%%%%%%%%%%%%%%%%%%%%%%%%%%%%%%%%%%%%%%%%%%
\section{Relevance of dimension-8 effective operators: An EWPO case study}
\label{sec:obsd8}
%%%%%%%%%%%%%%%%%%%%%%%%%%%%%%%%%%%%%%%%%%%%%%%%%%%%%%%%%%%%%%%%%%%%%%%%%%%%%%%%%%%%%%%%%

Different observables and precision measurements serve as powerful tools to differentiate between UV scenarios when analyzed through matched EFT frameworks. In this context, dimension-eight effects offer a diverse and rich phenomenology with significant information. We consider electroweak precision observables to explore EFT-based discrimination of the UV scenarios. The oblique electroweak precision observables \cite{Peskin:1991sw,Peskin:1990zt,Bagnaschi:2022whn,Asadi:2022xiy,Corbett:2021eux} are a relevant phenomenological sector having supreme sensitivity potential at future lepton colliders such as FCC-ee with a potential GigaZ option. The high precision that can be expected in these environments ties to the relevance of the dimension-eight terms discussed in this work. As we are considering different $SU(2)_L$ gauge representations for our extensions, it is clear that oblique corrections will be sensitive to the discrimination between the models of Secs.~\ref{subsec:triplet} and~\ref{subsec:doublet}. What is less clear is the precise sensitivity to new physics scales in these scenarios when considering relevant dimension-eight terms as part of the matching. Furthermore, the existence of blind direction serves as a key motivation to consider a multitude of differential measurements of collider processes~\cite{Belvedere:2024wzg} including those that carry a genuine dimension-eight sensitivity (e.g.~\cite{Corbett:2023qtg, Hamoudou:2022tdn,  Dawson:2022cmu,DasBakshi:2024krs}). We leave a more detailed investigation of these latter points for future work.

At the dimension-six level, these parameters are expressed in the Warsaw basis as:
\begin{equation}
	\frac{g g'}{16 \pi} \Delta S = \frac{v^2}{\Lambda^2} C_{\phi WB} , \quad  - \frac{g^2 g'^2}{2 \pi \left(g^2 + g'^2\right)} \Delta T  = \frac{v^2}{\Lambda^2} C_{\phi D} \, ,
\end{equation}
while in the SILH basis these relations are given by $\alpha \Delta T = \bar C_T$ and $\alpha \Delta S= 4 s_W^2 (\bar C_W + \bar C_B)$, at leading order.
Considering the parameters having effects up to dimension-eight operators, we have
\begin{align} \label{STmapWC}
	\frac{g g'}{16 \pi} \Delta S &= \frac{v^2}{\Lambda^2} \Big( C_{\phi WB} + \frac{v^2}{2 \Lambda^2} C_{WB \phi^4}^{(1)}\Big) \equiv \frac{v^2}{\Lambda^2}\tilde{C}_{\phi WB}  \nonumber,\\  - \frac{g^2 g'^2}{2 \pi \left(g^2 + g'^2\right)} \Delta T & = \frac{v^2}{\Lambda^2} \Big( C_{\phi D} + \frac{v^2}{\Lambda^2} C_{\phi^6}^{(1)} \Big) \equiv \frac{v^2}{\Lambda^2}\tilde{C}_{\phi D},  \nonumber \\ \frac{g^2}{16 \pi} \Delta U & = \frac{v^4}{\Lambda^4} C_{W^2 \phi^4}^{(3)}  .
\end{align}
The oblique parameter $U$ receives the first non-negligible contribution at dimension-eight.

%%%%%%%%%%%%%%%%%%%%%%%%%%%%%%%%%%%%%%%%%%%%%%%%%%%%%%%
\begin{figure}[!b]
	\centering~
	\includegraphics[width=4.8cm, height=4.2cm]{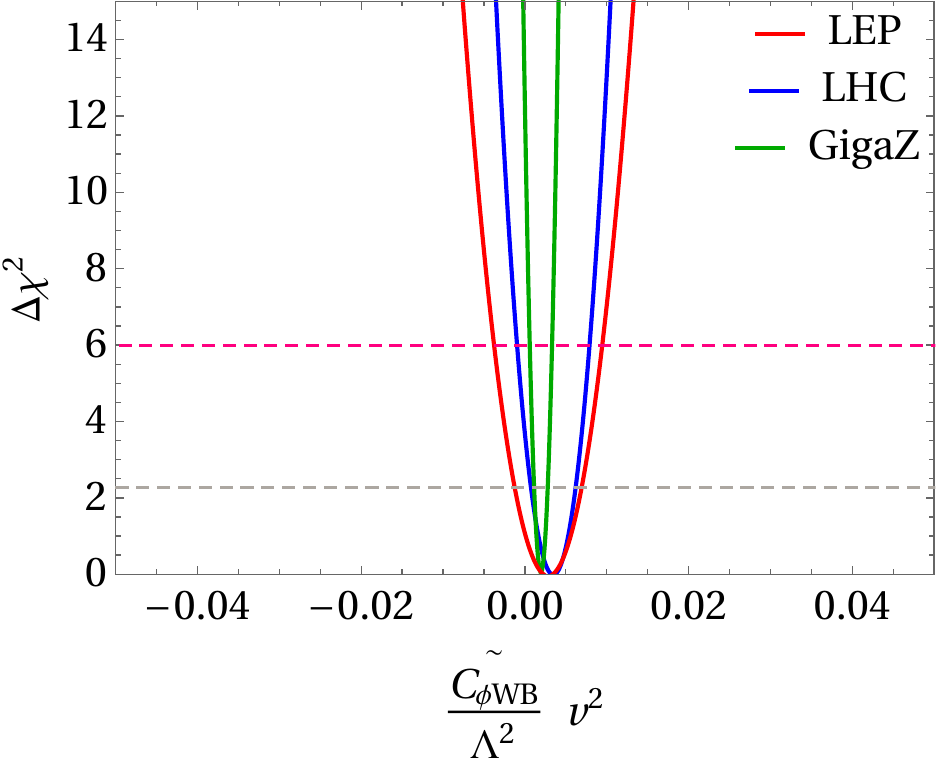}\hfill
	\includegraphics[width=4.8cm, height=4.2cm]{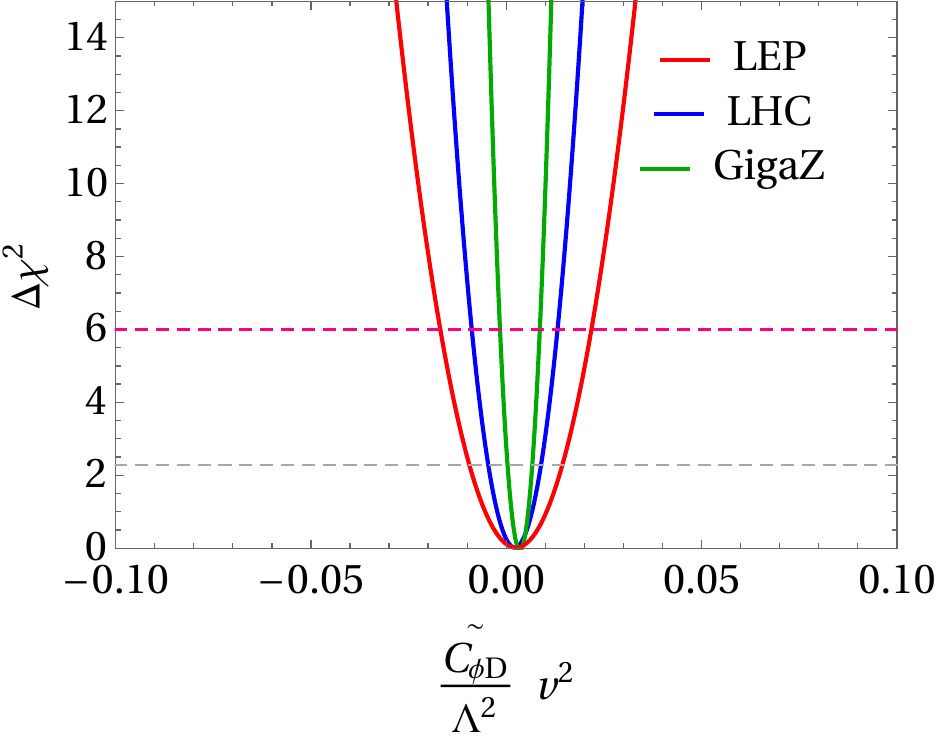}\hfill
	\includegraphics[width=4.8cm, height=4.2cm]{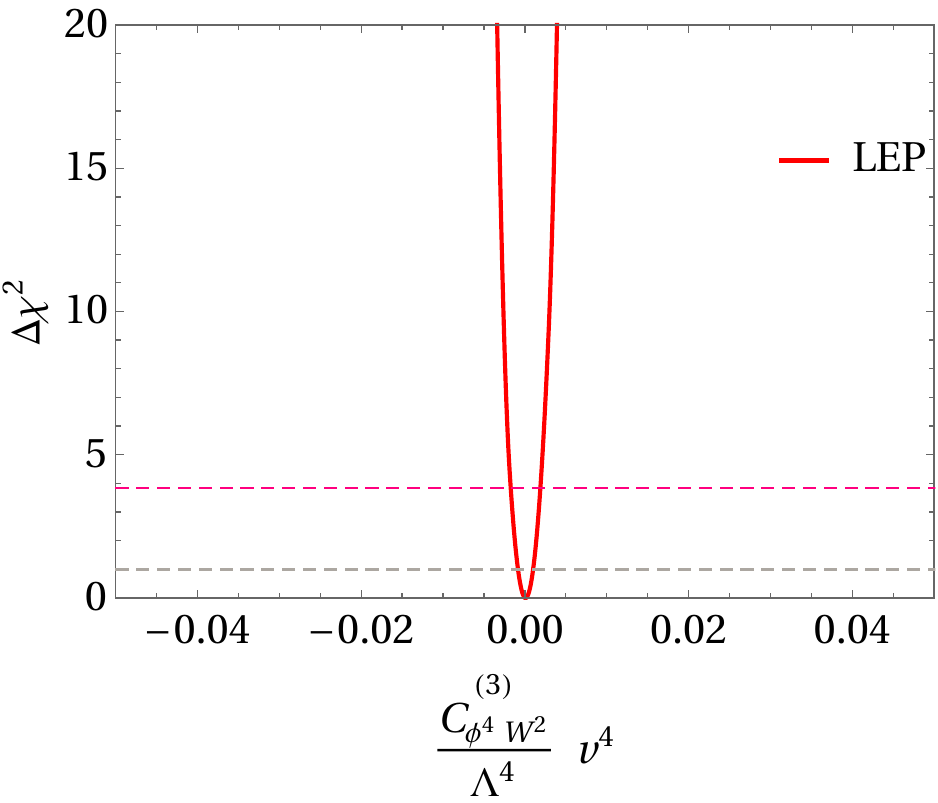}
	\caption{One-dimensional fit and projections for the oblique parameters expressed in terms of coefficients ${\tilde{C}_{\phi W B}^{(1)}}/{\Lambda^2},{\tilde{C}_{\phi D}}/{\Lambda^2},{C_{ \phi^4 W^2}^{(3)}}/{\Lambda^4}$.}
	\label{SnTparameters_WC}
\end{figure}
%%%%%%%%%%%%%%%%%%%%%%%%%%%%%%%%%%%%%%%%%%%%%%%%%%%%%%%

%%%%%%%%%%%%%%%%%%%%%%%%%%%%%%%%%%%%%%%%%%%%%%%%%%%%%%%
\begin{figure}[!t]
	\centering
	\parbox{7.5cm}{\includegraphics[width=7.2cm, height=6.5cm]{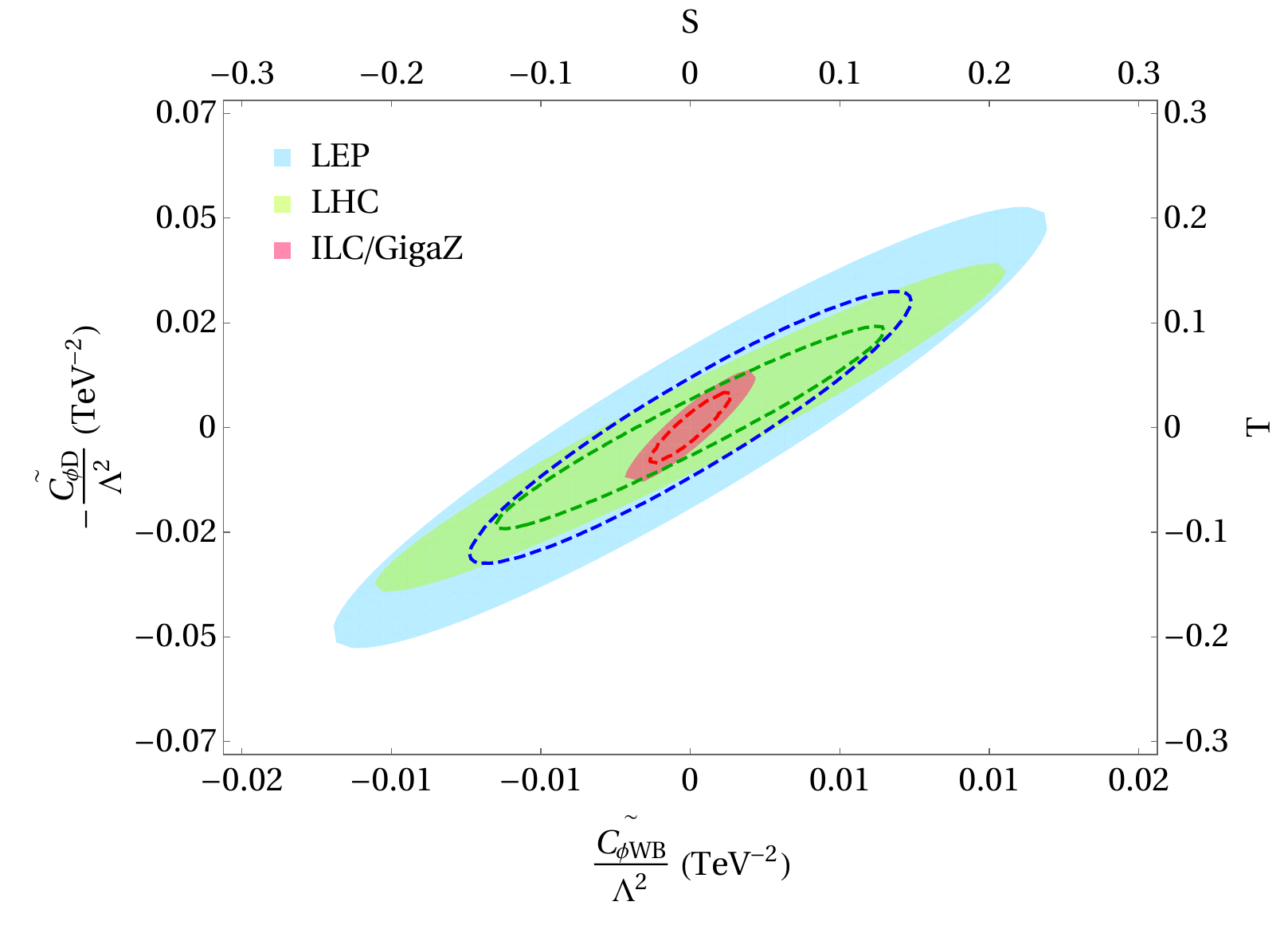}}\hfill
    \parbox{7.4cm}{
    \vspace{.20cm}
	\caption{Two-dimensional fit for the oblique parameters expressed in the $S-T$ plane for terms of the present uncertainties (blue), the LHC (green) and the ILC/GigaZ prospects (red). These are equivalent to the coefficients ${\tilde{C}_{\phi W B}^{(1)}}/{\Lambda^2}$ and ${\tilde{C}_{\phi D}}/{\Lambda^2}$ via the map in Eqn.~\ref{STmapWC}. The shaded regions show the $2\sigma$ preferred regions, while the dashed curves mark the $1\sigma$ contours.}
    \label{SnTparameters}}
\end{figure}
%%%%%%%%%%%%%%%%%%%%%%%%%%%%%%%%%%%%%%%%%%%%%%%%%%%%%%%

To quantify the relevance of our matching computations, we adopt the GFitter confidence levels (CLs) of~\cite{Baak:2014ora} and take on board the future collider option through GigaZ. Especially when constraints are high quality, the results of going beyond the leading order approximation in the matching computation become relevant. In parallel, the level of agreement and disagreement with the full theory computation then informs associated systematic effects quantitatively.

%%%%%%%%%%%%%%%%%%%%%%%%%%%%%%%%%%%%%%%%%%%%%%%%%%%%%%%
	\begin{table}[!t]
		\centering
		\resizebox{\linewidth}{!}{
			\renewcommand\arraystretch{1.1}{
			\begin{tabular}{|c|c|c|c|c|}
				\hline
				& \multicolumn{4}{c|}{EWPO 95\% CL allowed range} \\
				& \multicolumn{2}{c|}{LHC} & \multicolumn{2}{c|}{FCC-ee (GigaZ)} \\\hline
				$i$ & $\frac{v^2}{\Lambda^2}{C_i}$ & $\frac{v^2}{\Lambda^2}{\tilde{C}_i}$ &  $\frac{v^2}{\Lambda^2}{C_i}$& $\frac{v^2}{\Lambda^2}{\tilde{C}_i}$ \\\hline
				$\phi WB$   &$[-0.0004,0.00068]$ &$[-0.0014,0.0076]$ &    $[-0.0010,0.0029]$&$[-0.0015,0.0039]$   \\[+0.1cm] \hline
				$\phi D$   & $[-0.0076,0.01020]$ &$[-0.0098,0.0120]$ &     $[-0.0011,0.0063]$&$[-0.0019,0.0084]$  \\[+0.1cm]
                \hline 
		\end{tabular}}}
		\caption{95\% CL constraints on the model-independent effective couplings contributing to the oblique parameters: (left-column) only dimension-six, and  (right-column) both the dimensions -six and -eight contributions for both LHC and FCC-ee (GigaZ) scenarios. Constraints on each coefficient of the first column are obtained after marginalising over the other coefficients. The $U$ parameter constrains the dimension-eight operator coefficient $\frac{v^4}{\Lambda^4}\, C^{(3)}_{W^2\phi^4}$ to [-0.0021,0.0019] at 95\% CL from current GFitter results~\cite{Baak:2014ora}. \label{ranges_ewpo}}
	\end{table}
%%%%%%%%%%%%%%%%%%%%%%%%%%%%%%%%%%%%%%%%%%%%%%%%%%%%%%%

%%%%%%%%%%%%%%%%%%%%%%%%%%%%%%%%%%%%%%%%%%%%%%%%%%%%%%%
\begin{figure}[t!]
	\centering
	\subfigure[\label{fig:a}]{\includegraphics[width=7.4cm]%, height=6.2cm]
    {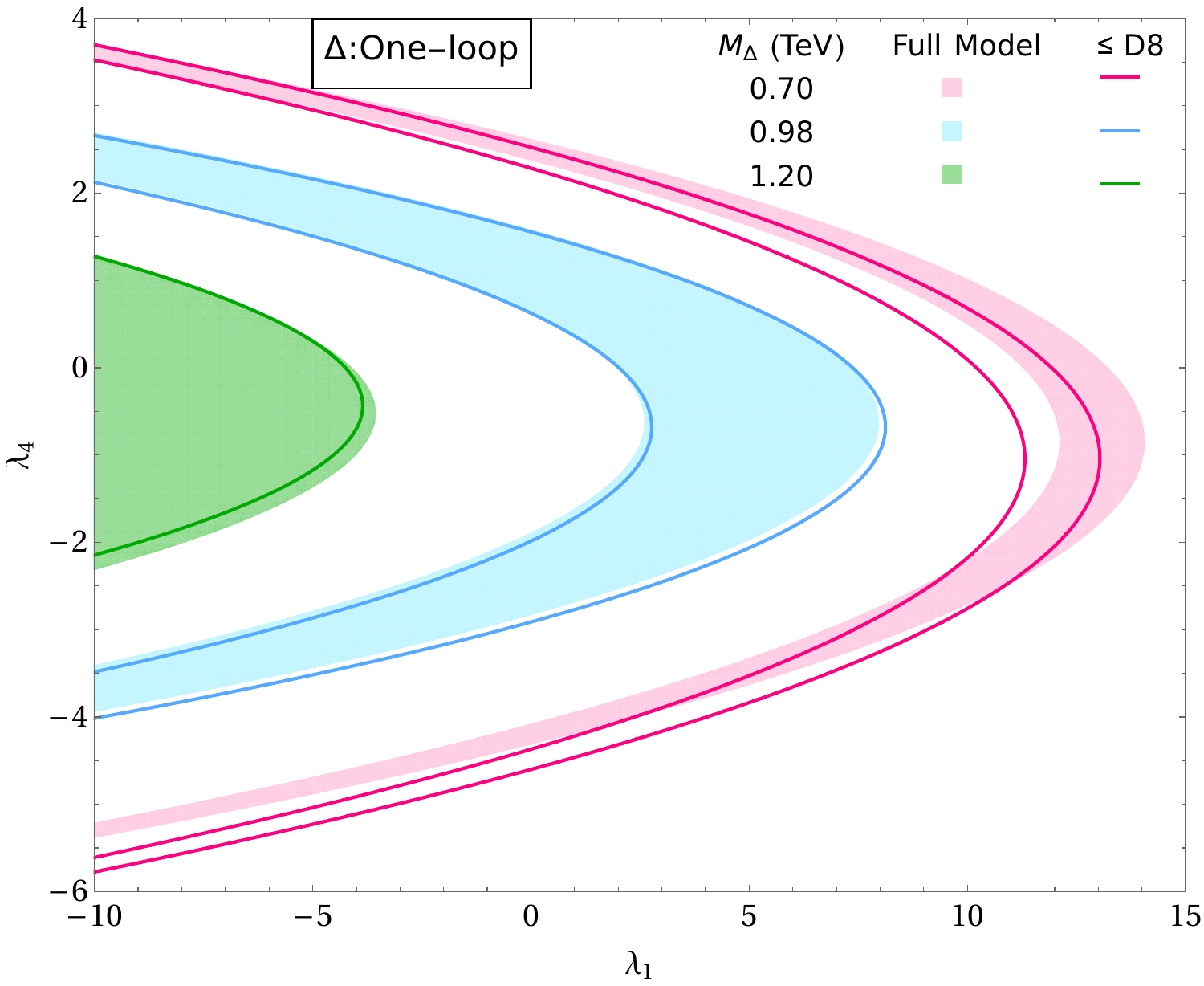}}\hfill
	\subfigure[\label{fig:b}]{\includegraphics[width=7.4cm]% height=6.2cm]
    {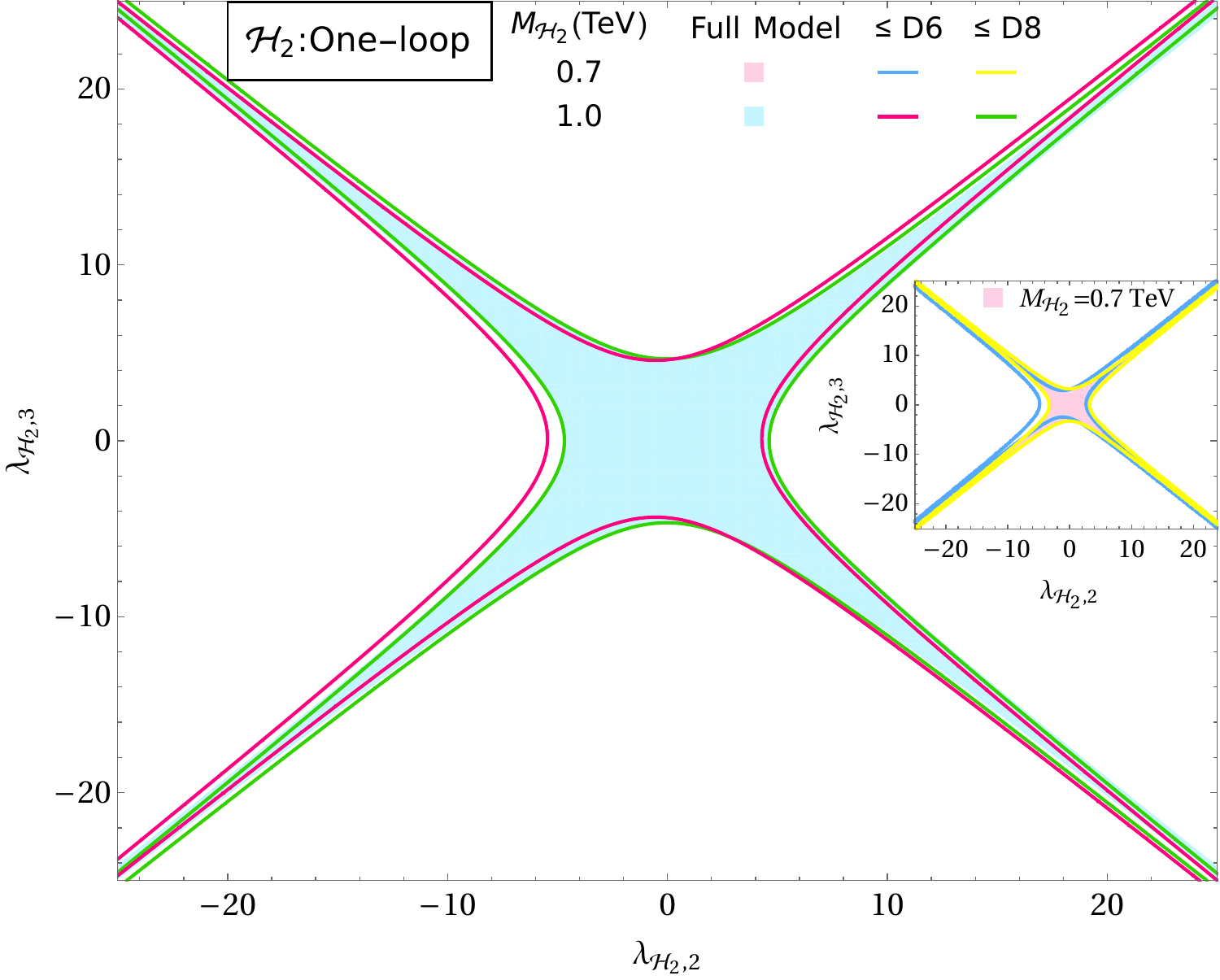}}
	\caption{(a) The 95\% CL GigaZ exclusion regions in $\lambda_1-\lambda_4$ plane for different values of heavy complex triplet masses, $M_\Delta$ (pink for $M_\Delta=0.7~{\rm TeV}$, blue for $M_\Delta=0.98~{\rm TeV}$, green for $M_\Delta=1~{\rm TeV}$), where we have fixed $\mu_\Delta/M = 0.1,\lambda_2 = \lambda_3 = 1.8$. The shaded region depicts the allowed region from the oblique parameter constraints corresponding to a given triplet mass in the full model, and the contours depict the SMEFT contribution up to dimension-eight after matching to the complex triplet model up to one-loop. (b) The 95\% CL GigaZ exclusion regions in $\lambda_{\mathcal{H}_2,2}-\lambda_{\mathcal{H}_2,3}$ plane for different values of heavy complex doublet masses, $M_{\mathcal{H}_2,2}$, where we have fixed $\eta_{\mathcal{H}}= 1.2,\eta_{\mathcal{H}_2} = 0.2,\lambda_{\mathcal{H}_2,1} = -1.4$. The blue and pink contours (for $M_{\mathcal{H}_2,2}=0.7~{\rm TeV}$ and $1~{\rm TeV}$ respectively) show the preferred regions for contributions from dimension-six operators, yellow and pink contours  (with $M_{\mathcal{H}_2,2}=0.7~{\rm TeV}$ and $1~{\rm TeV}$ respectively) represent contributions from dimension-eight operators.}
    \label{lam1_lam4_ctm}
\end{figure}
%%%%%%%%%%%%%%%%%%%%%%%%%%%%%%%%%%%%%%%%%%%%%%%%%%%%%%%

The results of a generic fit to the Wilson coefficients (WCs) at the LEP, expected LHC and ILC/GigaZ are shown in Figs.~\ref{SnTparameters_WC} and \ref{SnTparameters} for completeness. A modest enhancement in precision is observed for the LHC, represented by the green ellipse in Fig.~\ref{SnTparameters} when comparing direct measurements with the SM expectation. In contrast, the GigaZ option (red ellipse in Fig.~\ref{SnTparameters}) demonstrates a significantly greater improvement in precision and sensitivity by an order of magnitude. We present the $95\%$ marginalized limits on the dimension-six and eight operators for LHC, and GigaZ runs in Table~\ref{ranges_ewpo}.\footnote{We note that a TeraZ factory at the FCC-ee will produce $10^3$ more $Z$ bosons than GigaZ and can, therefore, in principle, constrain the oblique parameters at the level of $10^{-5}$. A plethora of electroweak contributions need to be under control within the SM to fully exploit the BSM sensitivity that such an environment provides. Hence, we will focus on the more conservative GigaZ option in the following.} Note that at dimension-six, we have only two WCs contributing, whereas at dimension-eight, we have three, which can be combined with two dimension-six coefficients, and their impact is captured in Eqn.~\ref{STmapWC}. Thus, in completely model-independent fits, it is difficult to assess the individual presence of dimension-eight operators, and marginalization leads to serious degrading of the sensitivity. This observation is not new: adding more free parameters to the analysis of a fixed data set introduces blind directions, and sensitivity, to some degree artificially and without motivation, is watered down. The relevant observation here is that moving forward to the precision frontier of the HL-LHC or a future Lepton collider, the contextualization of EFT constraints with matching calculations at high precision will be necessary to exploit the quality of available data fully. Generic fits that leave all parameters free can inform this program through the available likelihoods. Still, taken in isolation, they provide neither a realistic nor an adequate measure of precision. Only matching to specific models provides clarity on their scale and effects.

We investigate the deviation of the EFT with the full theory after performing the one-loop level matching to the SMEFT up to dimension-eight. When the relatively conservative projections of the GigaZ EW precision constraints are contrasted with the model-specific circumstances of the models discussed in this paper, we obtain Fig.~\ref{lam1_lam4_ctm}. There, the dimension-six contributions contain terms in the linear dimension-six WCs, where the corrections up to dimension-eight operators include diemsnions -six and -eight in quaidratic and linear, respectively. 

We first consider the complex triplet model and study the impact on the quartic couplings $\lambda_1$ and $\lambda_4$ that is illustrated in Fig.~\ref{fig:a}. Here, we assume the dimensionful trilinear coupling $\mu_\Delta$ is much less than the mass of the triplet scalar: $\mu_\Delta/M_\Delta=0.1$, and also set $\lambda_2=\lambda_3=1.8$. It is quite evident that the effective theory approaches the full theory for a larger mass of the integrated-out fields and for weaker couplings (note that scales and couplings are related in our parameter choice). For example, for lower mass $M_\Delta \sim 700$ GeV effective theory does not emerge as the correct approximate version of the full theory, especially within the limitations of the EWPD. Thus, we need to add higher mass dimensional operators. Including dimenson-eight interactions is also important to constrain the model couplings, e.g., $\lambda_1$, where no sufficient sensitivity on the model side is given for lower dimension effective operators. Similar observations hold for the 2HDM, see Fig.~\ref{fig:b}. Here, we see clearly that including only dimension-six terms fails to approximate the full theory for the precision offered by GigaZ. Only the inclusion of properly matched dimension-eight contributions enables an adequate comparison with the full model over a broad range of masses and couplings. The precise determination of the microscopic parameters of the BSM model using model-independent techniques, therefore, crucially rests on precise matching.

%%%%%%%%%%%%%%%%%%%%%%%%%%%%%%%%%%%%%%%%%%%%%%%%%%%%%%%%%%%%%%%%%%%%%%%%%%%%%%%%%%%%%%%%%
\section{Conclusions}\label{sec:conclusions}
%%%%%%%%%%%%%%%%%%%%%%%%%%%%%%%%%%%%%%%%%%%%%%%%%%%%%%%%%%%%%%%%%%%%%%%%%%%%%%%%%%%%%%%%%
We explored the implications of dimension-eight operators within the SMEFT framework by systematically deriving their contributions from two extensions of the Standard Model: the Complex Scalar Triplet Model and the Two Higgs Doublet Model. Using the Heat Kernel method, a one-loop matching approach was utilized to obtain the effective operators and their corresponding Wilson coefficients, shedding light on the interplay between heavy scalar fields and light-sector interactions.  The resulting effective interactions, computed up to dimension-eight and one-loop, are systematically displayed in Tables~\ref{tab:matching_triplet_complex_scalar_1} to \ref{tab:matching_doublet_complex_scalar_3}. Furthermore, Tables~\ref{Tab:nonredundant_basis_1} and~\ref{Tab:nonredundant_basis_2} provide a mapping of these interactions to a non-redundant basis, facilitating their application in further theoretical and phenomenological studies.

Our results show the critical role of dimension-eight operators in enhancing the precision of effective field theories, especially in cases where lower-dimensional operators fail to encapsulate the nuances of higher-order effects. Furthermore, we demonstrated the relevance of these contributions to electroweak precision observables, providing quantitative insights into their impact on parameters such as  $\Delta S$,  $\Delta T$, and  $\Delta U$. The analysis evidences how incorporating dimension-eight operators increases the sensitivity of potential deviations from the Standard Model predictions.

We also investigated the fermionic interactions mediated by heavy scalar fields. We identified and categorized the effective operators, including those affecting four-fermion interactions and neutrino mass terms. These operators include corrections to known dimension-six terms and generate new contributions that first appear at dimension-seven and -eight. We find that fermionic interactions mediated by heavy scalar fields enrich the structure of the effective field theory and introduce distinctive signatures in precision observables and collider processes, enabling potential discrimination between different ultraviolet-complete models. Thus, this shows a non-trivial interplay between scalar and fermionic sectors in effective descriptions of BSM physics.

The study provides a framework for distinguishing between UV-complete models utilising precision measurements and differential collider observables. As future experiments such as the high-luminosity LHC and prospective lepton colliders, e.g. the FCC-ee or ILC, promise unprecedented sensitivity, our findings lay a foundation for probing new physics through higher-order effects. It also serves as a stepping stone for extending EFT methodologies to include even higher-dimensional operators, further refining our ability to interpret physics signals beyond the Standard Model.

%\bigskip
%%%%%%%%%%%%%%%%%
\subsubsection* {\bf{Acknowledgments}}
%\noindent {\bf{Acknowledgments}} --- 
N.A. acknowledges the Indian Institute of Technology Kanpur (IITK) for an institute assistantship.
The work of T.B.~is supported by the Science and Engineering Research Board (SERB, Government of India) grant under the Project SERB/PHY/2023799. J.C. acknowledges support from CRG project, SERB, India.
C.E.~is supported by the STFC under grant ST/X000605/1 and by the Leverhulme Trust under Research Project Grant RPG-2021-031 and Research Fellowship RF-2024-300$\backslash$9. C.E. is further supported by the Institute for Particle Physics Phenomenology Associateship Scheme.
M.S.~is supported by the STFC under grant ST/P001246/1.
%%%%%%%%%%%%%%%%%%%%%%%%%%%%%%%%%%%%%%%%%%%%%%%%%%%%%

\appendix	
	\section{Matching Details}	
	\label{matching}

	\subsection{Case I: Integrating out Complex Triplet Scalars}	
	\label{matching_d8_ctm}
	In this appendix, we present the computation details after applying the generic procedure described in Sec.~\ref{sec:models} to the particular case of the complex triplet model.
	We can express terms in the Lagrangian Eqn.~\ref{triplet_potential} as
	\begin{equation}
		\begin{split}
			\text{Tr}[\Delta^\dagger \Delta] &= \text{Tr}[\Delta^*_i \Delta_j \tau^i \tau^j] = \dfrac{1}{2}\Delta^*_i \Delta_i = \, \dfrac{1}{2}|\Delta|^2, \\
			\left(\text{Tr}[\Delta^\dagger \Delta]\right)^2 &= \left(\text{Tr}[\Delta^*_i \Delta_j \tau^i \tau^j]\right)^2 = \dfrac{1}{4}\left(\Delta^*_i \Delta_i \right)^2 = \, \dfrac{1}{4}|\Delta|^4, \\
			\text{Tr}\left[\left(\Delta^\dagger \Delta\right)^2\right] & = \Delta^*_i \Delta_j \Delta^*_k \Delta_l \text{Tr} [\tau^i \tau^j \tau^k \tau^l] \\
			& = \dfrac{1}{8} \Delta^*_i \Delta_j \Delta^*_k \Delta_l (\delta_{ij} \delta_{kl} + \delta_{il} \delta_{jk} - \delta_{ik} \delta_{jl})\\
		%	& = \dfrac{1}{4} (\Delta^*_i \Delta_i)(\Delta^*_j \Delta_j) -\dfrac{1}{8}(\Delta^*_i \Delta^*_i)(\Delta_j \Delta_j)\\
			& = \dfrac{1}{4}|\Delta|^4  -\dfrac{1}{8}(\Delta^\ast)^2(\Delta)^2.
		\end{split}
        \label{triplet_decomposition}
	\end{equation}

We obtain the classical solution for the heavy field using the covariant derivative expansion method. In analogy to the basis in $\left(\Delta, \Delta^*\right)$ defined in Eqn.~\ref{tripletsol1}, we define a basis for $B$ such that :
	\begin{align}
		\begin{pmatrix}
			\Delta_i \\ \Delta^\ast_i 
		\end{pmatrix}
		=  &
		\frac{1}{M_\Delta^2}
		\begin{pmatrix}
			B_i \\ B^\ast_i 
		\end{pmatrix}
		+ 
		\frac{1}{M_\Delta^4}
		\left(  \left( P^2 \delta_{ij} \right) 
		\begin{pmatrix}
			\mathbb{1} & 0 \\
			0 & \mathbb{1} 
		\end{pmatrix}
		-
		\begin{pmatrix}
			\left(U_{11}\right)_{ij} & \left(U_{12}\right)_{ij} \\
			\left(U_{21}\right)_{ij} & \left(U_{22}\right)_{ij}
		\end{pmatrix} 
		\right)
		\begin{pmatrix}
			B_j \\ B^\ast_j 
		\end{pmatrix} \nonumber \\ 
		+ &
		\frac{1}{M_\Delta^6}
		\left(  \left( P^2 \delta_{ij} \right) 
		\begin{pmatrix}
			\mathbb{1}  & 0 \\
			0 & \mathbb{1}  
		\end{pmatrix}
		-
		\begin{pmatrix}
			\left(U_{11}\right)_{ij} & \left(U_{12}\right)_{ij} \\
			\left(U_{21}\right)_{ij} & \left(U_{22}\right)_{ij}
		\end{pmatrix} 
		\right)^2
		\begin{pmatrix}
			B_j \\ B^\ast_j 
		\end{pmatrix} + \cdots
		\label{tripletsol_1}.
	\end{align}
A few comments on the dimensions of $U$ and $\Delta$, required to generate the operators at a particular mass dimension, are in order:
	\begin{enumerate}[label=\roman*)]
		\item The scalar functional $U$, a double derivative of the Lagrangian {\it w.r.to} the heavy fields $\Delta_i$, always has a dimension two for our four-dimensional spacetime. However, it also contains the heavy field dependence in quadrature, as given in the underlined parts of $(U_{{11}})_{ij}$  and $(U_{{12}})_{ij}$ in Eqns. of~\ref{matrix_u111}.
	\begin{eqnarray}
		(U_{{11}})_{ij} &=& \dfrac{1}{4} (2 \lambda_1 + \lambda_4) (H^\dag H) \delta_{ij} + \dfrac{1}{2}\lambda_2 \underline{ \Big(\, \Delta_a^* \Delta_a~\delta_{ij} + \Delta_i \Delta^*_j \, \Big)} \nonumber \\
		& &+ \dfrac{1}{2} \lambda_3 \underline{\Big(\, \Delta_a^* \Delta_a~\delta_{ij} + \Delta_i \Delta^*_j - \Delta^* _i \Delta_j \, \Big)} - \frac{i}{2} \lambda_4 \epsilon_{ijk} ( H^\dagger \tau^k H) \nonumber,\label{matrix_u111}
		 \\   
		(U_{{12}})_{ij} & =& \dfrac{1}{2} \lambda_2  \underline{(\Delta^\ast_i \Delta^\ast_j)} + \dfrac{1}{4}\lambda_3 \underline{\left(2 \Delta^\ast_i \Delta^\ast_j -\Delta^\ast_a \Delta^\ast_a \delta_{ij}\right) },   \label{matrix_u121}
	\end{eqnarray}
        Thus, substituting the solution of the heavy field in the $U$ matrix, we find the local operators in light fields having various mass dimensions. 
		\item For dimension-six, we only need to consider the first two terms of the  heavy triplet field  solution (Eqn.~\ref{tripletsol_1})
		\begin{equation}
			\left(\Delta_c\right)_i = \underline{\frac{B_i}{M_\Delta^2}} + \uwave{\frac{(P^2 - U)_{ij}}{M_\Delta^4} B_j} + \underline{\underline{{\frac{(P^2 - U)_{ij}^2}{M_\Delta^6} B_j}}} + \cdots.
			\label{tripletsol3}
		\end{equation}
		As these elements of the series contribute at $\mathcal{O}(\Delta_c^2)$ via $U$, we will have contributions up to order $1/M^6$ ($U$ depends on the light fields only).
        
		\item In this work, first three terms in the heavy field expansion are sufficient and thus we truncate the series in $\Delta_c^\ast \Delta_c$ at order $1/M^8$. In the second term of Eqn.~\ref{tripletsol3} (wave-underlined), linearly proportional to $(P^2-U)$, $U$ depends on the heavy field, see Eqn.~\ref{matrix_u22}. However, as part of the solution for this heavy field, we must substitute the classical solution up to dimension-two in the third term in $\Delta_c^\ast \Delta_c$ (double underlined in Eqn.~\ref{tripletsol3}) to get dimension-eight operators.
	\end{enumerate}
        Keeping this strategy in mind, the dimension-eight operators are categorized into the following classes, depending on how they contribute in Eqn.~\ref{eq:finite}:
        \setlength{\leftmargini}{0.5cm} % Set the indentation for the first level    
	\begin{itemize}
		\item {\textbf{Contributions from $M^{-4}$ operators}}: 
		Each operator term in Eqn.~\ref{eq:finite} in blue colour has dimension-eight. The possibilities for dimension-eight operators in the Green’s basis~\cite{Chala:2021cgt} are :
		\begin{align}
			%\begin{center}
			\phi^8,\; \phi^6 D^2,\; \phi^4 D^4,\; X^3 \phi^2 ,\; X^2 \phi^4,\;  X \phi^2 D^4,  \nonumber \\
			X \phi^4 D^2, \; X^2 \phi^2 D^2, \; X^4, \; X^2 X'^2,  \; X^3 D^2 , \; X^2 D^4 .
			%   \end{center}
	    \end{align}
	where $D$ denotes covariant derivative $P_\mu$, and $X$ represents the field strength tensor $X_{\mu \nu}$.  
	\item {\textbf{Contributions from $M^{-2}$ operators}}: The total effective Lagrangian up to dimension-eight at one-loop has $\mathcal{O}(1/M^2)$ terms as shown in pink colour in Eqn.~\ref{eq:finite}. Here, the $U$-dependent terms are:
	\begin{equation}
		\frac{1}{6}  \,\Bigg[ -U^3 - \frac{1}{2} (P_\mu U)^2-\frac{1}{2}U\,(G_{\mu\nu})^2  \Bigg].
	\end{equation}
The heavy field solution truncated at $B_i/M^2$ generates dimension-eight operators when plugged in $U$-matrix, see Eqn.~\ref{matrix_u22}.
The possible dimension-eight operators in the Green’s basis~\cite{Chala:2021cgt} are:
	\begin{equation}
		\phi^8,\; \phi^6 D^2,\; \phi^4 X^2.
	\end{equation}
If we restrict $U$ to be a functional of light fields only, we generate dimension-six ones. The operators arising in this set (see Eqn.~\ref{eq:finite}; in pink) belong to the following SILH-basis operators ~\cite{Giudice:2007fh}:
	\begin{equation}
		\phi^6,\; \phi^4 D^2,\;  \phi^2 X^2,\; X^3.
	\end{equation}
	\item {\textbf{Contributions from $M^0$ operators}}: As $U$ is a functional of heavy fields, the higher order local expansion in heavy fields may generate higher mass dimensional operators. For example, from $\tr \Big[{1}/{2}  \left(- \ln\left[{M^2}/{\mu^2}\right] \, U^2\right)\Big]$, the dimension-eight operators
	\begin{equation}
		\phi^8,\; \phi^6 D^2 , 
	\end{equation}
	and the dimension-six operators
	\begin{equation}
		\phi^6,\; \phi^4 D^2 .
	\end{equation}
    can emerge. 
	\item {\textbf{Contributions from $M^2$ operators}}: Similarly, the term $\tr(U)$ in effective action offers  the dimension-eight 
	\begin{equation}
		\phi^8,\; \phi^6 D^2,\; \phi^4 D^4, 
	\end{equation}
	and the dimension-six 
	\begin{equation}
		\phi^6,\; \phi^4 D^2
	\end{equation}
    operator. In this case, $\mathcal{O}(B^\dag (P^2-U)^2 B) \subset \Delta_c$ is considered. 
\end{itemize}

Here, we enlist the higher mass-dimension effective operators, generated at the one-loop when $U$ is a functional of light fields only. The dimension-eight operators are displayed in Green's basis.
\setlength{\leftmargini}{0.67cm} 

\subsubsection*{\bf Dimension-eight operators from $\mathcal{O}(M^{-4})$ term:}
\begin{enumerate}
	\item \underline{Covariant operator: $U^4$} 
	\begin{align}
		\tr[U^4] \supset  c^{\llbracket U^{4} \rrbracket}_{\phi^8} \mathcal{O}_{\phi^8} .
	\end{align}
	The matching relation is:
	\begin{gather}
		c^{\llbracket U^{4} \rrbracket}_{\phi^8} =  \frac{3}{8} \lambda_4^4 + \frac{3}{4} \lambda_1^3 \lambda_4 + \frac{15}{16} \lambda_1^2 \lambda_4^2 + \frac{9}{16} \lambda_1 \lambda_4^3 + \frac{17}{128} \lambda_4^4 ~.
	\end{gather}
	%%%%%%%%%%%%%%%%%%%%%%%%%%%%%%%%%%%%%%%%%%%%%%%%%%%%%%%%%%%%%%%%%%%%%%%
	\item \underline{Covariant operator: $U^2 (P^2 U)$}\\
	\begin{align}
		\tr[U^2 (P^2 U)] &\supset c^{(1), \llbracket U^2 (P^2 U) \rrbracket}_{\phi^6} \mathcal{O}^{(1)}_{\phi^6} + c^{(2), \llbracket U^2 (P^2 U) \rrbracket}_{\phi^6} \mathcal{O}^{(2)}_{\phi^6} + c^{(3), \llbracket U^2 (P^2 U) \rrbracket}_{\phi^6} \mathcal{O}^{(3)}_{\phi^6}  .
	\end{align}
	The matching relations are: 
	\begin{align}
		c^{(1), \llbracket U^2 (P^2 U) \rrbracket}_{\phi^6} &=   \frac{3 \lambda_1^3}{2} + \frac{9 \lambda_1^2 \lambda_4}{4} + \frac{11 \lambda_1 \lambda_4^2}{8} + \frac{5 \lambda_4^3}{16} ~, &c^{(2), \llbracket U^2 (P^2 U) \rrbracket}_{\phi^6} &=  \frac{\lambda_1 \lambda_4^2}{2} + \frac{\lambda_4^3}{4} ~, \nonumber\\
		c^{(3), \llbracket U^2 (P^2 U) \rrbracket}_{\phi^6} & =  \frac{3 \lambda_1^3}{4} + \frac{9 \lambda_1^2 \lambda_4}{8} + \frac{15 \lambda_1 \lambda_4^2}{16} + \frac{9 \lambda_4^3}{32}~.  &&
	\end{align}
	%%%%%%%%%%%%%%%%%%%%%%%%%%%%%%%%%%%%%%%%%%%%%%%%%%%%%%%%%%%%%%%%%%%%%%%
	\item \underline{Covariant operator: $U^2 (G_{\mu \nu})^2$} \\
	\begin{align}
		\tr[U^2 (G_{\mu \nu})^2] &\supset  c^{(1), \llbracket U^2 (G_{\mu \nu})^2 \rrbracket}_{W^2 \phi^4} \mathcal{O}^{(1)}_{W^2 \phi^4} + c^{(3), \llbracket U^2 (G_{\mu \nu})^2 \rrbracket}_{W^2 \phi^4} \mathcal{O}^{(3)}_{W^2 \phi^4} + c^{(1), \llbracket U^2 (G_{\mu \nu})^2 \rrbracket}_{B^2  \phi^4} \mathcal{O}^{(1)}_{B^2  \phi^4} \nonumber \\
		& \hspace{1cm}+ c^{(1), \llbracket U^2 (G_{\mu \nu})^2 \rrbracket}_{W B \phi^4} \mathcal{O}^{(1)}_{W B \phi^4} .
	\end{align}
	The matching relations are: 
	\begin{align}
		c^{(1), \llbracket U^2 (G_{\mu \nu})^2 \rrbracket}_{W^2 \phi^4} &= - g^2 \Big(\frac{1}{4}  \lambda_4^2 +  \lambda_1^2 +  \lambda_1 \lambda_4 \Big) , &c^{(3), \llbracket U^2 (G_{\mu \nu})^2 \rrbracket}_{W^2 \phi^4} &= -g^2 \lambda_4^2~, \nonumber\\
		c^{(1), \llbracket U^2 (G_{\mu \nu})^2 \rrbracket}_{B^2  \phi^4} &=  - g'^2 \Big( \frac{3}{2} \lambda_1^2 + \frac{3}{2}  \lambda_1 \lambda_4   + \frac{3}{4}  \lambda_4^2 \Big) ,& c^{(1), \llbracket U^2 (G_{\mu \nu})^2 \rrbracket}_{W B \phi^4}  &= g g' \Big( \frac{3}{2}  \lambda_1^2 - \frac{3}{4}  \lambda_1 \lambda_4 - \frac{3}{4}  \lambda_4^2 \Big).
	\end{align}
	%%%%%%%%%%%%%%%%%%%%%%%%%%%%%%%%%%%%%%%%%%%%%%%%%%%%%%%%%%%%%%%%%%%%%%%
	\item \underline{Covariant operator: $(U G_{\mu \nu})^2$} \\
	\begin{align}
		\tr[(U G_{\mu \nu})^2] &\supset  c^{(1), \llbracket (U G_{\mu \nu})^2 \rrbracket}_{W^2 \phi^4} \mathcal{O}^{(1)}_{W^2 \phi^4} + c^{(3), \llbracket (U G_{\mu \nu})^2 \rrbracket}_{W^2 \phi^4} \mathcal{O}^{(3)}_{W^2 \phi^4} + c^{(1), \llbracket (U G_{\mu \nu})^2 \rrbracket}_{B^2  \phi^4} \mathcal{O}^{(1)}_{B^2  \phi^4}\nonumber \\
		& \hspace{1cm}+ c^{(1), \llbracket (U G_{\mu \nu})^2 \rrbracket}_{W B \phi^4} \mathcal{O}^{(1)}_{W B \phi^4}.
	\end{align}
	The matching relations are: 
	\begin{align}
		c^{(1), \llbracket (U G_{\mu \nu})^2 \rrbracket}_{W^2 \phi^4} &=  -\frac{1}{4} g^2 (2\lambda_1 + \lambda_4)^2  ~, &c^{(3), \llbracket (U G_{\mu \nu})^2 \rrbracket}_{W^2 \phi^4} &=   -g^2 \lambda_4^2 ~,\nonumber\\
		c^{(1), \llbracket (U G_{\mu \nu})^2 \rrbracket}_{B^2  \phi^4} &=  -\frac{1}{4} g'^2 \lambda_4^2 - \frac{3}{4} g'^2 (2\lambda_1 + \lambda_4)^2  ~,& c^{(1), \llbracket (U G_{\mu \nu})^2 \rrbracket}_{W B \phi^4}  &= 2 g g' \lambda_4 (2\lambda_1 + \lambda_4)~.
	\end{align}
	%%%%%%%%%%%%%%%%%%%%%%%%%%%%%%%%%%%%%%%%%%%%%%%%%%%%%%%%%%%%%%%%%%%%%%%
	\item \underline{Covariant operator: $(P^2 U)^2$} \\
	\begin{align}
		\tr[(P^2 U)^2] &\supset  c^{(3), \llbracket (P^2 U)^2 \rrbracket}_{\phi^4} \mathcal{O}^{(3)}_{\phi^4} + c^{(4), \llbracket (P^2 U)^2 \rrbracket}_{\phi^4} \mathcal{O}^{(4), \llbracket (P^2 U)^2 \rrbracket}_{\phi^4} + c^{(8), \llbracket (P^2 U)^2 \rrbracket}_{\phi^4} \mathcal{O}^{(8)}_{\phi^4}  \nonumber \\
		& \hspace{1cm}+ c^{(10), \llbracket (P^2 U)^2 \rrbracket}_{\phi^4} \mathcal{O}^{(10)}_{\phi^4} + c^{(11), \llbracket (P^2 U)^2 \rrbracket}_{\phi^4} \mathcal{O}^{(11)}_{\phi^4} .
	\end{align}
	The matching relations are: 
	\begin{align}
		c^{(3), \llbracket (P^2 U)^2 \rrbracket}_{\phi^4} &= 6 \lambda_1^2 + 6 \lambda_1 \lambda_4 + \frac{5}{2} \lambda_4^2 ~, &c^{(4), \llbracket (P^2 U)^2 \rrbracket}_{\phi^4} &=   6 \lambda_1^2 + 6 \lambda_1 \lambda_4 + \frac{5}{2} \lambda_4^2  ~,\nonumber\\
		c^{(8), \llbracket (P^2 U)^2 \rrbracket}_{\phi^4} &=  \frac{3}{2} \lambda_1^2 + \frac{3}{2}\lambda_1 \lambda_4 + \frac{5}{8} \lambda_4^2 ~,& c^{(10), \llbracket (P^2 U)^2 \rrbracket}_{\phi^4}  &=   \lambda_4^2  ~,\nonumber\\
		c^{(11), \llbracket (P^2 U)^2 \rrbracket}_{\phi^4} &=  3 \lambda_1^2+ 3 \lambda_1 \lambda_4 + \frac{1}{4} \lambda_4^2  ~. &&
	\end{align}
	%%%%%%%%%%%%%%%%%%%%%%%%%%%%%%%%%%%%%%%%%%%%%%%%%%%%%%%%%%%%%%%%%%%%%%%
	\item \underline{Covariant operator: $U (P_\mu U) J_\mu$} \\
	\begin{align}
		\tr[U (P_\mu U) J_\mu] &\supset c_{W\phi^4 D^2}^{(6), \llbracket U (P_\mu U) J_\mu \rrbracket}  \mathcal{O}_{W\phi^4 D^2}^{(6)} + c_{W\phi^4 D^2}^{(7), \llbracket U (P_\mu U) J_\mu \rrbracket} \mathcal{O}_{W\phi^4 D^2}^{(7)}  + c_{B\phi^4 D^2}^{(3), \llbracket U (P_\mu U) J_\mu \rrbracket} \mathcal{O}_{B\phi^4 D^2}^{(3)} .
	\end{align}
	The matching relations are: 
	\begin{align}
		c_{W\phi^4 D^2}^{(6), \llbracket U (P_\mu U) J_\mu \rrbracket} &=  \frac{-3}{4} g \lambda_4 \left( 2\lambda_1 + \lambda_4 \right) ~, &c_{W\phi^4 D^2}^{(7), \llbracket U (P_\mu U) J_\mu \rrbracket} &=    g \lambda_4^2~, \nonumber\\
		c_{B\phi^4 D^2}^{(3), \llbracket U (P_\mu U) J_\mu \rrbracket} &=  \frac{3}{8} g' (2\lambda_1 + \lambda_4)^2 -  g' \lambda_4^2~. & &
	\end{align}
	%%%%%%%%%%%%%%%%%%%%%%%%%%%%%%%%%%%%%%%%%%%%%%%%%%%%%%%%%%%%%%%%%%%%%%%
	\item \underline{Covariant operator: $U (J_\mu)^2$} \\
	\begin{align}
		\tr[U (J_\mu)^2] &\supset  c^{(6), \llbracket U (J_\mu)^2 \rrbracket}_{B^2 \phi^2 D^2} \mathcal{O}^{(6)}_{B^2 \phi^2 D^2} + c^{(8), \llbracket U (J_\mu)^2 \rrbracket}_{B^2 \phi^2 D^2} \mathcal{O}^{(8)}_{B^2 \phi^2 D^2} + c^{(10), \llbracket U (J_\mu)^2 \rrbracket}_{W B \phi^2 D^2} \mathcal{O}^{(10)}_{W B \phi^2 D^2} \nonumber \\
		& \hspace{1cm}+ c^{(13), \llbracket U (J_\mu)^2 \rrbracket}_{W^2 \phi^2 D^2} \mathcal{O}^{(13)}_{W^2 \phi^2 D^2} .
	\end{align}
	The matching relations are: 
	\begin{align}
		c^{(6), \llbracket U (J_\mu)^2 \rrbracket}_{B^2 \phi^2 D^2} &=  \frac{3}{2} g'^2 (2\lambda_1 + \lambda_4)  ~, &c^{(8), \llbracket U (J_\mu)^2 \rrbracket}_{B^2 \phi^2 D^2} &=  \frac{3}{2} g'^2 (2\lambda_1 + \lambda_4)~,  \nonumber\\
		c^{(10), \llbracket U (J_\mu)^2 \rrbracket}_{W B \phi^2 D^2} &=  -4 gg' \lambda_4 ~,&c^{(13), \llbracket U (J_\mu)^2 \rrbracket}_{W^2 \phi^2 D^2} &= - g^2 (2\lambda_1 + \lambda_4)~.
	\end{align}
	%%%%%%%%%%%%%%%%%%%%%%%%%%%%%%%%%%%%%%%%%%%%%%%%%%%%%%%%%%%%%%%%%%%%%%%%
	\item \underline{Covariant operator: $(P^2 U) (G_{\rho \sigma})^2$} \\
	\begin{align}
		\tr[(P^2 U) (G_{\rho \sigma})^2] &\supset  c^{(14), \llbracket  (P^2 U) (G_{\rho \sigma})^2  \rrbracket}_{W^2 \phi^2 D^2} \mathcal{O}^{(14)}_{W^2 \phi^2 D^2} + c^{(2), \llbracket  (P^2 U) (G_{\rho \sigma})^2  \rrbracket}_{B^2 \phi^2 D^2} \mathcal{O}^{(2)}_{B^2 \phi^2 D^2}  \nonumber \\
		&+ c^{(1), \llbracket  (P^2 U) (G_{\rho \sigma})^2  \rrbracket}_{W B \phi^2 D^2} \mathcal{O}^{(1)}_{W B \phi^2 D^2} + c^{\llbracket  (P^2 U) (G_{\rho \sigma})^2 \rrbracket}_1 (D^2 \phi^\dagger \phi+\phi^\dagger D^2\phi) B_{\rho \sigma} B^{\rho \sigma} \nonumber \\
		& + c^{\llbracket  (P^2 U) (G_{\rho \sigma})^2  \rrbracket}_2 (D^2 \phi^\dagger \sigma^a \phi+\phi^\dagger \sigma^a D^2\phi) W_{\rho \sigma}^a B^{\rho \sigma}. 
	\end{align}
	\begin{align}
		\tr[(P^2 U) (G_{\rho \sigma})^2] &\supset  c^{(14), \llbracket  (P^2 U) (G_{\rho \sigma})^2 \rrbracket}_{W^2 \phi^2 D^2} \mathcal{O}^{(14)}_{W^2 \phi^2 D^2} + c^{(2), \llbracket  (P^2 U) (G_{\rho \sigma})^2  \rrbracket}_{B^2 \phi^2 D^2} \mathcal{O}^{(2)}_{B^2 \phi^2 D^2} \nonumber \\ 
		&\hspace{1cm}+ c^{(1), \llbracket  (P^2 U) (G_{\rho \sigma})^2 \rrbracket}_{W B \phi^2 D^2} \mathcal{O}^{(1)}_{W B \phi^2 D^2}  + c^{(1), \llbracket (P^2 U) (G_{\rho \sigma})^2 \rrbracket}_{W B \phi^4} \mathcal{O}^{(1)}_{W B \phi^4} \nonumber \\
		& \hspace{1cm}+ c_{B^2 \phi^4}^{(1), \llbracket  (P^2 U) (G_{\rho \sigma})^2 \rrbracket} \mathcal{O}_{B^2 \phi^4}^{(1)} + c^{\llbracket (P^2 U) (G_{\rho \sigma})^2 \rrbracket}_{BB} \mathcal{O}_{BB} \nonumber \\
		&=    c^{(14), \llbracket (P^2 U) (G_{\rho \sigma})^2 \rrbracket}_{W^2 \phi^2 D^2} (D_\mu\phi^\dagger\phi+\phi^\dagger D_\mu\phi) W^{I\nu\rho} D^\mu W^{I}_{\nu\rho} \nonumber \\
		&\hspace{1cm} + c^{(2), \llbracket (P^2 U) (G_{\rho \sigma})^2 \rrbracket}_{B^2 \phi^2 D^2} (D^{\mu} \phi^{\dag} D_{\mu} \phi) B_{\nu\rho} B^{\nu\rho} \nonumber \\
		& \hspace{1cm}+ c^{(1), \llbracket (P^2 U) (G_{\rho \sigma})^2 \rrbracket}_{W B \phi^2 D^2} (D^{\mu} \phi^{\dag} \sigma^I D_{\mu} \phi) B_{\nu\rho} W^{I \nu\rho} \nonumber \\
		& \hspace{1cm}+ c^{(1), \llbracket (P^2 U) (G_{\rho \sigma})^2 \rrbracket}_{W B \phi^4}  (\phi^\dag \phi) (\phi^\dag \sigma^I \phi) W^I_{\mu\nu} B^{\mu\nu}  \nonumber \\ 
		&\hspace{1cm} + c^{\llbracket (P^2 U) (G_{\rho \sigma})^2 \rrbracket}_{WB} (\phi^{\dag} \sigma^I \phi) W_{\mu \nu}^I B^{\mu \nu} \nonumber \\
		& \hspace{1cm}+ c^{\llbracket (P^2 U) (G_{\rho \sigma})^2  \rrbracket}_{BB} (\phi^{\dag} \phi ) B_{\mu \nu} B^{\mu \nu} . 
	\end{align}
	The matching relations are: 
	\begin{align}
		c^{(14), \llbracket (P^2 U) (G_{\rho \sigma})^2 \rrbracket}_{W^2 \phi^2 D^2} &=  -g^2 (2 \lambda_1+\lambda_4)   ~, &c^{(2), \llbracket (P^2 U) (G_{\rho \sigma})^2 \rrbracket}_{B^2 \phi^2 D^2}  &=  {3} g'^2 (2\lambda_1 + \lambda_4) ~,  \nonumber\\
		c^{(1), \llbracket (P^2 U) (G_{\rho \sigma})^2 \rrbracket}_{W B \phi^2 D^2} &=  2 gg' \lambda_4 ~,&c^{(1), \llbracket  (P^2 U) (G_{\rho \sigma})^2 \rrbracket}_{B^2 \phi^4} &= - 3 g'^2  (2\lambda_1 + \lambda_4) \lambda_{\rm SM} ~, \nonumber\\
		c^{(1), \llbracket (P^2 U) (G_{\rho \sigma})^2 \rrbracket}_{W B \phi^4} &=  -2 \lambda_4 \lambda_{\rm SM} g g'~, & c^{\llbracket (P^2 U) (G_{\rho \sigma})^2 \rrbracket}_{WB} & =   \lambda_4 m_H^2 ~, \nonumber\\
		c^{\llbracket (P^2 U) (G_{\rho \sigma})^2 \rrbracket}_{BB} &= 3 g'^2 (2 \lambda_1+\lambda_4) m_H^2 ~.&&
	\end{align}
	%%%%%%%%%%%%%%%%%%%%%%%%%%%%%%%%%%%%%%%%%%%%%%%%%%%%%%%%%%%%%%%%%%%%%%%
	\item \underline{Covariant operator: $ U G_{\mu \nu} G_{\nu \rho} G_{\rho \mu}$} \\
	\begin{align}
		\tr[U G_{\mu \nu} G_{\nu \rho} G_{\rho \mu}] &\supset  c^{(1), \llbracket U G_{\mu \nu} G_{\nu \rho} G_{\rho \mu} \rrbracket}_{W^3 \phi^2} \mathcal{O}^{(1)}_{W^3 \phi^2}~.
	\end{align}
	The matching relation is: 
	\begin{align}
		c^{(1), \llbracket U G_{\mu \nu} G_{\nu \rho} G_{\rho \mu} \rrbracket}_{W^3 \phi^2} & =  \frac{3}{2}(2 \lambda_1 + \lambda_4) g^3~.
	\end{align}
	%%%%%%%%%%%%%%%%%%%%%%%%%%%%%%%%%%%%%%%%%%%%%%%%%%%%%%%%%%%%%%%%%%%%%%%
	\item \underline{Covariant operator: $(P_{\mu} P_{\nu} U) G_{\rho \mu} G_{\rho \nu}$} \\
	\begin{align}
		\tr[(P_{\mu} P_{\nu} U) G_{\rho \mu} G_{\rho \nu}] &\supset  c^{(4), \llbracket (P_{\mu} P_{\nu} U) G_{\rho \mu} G_{\rho \nu} \rrbracket}_{W^2 \phi^2 D^2} \mathcal{O}^{(4)}_{W^2 \phi^2 D^2} + c^{(11), \llbracket (P_{\mu} P_{\nu} U) G_{\rho \mu} G_{\rho \nu} \rrbracket}_{W^2 \phi^2 D^2} \mathcal{O}^{(11)}_{W^2 \phi^2 D^2} \nonumber \\
		& \hspace{0.5cm}+ c^{(14), \llbracket (P_{\mu} P_{\nu} U) G_{\rho \mu} G_{\rho \nu} \rrbracket}_{W^2 \phi^2 D^2} \mathcal{O}^{(14)}_{W^2 \phi^2 D^2} + c^{(4), \llbracket (P_{\mu} P_{\nu} U) G_{\rho \mu} G_{\rho \nu} \rrbracket}_{B^2 \phi^2 D^2} \mathcal{O}^{(4)}_{B^2 \phi^2 D^2} \nonumber \\
		& \hspace{0.5cm}+c^{(8), \llbracket (P_{\mu} P_{\nu} U) G_{\rho \mu} G_{\rho \nu} \rrbracket}_{B^2 \phi^2 D^2} \mathcal{O}^{(8)}_{B^2 \phi^2 D^2}  + c^{(8), \llbracket (P_{\mu} P_{\nu} U) G_{\rho \mu} G_{\rho \nu} \rrbracket}_{W B \phi^2 D^2} \mathcal{O}^{(8)}_{W B \phi^2 D^2} \nonumber \\ 
		& \hspace{0.5cm}+ c^{(11), \llbracket (P_{\mu} P_{\nu} U) G_{\rho \mu} G_{\rho \nu} \rrbracket}_{W B \phi^2 D^2} \mathcal{O}^{(11)}_{W B \phi^2 D^2}. 
	\end{align}
	The matching relations are: 
	\begin{align}
		c^{(4), \llbracket (P_{\mu} P_{\nu} U) G_{\rho \mu} G_{\rho \nu} \rrbracket}_{W^2 \phi^2 D^2} & =  - g^2 \lambda_4  ~, &c^{(11), \llbracket (P_{\mu} P_{\nu} U) G_{\rho \mu} G_{\rho \nu} \rrbracket}_{W^2 \phi^2 D^2}  & =  ~ g^2 (2\lambda_1 + \lambda_4) ~,  \nonumber\\
		c^{(14), \llbracket (P_{\mu} P_{\nu} U) G_{\rho \mu} G_{\rho \nu} \rrbracket}_{W^2 \phi^2 D^2} & = g^2 (2 \lambda_1+\lambda_4)   ~, &c^{(4), \llbracket (P_{\mu} P_{\nu} U) G_{\rho \mu} G_{\rho \nu} \rrbracket}_{B^2 \phi^2 D^2}  &= \frac{3}{2} g'^2 (2\lambda_1 + \lambda_4) ~,  \nonumber\\
		c^{(8), \llbracket (P_{\mu} P_{\nu} U) G_{\rho \mu} G_{\rho \nu} \rrbracket}_{B^2 \phi^2 D^2}  &=  \frac{3}{2} g'^2 (2\lambda_1 + \lambda_4) ~, & c^{(8), \llbracket (P_{\mu} P_{\nu} U) G_{\rho \mu} G_{\rho \nu} \rrbracket}_{W B \phi^2 D^2} &=  - 2 gg' \lambda_4 ~, \nonumber\\
		c^{(11), \llbracket (P_{\mu} P_{\nu} U) G_{\rho \mu} G_{\rho \nu} \rrbracket}_{W B \phi^2 D^2} &=  -2 gg' \lambda_4 ~.&&
	\end{align}
	%%%%%%%%%%%%%%%%%%%%%%%%%%%%%%%%%%%%%%%%%%%%%%%%%%%%%%%%%%%%%%%%%%%%%%%
	\item \underline{Covariant operator: $(P_\nu J_\mu)^2$} \\
	\begin{align}
		\tr[(P_\nu J_\mu)^2] &\supset  c^{\llbracket (P_\nu J_\mu)^2 \rrbracket}_{B^2 D^4}  \mathcal{O}_{B^2 D^4}  + c^{\llbracket (P_\nu J_\mu)^2 \rrbracket}_{W^2 D^4} \mathcal{O}_{W^2 D^4}  .
	\end{align}
	The matching relations are: 
	\begin{align}
		c^{\llbracket (P_\nu J_\mu)^2 \rrbracket}_{B^2 D^4} &=  - 3 g'^2 ~, &c^{\llbracket (P_\nu J_\mu)^2 \rrbracket}_{W^2 D^4} &= 2 g^2~.  
	\end{align}
	%%%%%%%%%%%%%%%%%%%%%%%%%%%%%%%%%%%%%%%%%%%%%%%%%%%%%%%%%%%%%%%%%%%%%%%
	\item \underline{Covariant operator: $(G_{\mu \nu} G_{\rho \sigma})^2$} 
	\begin{align}
		\tr[(G_{\mu \nu} G_{\rho \sigma})^2] &\supset  c_{W^4}^{(3), \llbracket (G_{\mu \nu} G_{\rho \sigma})^2 \rrbracket} Q_{W^4}^{(3)} + c_{B^4}^{(1), \llbracket (G_{\mu \nu} G_{\rho \sigma})^2 \rrbracket} Q_{B^4}^{(1)}  \nonumber \\
		&\hspace{1cm} + c_{W^2 B^2}^{(1), \llbracket (G_{\mu \nu} G_{\rho \sigma})^2 \rrbracket} Q_{W^2 B^2}^{(1)} + c_{W^2 B^2}^{(3), \llbracket (G_{\mu \nu} G_{\rho \sigma})^2 \rrbracket} Q_{W^2 B^2}^{(3)} .
	\end{align}
	The matching relations are: 
	\begin{align}
		c_{W^4}^{(3), \llbracket (G_{\mu \nu} G_{\rho \sigma})^2 \rrbracket}  &=  2 g^4 ~, & c_{B^4}^{(1), \llbracket (G_{\mu \nu} G_{\rho \sigma})^2 \rrbracket} &=  3 g'^4 ~, \nonumber\\
		c_{W^2 B^2}^{(1), \llbracket (G_{\mu \nu} G_{\rho \sigma})^2 \rrbracket} &=  4 g^2 g'^2 ~,&c_{W^2 B^2}^{(3), \llbracket (G_{\mu \nu} G_{\rho \sigma})^2 \rrbracket} &=  8 g^2 g'^2~.
	\end{align}
	%%%%%%%%%%%%%%%%%%%%%%%%%%%%%%%%%%%%%%%%%%%%%%%%%%%%%%%%%%%%%%%%%%%%%%%
	\item \underline{Covariant operator: $(G_{\mu \nu})^2 (G_{\rho \sigma})^2$} 
	\begin{align}       
		\tr[(G_{\mu \nu})^2 (G_{\rho \sigma})^2] &\supset c_{W^4}^{(3), \llbracket (G_{\mu \nu})^2 (G_{\rho \sigma})^2 \rrbracket} Q_{W^4}^{(3)} + c_{B^4}^{(1), \llbracket (G_{\mu \nu})^2 (G_{\rho \sigma})^2 \rrbracket} Q_{B^4}^{(1)}   \nonumber \\
		& \hspace{1cm}+ c_{W^2 B^2}^{(1), \llbracket (G_{\mu \nu})^2 (G_{\rho \sigma})^2 \rrbracket} Q_{W^2 B^2}^{(1)} + c_{W^2 B^2}^{(3), \llbracket (G_{\mu \nu})^2 (G_{\rho \sigma})^2 \rrbracket} Q_{W^2 B^2}^{(3)}   .
	\end{align}
	The matching relations are: 
	\begin{align}
		c_{W^4}^{(3), \llbracket (G_{\mu \nu})^2 (G_{\rho \sigma})^2 \rrbracket}  &= 2 g^4 ~, & c_{B^4}^{(1), \llbracket (G_{\mu \nu})^2 (G_{\rho \sigma})^2 \rrbracket} &=  3 g'^4 ~, \nonumber\\
		c_{W^2 B^2}^{(1), \llbracket (G_{\mu \nu})^2 (G_{\rho \sigma})^2 \rrbracket} & =  4 g^2 g'^2 ~,&c_{W^2 B^2}^{(3), \llbracket (G_{\mu \nu})^2 (G_{\rho \sigma})^2 \rrbracket} & = 8 g^2 g'^2~.
	\end{align}
	%%%%%%%%%%%%%%%%%%%%%%%%%%%%%%%%%%%%%%%%%%%%%%%%%%%%%%%%%%%%%%%%%%%%%%%%%%%%%%%%%%%%%%%%%
	\item \underline{Covariant operators: $(G_{\mu \nu} G_{\nu \rho})^2$, $G_{\mu \nu} G_{\nu \rho} G_{\rho \sigma} G_{\sigma \mu}$, $G_{\mu \nu} J_{\mu} J_{\nu}$, $P_{\mu} J_{\nu} G_{\nu \sigma} G_{\sigma \mu}$} 
	\begin{align}
		\tr[(G_{\mu \nu} G_{\nu \rho})^2] &\supset - (D_\nu G_{ \nu \rho}) (D_\mu G_{\mu \nu}) G_{\nu \rho} - (D_\nu G_{ \nu \rho}) G_{\mu\nu} (D_\mu G_{\nu \rho}) \nonumber \\
		& \hspace{1cm}+ (D_\mu G_{ \nu \rho})  (D_\nu G_{\mu \nu})G_{\nu\rho} +  (D_\mu G_{ \nu \rho}) G_{\mu \nu} (D_\nu G_{\nu\rho})~.
		\label{Gmununurho}
	\end{align}
    \begin{align}
		\tr[G_{\mu \nu} J_{\mu} J_{\nu}] & = g^3 \epsilon^{IJK} W_{\mu \nu}^I D_\rho W_{\rho \mu}^J D_{\sigma} W_{\sigma \nu}^K \nonumber \\
		& + 4 i g^2 g' W_{\mu \nu}^I D_{\rho} W_{\rho \mu}^I D_\sigma B_{\sigma \nu} + 2 i g^2 g' W_{\mu \nu}^I D_{\rho} B_{\rho \mu} D_{\sigma} W_{\sigma \nu}^I~.
	\end{align}
    
	Employing the classical EOM for the field strength tensor, each of the terms in the above equations, we can construct the non-redundant dimension-eight operators involving light scalar, and fermionic fields.
	%%%%%%%%%%%%%%%%%%%%%%%%%%%%%%%%%%%%%%%%%%%%%%%%%%%%%%%%%%%%%%%%%%%%%%%%%%%%%%%%%%%%%%%%%

	%%%%%%%%%%%%%%%%%%%%%%%%%%%%%%%%%%%%%%%%%%%%%%%%%%%%%%%%%%%%%%%%%%%%%%%%%%%%%%%%%%%%%%%%%
\end{enumerate}
\subsubsection*{\bf Dimension-eight and dimension-six operators from $\mathcal{O}(M^{-2})$ term :}
\begin{enumerate}
	\item \underline{Covariant operator: $U^3$} 
	\begin{align}
		\tr[U^3] \supset  c_{\phi^8}^{\llbracket U^{3} \rrbracket} \mathcal{O}_{\phi^8} .
	\end{align}
	The matching relation at dimension-eight is:
	\begin{gather}
		c_{\phi^8}^{\llbracket U^{3} \rrbracket} =  \frac{3 \mu_\Delta^2}{4 M_\Delta^4} \Big( \lambda_1^2 \lambda_2 + \frac{3}{4} \lambda_1^2 \lambda_3 + \lambda_1 \lambda_2 \lambda_4 + \frac{3}{4} \lambda_1 \lambda_2 \lambda_4 + \frac{5}{16} \lambda_2 \lambda_4^2 + \frac{1}{4} \lambda_3 \lambda_4^2 \Big) .
	\end{gather}
	\begin{eqnarray}
		\text{Tr} [U^3] & \supset  c_6^{\llbracket U^{3} \rrbracket} \mathcal{O}_6 =  c^{\llbracket U \rrbracket}_{6} (\phi^\dagger \phi)^3 ~,
	\end{eqnarray}  
	with the matching relation at dimension-six as:
	\begin{eqnarray}
		c_6^{\llbracket U^{3} \rrbracket} =   \dfrac{3}{32} \Big( \, 8 \lambda^3_1 + 3 \lambda^3_4 + 12 \lambda^2_1 \lambda_4 + 10 \lambda_1 \lambda^2_4 \, \Big).
	\end{eqnarray}  
	%%%%%%%%%%%%%%%%%%%%%%%%%%%%%%%%%%%%%%%%%%%%%%%%%%%%%%%%%%%%%%%%%%%%%%%%%%%%%%%%%%%%%%%%%
	\item \underline{Covariant operator: $(P_\mu U)^2$}
	\begin{align}
		\tr[(P_\mu U)^2] & \supset c^{(3),\llbracket (P_\mu U)^2 \rrbracket}_{\phi^6} \mathcal{O}^{(3)}_{\phi^6}  
	\end{align}
	The matching relation at dimension-eight is: 
	\begin{align}
		c^{(3),\llbracket (P_\mu U)^2 \rrbracket}_{\phi^6} & = - \frac{\mu_\Delta^2}{M_\Delta^4} \Big( \lambda_1 \lambda_2 + \frac{15}{8} \lambda_1 \lambda_3 + \frac{1}{2} \lambda_2 \lambda_4  +  \frac{15}{16} \lambda_3 \lambda_4 \Big) .
	\end{align}
	\begin{align}
		\tr[(P_\mu U)^2] & \supset    c_{H}^{\llbracket (P_\mu U)^2 \rrbracket} \mathcal{O}_{H} + c_{T}^{\llbracket (P_\mu U)^2 \rrbracket} \mathcal{O}_{T} + c_{R}^{\llbracket (P_\mu U)^2 \rrbracket} \mathcal{O}_{R}  \nonumber\\
			& = c_{H} \frac{1}{2} \left( \partial_\mu |\phi|^2 \right) + c_{T} \frac{1}{2} \big( \phi^{\dag} \Dfb \phi \big)^2  + c_{R}  |\phi|^2 |D_\mu \phi|^2
	\end{align}  
	with the matching relation at dimension-six as:
	\begin{align}
		c_{H}^{\llbracket (P_\mu U)^2 \rrbracket} &= - \frac{3}{4} \left( 4 \lambda_1^2 + \lambda_4^2 + 4 \lambda_1 \lambda_4 \right)  ~, &
		c_{T}^{\llbracket (P_\mu U)^2 \rrbracket}  &=  - \frac{1}{2} \left( \lambda_4^2 \right)~,  &
		c_{R}^{\llbracket (P_\mu U)^2 \rrbracket}  &=  - \lambda_4 ^4~.
	\end{align}  
	%%%%%%%%%%%%%%%%%%%%%%%%%%%%%%%%%%%%%%%%%%%%%%%%%%%%%%%%%%%%%%%%%%%%%%%%%%%%%%%%%%%%%%%%%
	\item \underline{Covariant operator: $U (G_{\mu \nu})^2$} 
	\begin{align}
		\tr[U (G_{\mu \nu})^2] & \supset c^{(1),\llbracket U (G_{\mu \nu})^2 \rrbracket}_{W^2 \phi^4} \mathcal{O}^{(1)}_{W^2 \phi^4} + c^{(3),\llbracket U (G_{\mu \nu})^2 \rrbracket}_{W^2 \phi^4} \mathcal{O}^{(3)}_{W^2 \phi^4} + c^{(1),\llbracket U (G_{\mu \nu})^2 \rrbracket}_{B^2  \phi^4} \mathcal{O}^{(1)}_{B^2  \phi^4}.
	\end{align}
	The matching relations at dimension-eight are: 
	\begin{align}
		c^{(1),\llbracket U (G_{\mu \nu})^2 \rrbracket}_{W^2 \phi^4} & =  -\frac{1}{M_\Delta^4} \mu_\Delta^2  g^2 \left(\lambda_2 + \lambda_3 \right) -\frac{1}{2 M_\Delta^4} g^2 \mu_\Delta^2 \left(\lambda_2 + 2 \lambda_3 \right) \nonumber\\
		c^{(3),\llbracket U (G_{\mu \nu})^2 \rrbracket}_{W^2  \phi^4} & = \frac{1}{M_\Delta^4} \mu_\Delta^2 g^2 \left(\lambda_2 + \lambda_3 \right) + \frac{1}{2 M_\Delta^4} \mu_\Delta^2  g^2 \lambda_3  ~,\nonumber\\
		c^{(1),\llbracket U (G_{\mu \nu})^2 \rrbracket}_{B^2 \phi^4} & =   - \frac{2}{M_\Delta^4}\mu_\Delta^2 g'^2 \lambda_2  - \frac{3}{2 M_\Delta^4} \mu_\Delta^2 g'^2 \lambda_3~.
	\end{align}
	\begin{align}
		\tr[U (G_{\mu \nu})^2] & \supset    c_{BB}^{\llbracket U (G_{\mu \nu})^2 \rrbracket} \mathcal{O}_{BB} + c_{WW}^{\llbracket U (G_{\mu \nu})^2 \rrbracket} \mathcal{O}_{WW} + c_{WB}^{\llbracket U (G_{\mu \nu})^2 \rrbracket} \mathcal{O}_{WB} \nonumber\\
		& = c_{BB}^{\llbracket U (G_{\mu \nu})^2 \rrbracket} g'^2 (\phi^\dag \phi) B_{\mu\nu} B^{\mu\nu}  + c_{WW}^{\llbracket U (G_{\mu \nu})^2 \rrbracket}  g^2 (\phi^\dag \phi) W^I_{\mu\nu} W^{I\mu\nu} \nonumber \\
		& \hspace{1cm}+ c_{WB}^{\llbracket U (G_{\mu \nu})^2 \rrbracket}  2 g g' (\phi^\dag \sigma^I \phi)  W^I_{\mu\nu} B^{\mu\nu}~, 
	\end{align}  
	with the matching relation at dimension-six as:
	\begin{align}
		c_{BB}^{\llbracket U (G_{\mu \nu})^2 \rrbracket} & =  - \frac{3}{2} \left( 2 \lambda_1 + \lambda_4 \right)  ~, &
		c_{WW}^{\llbracket U (G_{\mu \nu})^2 \rrbracket} & =  - \left(2 \lambda_1 + \lambda_4 \right)~,  &
		c_{WB}^{\llbracket U (G_{\mu \nu})^2 \rrbracket} & = 2 \lambda_4 ~.
	\end{align}  
	%%%%%%%%%%%%%%%%%%%%%%%%%%%%%%%%%%%%%%%%%%%%%%%%%%%%%%%%%%%%%%%%%%%%%%%%%%%%%%%%%%%%%%%%%
\end{enumerate}
\subsubsection*{\bf Dimension-eight and dimension-six operators from $\mathcal{O}(M^{0})$ term:}
\begin{enumerate}
	\item \underline{Covariant operator: $U^2$} 
	
	\begin{align}
		\tr[U^2] \supset  c_{\phi^8}^{\llbracket U^{2} \rrbracket} \mathcal{O}_{\phi^8}.
	\end{align}
	The matching relation at dimension-eight is:
	\begin{align}
		c_{\phi^8}^{\llbracket U^{2} \rrbracket} & = \frac{1}{2} \frac{\mu_\Delta^2}{M_\Delta^6} \lambda_1^2 \lambda_2  + \frac{3}{8} \frac{\mu_\Delta^2}{M_\Delta^6} \lambda_1^2 \lambda_3 + \frac{1}{2}\frac{\mu_\Delta^2}{M_\Delta^6}  \lambda_1 \lambda_2 \lambda_4+ \frac{3}{8}\frac{\mu_\Delta^2}{M_\Delta^6}  \lambda_1 \lambda_3 \lambda_4 + \frac{7}{8} \frac{\mu_\Delta^4}{M_\Delta^8} \lambda_2^2 \nonumber \\
		&\hspace{1cm}+ \frac{11}{8} \frac{\mu_\Delta^4}{M_\Delta^8} \lambda_2 \lambda_3 + \frac{1}{8} \frac{\mu_\Delta^2}{M_\Delta^6} \lambda_2 \lambda_4^2 + \frac{33}{32} \frac{\mu_\Delta^4}{M_\Delta^8} \lambda_3^2 + \frac{3}{32} \frac{\mu_\Delta^2}{M_\Delta^6} \lambda_3 \lambda_4^2 ~.
	\end{align}

	\begin{eqnarray}
		\text{Tr} \, [U^2] & \supset  c_6^{\llbracket U^{2} \rrbracket} \mathcal{O}_6 ,
	\end{eqnarray}  
	with the matching relation at dimension-six as:
	\begin{eqnarray}
		c_6^{\llbracket U^{2} \rrbracket} =   \dfrac{\mu_\Delta^2}{M_\Delta^4} \Big(\, 2 \lambda_1 \lambda_2 + \frac{3}{2} \lambda_1 \lambda_3 + \lambda_2  \lambda_4 + \frac{3}{4} \lambda_3 \lambda_4 \, \Big)~.
	\end{eqnarray}  
	%%%%%%%%%%%%%%%%%%%%%%%%%%%%%%%%%%%%%%%%%%%%%%%%%%%%%%%%%%%%%%%%%%%%%%%%%%%%%%%%%%%%%%%%%
\end{enumerate}
\subsubsection*{\bf Dimension-eight and dimension-six operators arising from $\mathcal{O}(M^{2})$ term:}

\begin{enumerate}
	\item \underline{Covariant operator: $U$}	
	\begin{align}
		\tr[U] & \supset c_{\phi^8}^{\llbracket U \rrbracket} \mathcal{O}_{\phi^8} + c^{(1),{\llbracket U \rrbracket}}_{\phi^6} \mathcal{O}^{(1)}_{\phi^6} + c^{(3),{\llbracket U \rrbracket}}_{\phi^6} \mathcal{O}^{(3)}_{\phi^6} + c^{(10),{\llbracket U \rrbracket}}_{\phi^4} \mathcal{O}^{(10)}_{\phi^4} + c^{(11),{\llbracket U \rrbracket}}_{\phi^4} \mathcal{O}^{(11)}_{\phi^4}.
	\end{align}
	The matching relations at dimension-eight are:
	\begin{align}
	    c_{\phi^8}^{\llbracket U \rrbracket}&=\frac{\mu_\Delta^2}{M_\Delta^8}\Big( 2 \lambda_1^2 \lambda_4^2 + 2 \lambda_1 \lambda_4^3 + \frac{1}{2} \lambda_4^4 \Big) - \frac{\mu_\Delta^4}{M_\Delta^{10}}\Big( \lambda_2^2 + \frac{11}{4} \lambda_2 \lambda_3 - \frac{3}{4} \lambda_3^2 \Big)~, \nonumber \\
		c^{(1),{\llbracket U \rrbracket}}_{\phi^6} &=  -\frac{\mu_\Delta^2}{M_\Delta^8}  \Big( 6 \lambda_1 \lambda_2 + 9 \lambda_1 \lambda_3 + 3 \lambda_2 \lambda_4 + \frac{9}{4} \lambda_3 \lambda_4 \Big) ~,\nonumber \\
		c^{(3),{\llbracket U \rrbracket}}_{\phi^6} &= -\frac{\mu_\Delta^2}{M_\Delta^8}  \Big( 6 \lambda_1 \lambda_2 + \frac{9}{2} \lambda_1 \lambda_3 + 3 \lambda_2 \lambda_4 + \frac{9}{4} \lambda_3 \lambda_4 \Big) ~, \nonumber \\
		c^{(10),{\llbracket U \rrbracket}}_{\phi^4} & = -\frac{\mu_\Delta^2}{M_\Delta^8}  \Big( 8 \lambda_2 + 6 \lambda_3 \Big)~,\nonumber \\
		c^{(11),{\llbracket U \rrbracket}}_{\phi^4} & =-\frac{\mu_\Delta^2}{M_\Delta^8}  \Big( 8 \lambda_2 + 6 \lambda_3 \Big)~.
	\end{align}

	\begin{eqnarray}
		\text{Tr} \, [U] & \supset c_H^{\llbracket U \rrbracket} \mathcal{O}_H + c_6^{\llbracket U \rrbracket} \mathcal{O}_6 ,
	\end{eqnarray}  
	with the matching relations at dimension-six as:
	\begin{eqnarray}
		c_H^{\llbracket U \rrbracket} = \dfrac{\mu_\Delta^2}{M^6} \Big(\, 8 \lambda_2 + 6 \lambda_3\Big)~,~~  c_6^{\llbracket U \rrbracket} =   - \dfrac{\mu_\Delta^2}{M^6} \Big(\, 2 \lambda_1 \lambda_2 + \lambda_2 \lambda_4 + \frac{3}{2}\lambda_1 \lambda_3  + \frac{3}{4}\lambda_3 \lambda_4 \, \Big)~.
	\end{eqnarray}  
	%%%%%%%%%%%%%%%%%%%%%%%%%%%%%%%%%%%%%%%%%%%%%%%%%%%%%%%%%%%%%%%%%%%%%%%%%%%%%%%%%%%%%%%%%
\end{enumerate}
The results discussed until now are summarized in Tables~\ref{tab:matching_triplet_complex_scalar_1} and ~\ref{tab:matching_triplet_complex_scalar_2}.

\subsubsection{Incorporating the effect of fermion Interactions: Complex Scalar Triplet}
\label{fermi_int_matching_ctm}
The Yukawa interaction between the SM fermions and the triplet scalar is depicted in Eqn.~\ref{triplet_fermi}.
The functionals $\widehat{B}$ and $\widehat{B}^\dag$, derived from the relevant part of the Lagrangian, Eqn.~\ref{lbsmCTS}, are given as
\begin{eqnarray}
	\widehat{B}_i = - \mu_\Delta \tilde{H}^\dagger \tau^i H + Y_{\Delta_{\alpha \beta}}^\ast \ell_{L_\alpha}^{r \dag} \tau_i i \sigma_2 C \ell_{L_\beta}^{s \ast}  = - \mu_\Delta \tilde{H}^\dagger \tau^i H  + Y_{\Delta_{\alpha \beta}}^\ast \bar{\ell}_{L_\alpha}^{r} \tau_i \tilde{\ell}_{L_\beta}^{s} ,\\
	\widehat{B}^\dag_i = - \mu_\Delta H^\dagger \tau^i \tilde{H} + Y_{\Delta_{\gamma \delta}} \ell_{L_\gamma}^{p T} C i \sigma_2 \tau_i \ell_{L_\delta}^{q} = - \mu_\Delta H^\dagger \tau^i \tilde{H} + Y_{\Delta_{\gamma \delta}} \bar{\tilde{\ell}}_{L_\gamma}^{p} \tau_i \ell_{L_\delta}^{q} . 
\end{eqnarray}
These functionals have mass dimension three.
Encapsulating these additional contributions, we find 
	\begin{equation}
		\Delta^\ast \Delta =  \widehat{B}^\dagger\frac{1}{M_\Delta^4}\widehat{B} + \widehat{B}^\dagger\frac{1}{M_\Delta^4}(P^2-U)\frac{1}{M_\Delta^2}\widehat{B}+ \widehat{B}^\dagger\frac{1}{M_\Delta^4}(P^2-U)\frac{1}{M_\Delta^2}(P^2-U)\frac{1}{M_\Delta^2}\widehat{B}+\cdots.
		\label{eq:delta_fermi_eff}
	\end{equation}
where $\widehat{B}_i^\dagger \widehat{B}_i$ is given by\footnote{We used the completeness relation for Pauli matrices $\tau^i_{\alpha \beta} \tau^i_{\gamma \delta} = \frac{1}{2} \delta_{\alpha \delta} \delta_{\beta \gamma} - \frac{1}{4} \delta_{\alpha \beta} \delta_{\gamma \delta}$.}:
\begin{align}
	\widehat{B}_i^\dagger \widehat{B}_i & = \frac{1}{4} \mu_{\Delta}^2 (H^\dagger H)^2 + \frac{1}{2} Y_{\Delta \gamma \delta} Y_{\Delta_{\alpha \beta}}^\ast \left( \bar{\ell}_{L_\alpha}^{r}  \ell_{L_\delta}^{q} \right)_i \left( \bar{\tilde \ell}_{L_\gamma}^{p}  \tilde{\ell}_{L_\beta}^{s} \right)_i \nonumber \\ & \hspace{1cm}- \frac{1}{4} Y_{\Delta \gamma \delta} Y_{\Delta_{\alpha \beta}}^\ast  \left( \bar{\ell}_{L_\alpha}^{r} \tilde{\ell}_{L_\beta}^{s} \right)_i \left( \bar{\tilde \ell}_{L_\gamma}^{p} \ell_{L_\delta}^{q}  \right)_i \nonumber  - \mu_\Delta Y_{\Delta_{\gamma \delta}} \tilde{H}^\dagger \tau^i H \ell_{L_\gamma}^{p T} C i \sigma_2 \tau_i \ell_{L_\delta}^{q} + \rm{h.c.} \nonumber  \\
	& = \frac{1}{4} \mu_{\Delta}^2 (H^\dagger H)^2 + Y_{\Delta \gamma \delta} Y_{\Delta_{\alpha \beta}}^\ast  \left( \bar{\ell}_{L_\alpha}^{r}  \ell_{L_\delta}^{q} \right)_i \left(  \bar{\ell}_{L_\beta}^{s} { \ell}_{L_\gamma}^{p} \right)_i - \frac{1}{2} \mu_\Delta Y_{\Delta_{\alpha \beta}} (\tilde{H}^\dagger \ell_{L_\gamma}^{p T}) C (\tilde{H}^\dagger \ell_{L_\delta}^q).     
	\label{reduce_bb}
\end{align}

After Fierz transformations\footnote{These read $\left(\bar{\psi}_{1L} \gamma_\mu \psi_{2L} \right) \left(\bar{\psi}_{3L} \gamma^\mu \psi_{4L} \right) = \left(\bar{\psi}_{1L} \gamma_\mu \psi_{4L} \right) \left(\bar{\psi}_{3L} \gamma^\mu \psi_{2L} \right)$.}, the four-fermion part in Eqn.~\ref{reduce_bb} can be further reduced to the form: 
\begin{align}
	(\widehat{B}_i^\dagger \widehat{B}_i)^{\rm lep} & = - \frac{1}{2} Y_{\Delta \gamma \delta} Y_{\Delta_{\alpha \beta}}^\ast   \left( \bar{\ell}_{L_\alpha}^{r} \gamma_\mu { \ell}_{L_\gamma}^{p} \right)_i \left( \bar{\ell}_{L_\beta}^{s} \gamma^\mu  \ell_{L_\delta}^{q}  \right)_i .   
	\label{reduce_bb_2}
\end{align}
Below, we systematically recollect all the additonal fermionic effective operators following the earlier principle. 

\subsubsection*{\bf {Dimension-five operator}}
We have only one dimension-five operator arising from the trace of $U$ in the effective Lagrangian~\ref{eq:finite}, which is the Weinberg operator: 
\begin{equation}
	\tr[U] \supset c_{\phi^2 L^2}^{\llbracket U \rrbracket} \mathcal{Q}_{\phi^2 L^2} = c_{\phi^2 L^2}^{\llbracket U \rrbracket} (\tilde{\phi}^\dagger \ell_{L_\gamma}^{p T}) C (\tilde{\phi}^\dagger \ell_{L_\delta}^q) \equiv c_{\phi^2 L^2}^{\llbracket U \rrbracket} \epsilon^{ij} \epsilon^{mn} \phi_i \phi_m (\ell_j^p)^T C \ell_n^q \nonumber
\end{equation}
with 
\begin{equation}
	c_{\phi^2 L^2}^{\llbracket U \rrbracket} = \frac{-1}{4 M_\Delta^4} \mu_\Delta Y_{\Delta_{pq}}  (4 \lambda_2 + 3 \lambda_3).
\end{equation}
It is worthy to mention that the operator $Q_{\phi^2 L^2}$ viotates the lepton number $L$ by two units. 
\subsubsection*{\bf {Dimension-six operator}}

\begin{equation}
	\tr[U] \supset c_{ll}^{\llbracket U \rrbracket} \mathcal{Q}_{ll} = c_{ll}^{\llbracket U \rrbracket} \left( \bar{\ell}_{L_\alpha}^{r} \gamma_\mu { \ell}_{L_\gamma}^{p} \right)_i \left( \bar{\ell}_{L_\beta}^{s} \gamma^\mu  \ell_{L_\delta}^{q}  \right) ~~ {\rm with} ~~
	c_{ll}^{\llbracket U \rrbracket} = - \frac{1}{4M_\Delta^2} Y_{\Delta_{p q}} Y_{\Delta_{r s}}^\ast  (4 \lambda_2 + 3 \lambda_3).
\end{equation}
\subsubsection*{\bf {Dimension-seven operators}}

\begin{align}
	\tr[U] & \supset c^{(1),\llbracket U \rrbracket}_{L \phi D} \mathcal{Q}^{(1)}_{L \phi D} + c^{(2),\llbracket U \rrbracket}_{L \phi D} \mathcal{Q}^{(2)}_{L \phi D} + c^{\llbracket U \rrbracket}_{L \phi} \mathcal{Q}_{L \phi} \nonumber \\
	& = c^{(1),\llbracket U \rrbracket}_{L \phi D}  \epsilon^{ij} \epsilon^{mn} \ell_i^p C (D^\mu \ell_j^q) \phi_m (D_\mu \phi_n)+ c^{(2),\llbracket U \rrbracket} \epsilon^{im} \epsilon^{jn} \ell_i^p C (D^\mu \ell_j^q) \phi_m (D_\mu \phi_n) \nonumber\\
	& \hspace{1cm} + c^{\llbracket U \rrbracket}_{L \phi} \epsilon^{ij} \epsilon^{mn} (\ell_i^p C \ell_m^q )  \phi_j \phi_n (\phi^\dagger \phi).
\end{align}
The matching relations are: 
\begin{align}
	&c^{(1),\llbracket U \rrbracket}_{L \phi D} =   \frac{-1}{2 M_\Delta^6} \mu_\Delta Y_{\Delta_{pq}}  ( 4 \lambda_2 + 3 \lambda_3)~, \hspace{1cm} c^{(2),\llbracket U  \rrbracket}_{L \phi D} =  \frac{-1}{2 M_\Delta^6} \mu_\Delta Y_{\Delta_{pq}} ( 4 \lambda_2 + 3 \lambda_3 ) ~,\nonumber\\
	%\end{align}\vspace{-1mm}
	%\begin{align}
	&c^{\llbracket U \rrbracket}_{L \phi} =   \frac{-1}{2 M_\Delta^6}  \mu_\Delta Y_{\Delta_{pq}} \Big( \lambda_2 \lambda_4 -\frac{3}{2} (\lambda_2+ \lambda_3) \lambda_4 + \lambda_2 (2 \lambda_1+\lambda_4) + \frac{3}{2} (\lambda_2 + \lambda_3)(2 \lambda_1 + \lambda_4)\Big) ~.
\end{align}

\begin{align}
	\tr[U^2] \supset  c^{\llbracket U^{2} \rrbracket}_{L \phi} \mathcal{Q}_{L \phi} .
\end{align}
The matching relation is:
\begin{gather}
	c^{\llbracket U^{2} \rrbracket}_{L \phi} =  \frac{1}{2 M_\Delta^4}  \mu_\Delta Y_{\Delta_{pq}} \Big( 2 \lambda_1 \lambda_2 + \frac{3}{2} \lambda_1 \lambda_3 + \lambda_2 \lambda_4 + \frac{3}{4} \lambda_3 \lambda_4 \Big).\nonumber
\end{gather}
\subsubsection*{\bf {Dimension-eight operators}}

\begin{align}
	\tr[U] &\supset  c^{(1), \llbracket U \rrbracket}_{l^4 \phi^2} \mathcal{Q}_{l^4 \phi^2}^{(1)} + c^{(2), \llbracket U \rrbracket}_{l^4 \phi^2} \mathcal{Q}_{l^4 \phi^2}^{(2)} + c^{(1), \llbracket U \rrbracket}_{l^4 D^2} \mathcal{Q}_{l^4 D^2}^{(1)}.
\end{align}
The matching relations are: 
\begin{align}
	c^{(1),\llbracket U \rrbracket}_{l^4 \phi^2} &=   \frac{1}{4 M_\Delta^6} Y_{\Delta_{rt}} Y_{\Delta_{ps}}^\ast \Big( 10 \lambda_1 \lambda_2 + 6 \lambda_1 \lambda_3 + 5 \lambda_2 \lambda_4 + 3 \lambda_3 \lambda_4 \Big) ~,\nonumber\\  
	c^{(2),\llbracket U  \rrbracket}_{l^4 \phi^2} &=  \frac{1}{4 M_\Delta^6}  Y_{\Delta_{rt}} Y_{\Delta_{ps}}^\ast \Big( 6 \lambda_1 \lambda_4 + 2 \lambda_2 \lambda_4 + 3 \lambda_3 \lambda_4 \Big)    ~,\nonumber\\
	c^{\llbracket U \rrbracket}_{l^4 D^2} & =  \frac{1}{4 M_\Delta^6}  Y_{\Delta_{rt}} Y_{\Delta_{ps}}^\ast \Big( 20 \lambda_2 + 12 \lambda_3 \Big) ~.
\end{align}

\begin{align}
	\tr[U^2] &\supset  c^{(1), \llbracket U^{2} \rrbracket}_{l^4 \phi^2} \mathcal{Q}_{l^4 \phi^2}^{(1)} + c^{(2), \llbracket U^{2} \rrbracket}_{l^4 \phi^2} \mathcal{Q}_{l^4 \phi^2}^{(2)} .
\end{align}
The matching relations are:
\begin{gather}
	c^{(1), \llbracket U^{2} \rrbracket}_{l^4 \phi^2} = - \frac{1}{M_\Delta^4}  Y_{\Delta_{rt}} Y_{\Delta_{ps}}^\ast \Big(  \lambda_1 \lambda_2 + \frac{3}{4} \lambda_1 \lambda_3 + \frac{1}{8} \lambda_3 \lambda_4 +  \frac{1}{2} \lambda_2 \lambda_4 \Big)~, ~c^{(2), \llbracket U^{2} \rrbracket}_{l^4 \phi^2} = \frac{1}{4 M_\Delta^4}  Y_{\Delta_{rt}} Y_{\Delta_{ps}}^\ast \lambda_4 \lambda_2.
\end{gather}
The effective operators for fermionic interactions discussed so far are summarized in Table~\ref{tab:matching_triplet_complex_scalar_3}.
\vskip 1 \baselineskip
\subsection{Case II: Integrating Out Complex Doublet Scalar}
\label{app:2hdm}

In this section, we discuss the integration out of the heavy complex doublet. We follow the same strategy similar to the complex triplet case.  We define a basis for the heavy doublet field $\left(\mathcal{H}_2, \mathcal{H}_2^*\right)$, and the classical solutions as
\begin{align}
	\begin{pmatrix}
		{\mathcal{H}_2}_i \\ {\mathcal{H}_2}^\ast_i 
	\end{pmatrix}
	&=  
	\frac{1}{M_{\mathcal{H}_{2}}^2}
	\begin{pmatrix}
		B_i \\ B^\ast_i 
	\end{pmatrix}
	+ 
	\frac{1}{M_{\mathcal{H}_{2}}^4}
	\left(  \left( P^2 \delta_{ij} \right) 
	\begin{pmatrix}
		\mathbb{1} & 0 \\
		0 & \mathbb{1} 
	\end{pmatrix}
	-
	\begin{pmatrix}
		\left(U_{11}\right)_{ij} & \left(U_{12}\right)_{ij} \\
		\left(U_{21}\right)_{ij} & \left(U_{22}\right)_{ij}
	\end{pmatrix} 
	\right)
	\begin{pmatrix}
		B_j \\ B^\ast_j 
	\end{pmatrix} \nonumber \\ 
	&\hspace{1cm}+
	\frac{1}{M_{\mathcal{H}_{2}}^6}
	\left(  \left( P^2 \delta_{ij} \right) 
	\begin{pmatrix}
		\mathbb{1}  & 0 \\
		0 & \mathbb{1}  
	\end{pmatrix}
	-
	\begin{pmatrix}
		\left(U_{11}\right)_{ij} & \left(U_{12}\right)_{ij} \\
		\left(U_{21}\right)_{ij} & \left(U_{22}\right)_{ij}
	\end{pmatrix} 
	\right)^2
	\begin{pmatrix}
		B_j \\ B^\ast_j 
	\end{pmatrix} + \cdots.
	\label{doubletsol111}
\end{align}
The heavy field is gone through the covariant derivative expansion and we truncate that exapnsion suitably tracking their contributions up to dimension-eight effective operators. 
		\begin{itemize}
			\item The quadratic contribution of the heavy field in $U$ as underlined in the following equation for $U_{11}$ receives contribution of the first term of the solution of $\mathcal{H}_2$ from Eqn.~\ref{doubletsol111}, i.e. $\mathcal{O}(B/M_{\mathcal{H}_2}^2)$: 
			\begin{eqnarray}
				\left(U_{11}^{\mathcal{H}_2}\right)_{i j}  &=& \frac{\lambda_{\mathcal{H}_2}}{2} \underline{ \left[\delta_{ij} ({\mathcal{H}_2}_k^* {\mathcal{H}_2}_k) + {\mathcal{H}_2}_j^* {\mathcal{H}_2}_i \right] }\nonumber \\
				&&- \eta_{\mathcal{H}_2} \uwave{\left[ \delta_{ij} \left( \tilde{H}_k^* {\mathcal{H}_2}_k + {\mathcal{H}_2}_k^* \tilde{H}_k \right) + {\mathcal{H}_2}_i \tilde{H}_j^* + {\mathcal{H}_2}_j^* \tilde{H}_i \right]} \nonumber \\
				&& + \lambda_{\mathcal{H}_2,1} \delta_{ij} \left( H_k^* H_k \right) + \lambda_{\mathcal{H}_2,2} \left(\tilde{H}_j \tilde{H}_i^* \right) .
				\label{2hdmmatrix_u11}
			\end{eqnarray}
			\item For the wave-underlined part of $U_{11}$, we can get dimension-eight operators in two ways. The solution for the heavy field of $\mathcal{H}_2$  needs to be truncated at $\mathcal{O}(B^\dag (P^2-U)^2 B)$ term. First, we consider the linear term in $U$ of this solution $( \mathcal{O}(B^\dag (P^2-U) B))$ wherein $U$ receives the contribution from this wave-underlined part of $U$ and we restrict to $\mathcal{O}(B/M_{\mathcal{H}_2}^2)$ in solution of $\mathcal{H}_2$. The second way in which the dimension-eight operators can be generated is from the quadratic term in $U$, i.e. $\mathcal{O}(B^\dag (P^2-U)^2 B)$, where $U$ gets contribution from the light fields only.
            \item Trace of $U^2$ can give rise to dimension-eight operators in two ways: (i) One of the $U's$ will carry dependence from heavy field solution up to dimension-six, i.e. up to $B^\dagger U B$ term while the other $U$ will have dimension-two terms from the light fields only. (ii) The other way is that both the $U's$ interact via the heavy fields only with the heavy field solution being truncated up to first term $\frac{B}{M^2}$, so that each $U$ having quadratic dependence on $|{H_2}_c|^2$ will have dimension-four operators, from the $B^\dagger B$ term. Similarly, we can compute dimension-eight operators from terms $U^3$, $(P_\mu U)^2$ and $U(G_{\mu \nu})^2$ of the total effective Lagrangian.
		\end{itemize} 

The WCs associated with the effective operators are functions of the following BSM parameters: $\lambda_{\mathcal{H}_2}, \lambda_{\mathcal{H}_2,1}, \lambda_{\mathcal{H}_2,2}, \lambda_{\mathcal{H}_2,3}, \eta_{\mathcal{H}_2}, \eta_H, Y_{\mathcal{H}_2}^{(e)}, Y_{\mathcal{H}_2}^{(u)}$, $Y_{\mathcal{H}_2}^{(d)}$ along with the SM ones. 
\vskip 1 \baselineskip

\subsubsection*{\bf Dimension-eight operators from $\mathcal{O}(M^{-4})$ term:}
\begin{enumerate}
	\item \underline{Covariant operator: $U^4$} \\
	\begin{align}
		\tr[U^4] \supset  c^{\llbracket U^4 \rrbracket}_{\phi^8} \mathcal{O}_{\phi^8} .
	\end{align}
	\begin{align}
		c^{\llbracket U^4 \rrbracket}_{\phi^8} &= 4 \lambda_{\mathcal{H}_2,1}^4 + 8 \lambda_{\mathcal{H}_2,1}^3 \lambda_{\mathcal{H}_2,2} + 12 \lambda_{\mathcal{H}_2,1}^2 \lambda_{\mathcal{H}_2,2}^2 + 48 \lambda_{\mathcal{H}_2,1}^2 \lambda_{\mathcal{H}_2,3}^2 + 8 \lambda_{\mathcal{H}_2,1} \lambda_{\mathcal{H}_2,2}^3 \nonumber \\
		&+ 96 \lambda_{\mathcal{H}_2,1} \lambda_{\mathcal{H}_2,2} \lambda_{\mathcal{H}_2,3}^2 + 4 \lambda_{\mathcal{H}_2,2}^4 + 48 \lambda_{\mathcal{H}_2,2}^2 \lambda_{\mathcal{H}_2,3}^2 + 32 \lambda_{\mathcal{H}_2,3}^4.
	\end{align}
	%%%%%%%%%%%%%%%%%%%%%%%%%%%%%%%%%%%%%%%%%%%%%%%%%%%%%%%%%%%%%%%%%%%%%%%%%%%%%%%%%%%%%%%%%
	\item \underline{Covariant operator: $U^2 (P^2 U)$}\\
	\begin{align}
		\tr[U^2 (P^2 U)] &\supset c^{(3), \llbracket U^2 (P^2 U) \rrbracket}_{\phi^6} \mathcal{O}^{(3)}_{\phi^6}  .
	\end{align}
	\begin{align}
		c^{(3), \llbracket U^2 (P^2 U) \rrbracket}_{\phi^6} & = 4 \lambda_{\mathcal{H}_2,1}^3 + 6 \lambda_{\mathcal{H}_2,1}^2 \lambda_{\mathcal{H}_2,2} + 6 \lambda_{\mathcal{H}_2,1} \lambda_{\mathcal{H}_2,2}^2 + 24 \lambda_{\mathcal{H}_2,1} \lambda_{\mathcal{H}_2,3}^2 + 4 \lambda_{\mathcal{H}_2,2}^3 + 24 \lambda_{\mathcal{H}_2,2} \lambda_{\mathcal{H}_2,3}^2.
	\end{align}
	%%%%%%%%%%%%%%%%%%%%%%%%%%%%%%%%%%%%%%%%%%%%%%%%%%%%%%%%%%%%%%%%%%%%%%%%%%%%%%%%%%%%%%%%%
	\item \underline{Covariant operator: $U^2 (G_{\mu \nu})^2$} \\
	\begin{align}
		\tr[U^2 (G_{\mu \nu})^2] & \supset c^{(1), \llbracket U^2 (G_{\mu \nu})^2 \rrbracket}_{W^2 \phi^4} \mathcal{O}^{(1)}_{W^2 \phi^4} + c^{(1), \llbracket U^2 (G_{\mu \nu})^2 \rrbracket}_{B^2  \phi^4} \mathcal{O}^{(1)}_{B^2  \phi^4} + c^{(1), \llbracket U^2 (G_{\mu \nu})^2 \rrbracket}_{W B \phi^4} \mathcal{O}^{(1)}_{W B \phi^4}  .
	\end{align}
	\begin{align}
		c^{(1), \llbracket U^2 (G_{\mu \nu})^2 \rrbracket}_{W^2 \phi^4} & =   -  g^2 \left(  \lambda_{\mathcal{H}_2,1}^2 + \lambda_{\mathcal{H}_2,1} \lambda_{\mathcal{H}_2,2} + \frac{1}{2} \lambda_{\mathcal{H}_2,2}^2 + \lambda_{\mathcal{H}_2,3}^2\right)  ~,\nonumber \\
		 c^{(1), \llbracket U^2 (G_{\mu \nu})^2 \rrbracket}_{B^2  \phi^4} & =  - 2 g'^2 \left( 2 \lambda_{\mathcal{H}_2,1}^2 + \lambda_{\mathcal{H}_2,2}^2 + 2 \lambda_{\mathcal{H}_2,1} \lambda_{\mathcal{H}_2,2} + 2 \lambda_{\mathcal{H}_2,3}^2\right) ~,\nonumber\\
		c^{(1), \llbracket U^2 (G_{\mu \nu})^2 \rrbracket}_{W B \phi^4}  & = -8 g g' \lambda_{\mathcal{H}_2,1} \lambda_{\mathcal{H}_2,2} .
	\end{align}
	%%%%%%%%%%%%%%%%%%%%%%%%%%%%%%%%%%%%%%%%%%%%%%%%%%%%%%%%%%%%%%%%%%%%%%%%%%%%%%%%%%%%%%%%%
	\item \underline{Covariant operator: $(U G_{\mu \nu})^2$} \\
	\begin{align}
		\tr[(U G_{\mu \nu})^2] &\supset c^{(1), \llbracket (U G_{\mu \nu})^2 \rrbracket}_{W^2 \phi^4} \mathcal{O}^{(1)}_{W^2 \phi^4}  + c^{(1), \llbracket (U G_{\mu \nu})^2 \rrbracket}_{B^2  \phi^4} \mathcal{O}^{(1), \llbracket (U G_{\mu \nu})^2 \rrbracket}_{B^2  \phi^4} + c^{(1)}_{W B \phi^4} \mathcal{O}^{(1)}_{W B \phi^4}  .
	\end{align}
	\begin{align}
		c^{(1), \llbracket (U G_{\mu \nu})^2 \rrbracket}_{W^2 \phi^4} & =  -  g^2 \left(  \lambda_{\mathcal{H}_2,1}^2 + \lambda_{\mathcal{H}_2,1} \lambda_{\mathcal{H}_2,2} + \frac{1}{2} \lambda_{\mathcal{H}_2,2}^2 + \lambda_{\mathcal{H}_2,3}^2\right)  ~,\nonumber \\
		c^{(1), \llbracket (U G_{\mu \nu})^2 \rrbracket}_{B^2  \phi^4} & =  - 2 g'^2 \left( 2 \lambda_{\mathcal{H}_2,1}^2 + \lambda_{\mathcal{H}_2,2}^2 + 2 \lambda_{\mathcal{H}_2,1} \lambda_{\mathcal{H}_2,2} + 2 \lambda_{\mathcal{H}_2,3}^2\right) ~,\nonumber\\
		c^{(1), \llbracket (U G_{\mu \nu})^2 \rrbracket}_{W B \phi^4}  & =-8 g g' \lambda_{\mathcal{H}_2,1} \lambda_{\mathcal{H}_2,2}. 
	\end{align}
	%%%%%%%%%%%%%%%%%%%%%%%%%%%%%%%%%%%%%%%%%%%%%%%%%%%%%%%%%%%%%%%%%%%%%%%%%%%%%%%%%%%%%%%%%
	\item \underline{Covariant operator: $(P^2 U)^2$} \\ 
	\begin{align}
		\tr[(P^2 U)^2] & \supset c^{(4),\llbracket (P^2 U)^2 \rrbracket}_{\phi^4} \mathcal{O}^{(4)}_{\phi^4} + c^{(8),\llbracket (P^2 U)^2 \rrbracket}_{\phi^4} \mathcal{O}^{(8)}_{\phi^4} + c^{(10),\llbracket (P^2 U)^2 \rrbracket}_{\phi^4} \mathcal{O}^{(10)}_{\phi^4}\nonumber \\
		&\hspace{1cm} + c^{(11),\llbracket (P^2 U)^2 \rrbracket}_{\phi^4} \mathcal{O}^{(11)}_{\phi^4} + c^{(12),\llbracket (P^2 U)^2 \rrbracket}_{\phi^4} \mathcal{O}^{(12)}_{\phi^4} .
	\end{align}
	\begin{align}
		c^{(4), \llbracket (P^2 U)^2 \rrbracket}_{\phi^4} &= 8 \left( 4 \lambda_{\mathcal{H}_2,1}^2 + 4 \lambda_{\mathcal{H}_2,1} \lambda_{\mathcal{H}_2,2} + \lambda_{\mathcal{H}_2,2}^2 \right)~,\nonumber \\
		 c^{(8), \llbracket (P^2 U)^2 \rrbracket}_{\phi^4}& =2 \left( 4 \lambda_{\mathcal{H}_2,1}^2 + 4 \lambda_{\mathcal{H}_2,1} \lambda_{\mathcal{H}_2,2} + \lambda_{\mathcal{H}_2,2}^2 \right)~,\nonumber \\
		 c^{(10), \llbracket (P^2 U)^2 \rrbracket}_{\phi^4}& = 8 \lambda_{\mathcal{H}_2,3}^2 ~,\nonumber \\
		c^{(11), \llbracket (P^2 U)^2 \rrbracket}_{\phi^4}& = 4 \left( 4 \lambda_{\mathcal{H}_2,1}^2 + 4 \lambda_{\mathcal{H}_2,1} \lambda_{\mathcal{H}_2,2} + \lambda_{\mathcal{H}_2,2}^2 + \lambda_{\mathcal{H}_2,3}^2  \right)~,\nonumber \\
	   c^{(12), \llbracket (P^2 U)^2 \rrbracket}_{\phi^4} & = - 8 \left( 4 \lambda_{\mathcal{H}_2,1}^2 + 4 \lambda_{\mathcal{H}_2,1} \lambda_{\mathcal{H}_2,2} + \lambda_{\mathcal{H}_2,2}^2 \right).
	\end{align}
	%%%%%%%%%%%%%%%%%%%%%%%%%%%%%%%%%%%%%%%%%%%%%%%%%%%%%%%%%%%%%%%%%%%%%%%%%%%%%%%%%%%%%%%%%
	\item \underline{Covariant operator: $U (J_\mu)^2$} \\
	\begin{align}
		\tr[U (J_\mu)^2] & \supset c^{(6), \llbracket U (J_\mu)^2 \rrbracket}_{B^2 \phi^2 D^2} \mathcal{O}^{(6)}_{B^2 \phi^2 D^2} + c^{(8), \llbracket U (J_\mu)^2 \rrbracket}_{B^2 \phi^2 D^2} \mathcal{O}^{(8)}_{B^2 \phi^2 D^2} + c^{(10), \llbracket U (J_\mu)^2 \rrbracket}_{W B \phi^2 D^2} \mathcal{O}^{(10)}_{W B \phi^2 D^2} \nonumber \\
		& \hspace{1cm}+ c^{(13), \llbracket U (J_\mu)^2 \rrbracket}_{W^2 \phi^2 D^2} \mathcal{O}^{(13)}_{W^2 \phi^2 D^2}  .
	\end{align}
	\begin{align}
		c^{(6), \llbracket U (J_\mu)^2 \rrbracket}_{B^2 \phi^2 D^2} & = -4 g'^2 (2\lambda_{\mathcal{H}_2,1} + \lambda_{\mathcal{H}_2,2})  ~, &c^{(8), \llbracket U (J_\mu)^2 \rrbracket}_{B^2 \phi^2 D^2} & = -4 g'^2 (2\lambda_{\mathcal{H}_2,1} + \lambda_{\mathcal{H}_2,2})~,   \nonumber\\
		c^{(10), \llbracket U (J_\mu)^2 \rrbracket}_{W B \phi^2 D^2} & = -4  gg' \lambda_{\mathcal{H}_2,2} ~,&c^{(13), \llbracket U (J_\mu)^2 \rrbracket}_{W^2 \phi^2 D^2} & =- g^2 (2\lambda_{\mathcal{H}_2,1} + \lambda_{\mathcal{H}_2,2})~.
	\end{align}
	%%%%%%%%%%%%%%%%%%%%%%%%%%%%%%%%%%%%%%%%%%%%%%%%%%%%%%%%%%%%%%%%%%%%%%%%%%%%%%%%%%%%%%%%%
	\item \underline{Covariant operator: $U (P_\mu U) J_\mu $} \\
	$\tr[U (P_\mu U) J_\mu]$ gets contributions from $\mathcal{O}^{(3)}_{B^2 \phi^2 D^2}$ of the Green's basis
	\begin{align}
		\tr[U (P_\mu U) J_\mu] &\supset c^{(3), \llbracket U (P_\mu U) J_\mu \rrbracket}_{B^2 \phi^2 D^2} \mathcal{O}^{(3)}_{B^2 \phi^2 D^2} .
	\end{align}
	The matching relation is: 
	\begin{align}
		c^{(3), \llbracket U (P_\mu U) J_\mu \rrbracket}_{B \phi^4 D^2}  & = -2 g' (2 \lambda_{\mathcal{H}_2,1}^2 + \lambda_{\mathcal{H}_2,2}^2 - \lambda_{\mathcal{H}_2,1} \lambda_{\mathcal{H}_2,2} - 2 \lambda_{\mathcal{H}_2,3}^2).  
	\end{align}
	%%%%%%%%%%%%%%%%%%%%%%%%%%%%%%%%%%%%%%%%%%%%%%%%%%%%%%%%%%%%%%%%%%%%%%%%%%%%%%%%%%%%%%%%%
	\item \underline{Covariant operator: $(P^2 U) (G_{\rho \sigma})^2$} \\
	\begin{align}
		\tr[(P^2 U) (G_{\rho \sigma})^2]  &\supset c^{(2), \llbracket (P^2 U) (G_{\rho \sigma})^2 \rrbracket}_{W^2 \phi^2 D^2} \mathcal{O}^{(4)}_{W^2 \phi^2 D^2} + c^{(4), \llbracket (P^2 U) (G_{\rho \sigma})^2 \rrbracket}_{W^2 \phi^2 D^2} \mathcal{O}^{(2)}_{W^2 \phi^2 D^2}  \nonumber \\ & + c^{(2), \llbracket (P^2 U) (G_{\rho \sigma})^2 \rrbracket}_{B^2 \phi^2 D^2} \mathcal{O}^{(2)}_{B^2 \phi^2 D^2} + c^{(1), \llbracket  (P^2 U) (G_{\rho \sigma})^2 \rrbracket}_{W B \phi^2 D^2} \mathcal{O}^{(1)}_{W B \phi^2 D^2}  \nonumber \\ 
		& -\frac{g^2}{2} (2 \lambda_{\mathcal{H}_2,1}+\lambda_{\mathcal{H}_2,2}) \underline{(D^2 \phi^\dag \phi+\phi^\dag D^2 \phi) W_{\rho \sigma}^a W^{a, \rho \sigma}} \nonumber \\ 
		& - i g^2 \lambda_{\mathcal{H}_2,2} \epsilon_{abc} \underline{(D^2 \phi^\dag \tau^c \phi + \phi^\dag \tau^c D^2 \phi) W_{\rho \sigma}^a W^{b, \rho \sigma} } \nonumber \\ \label{eomphi}
		& + 4 g g' \lambda_{\mathcal{H}_2,2} \underline{ (D^2 \phi^\dag \tau^a \phi + \phi^\dag \tau^a D^2 \phi) B_{\rho \sigma} W_{\rho \sigma}^a } \nonumber \\
		& - 2 (2 \lambda_{\mathcal{H}_2,1} + \lambda_{\mathcal{H}_2,2}) g'^2 \underline{(D^2 \phi^\dag \phi+\phi^\dag D^2 \phi) B_{\rho \sigma} B^{\rho \sigma}} \\
		= &~c^{(2), \llbracket  (P^2 U) (G_{\rho \sigma})^2 \rrbracket}_{W^2 \phi^2 D^2}  (D^\mu\phi^\dagger D_\mu\phi) W_{\nu\rho}^I W^{I\nu\rho} \nonumber \\
		& + c^{(4), \llbracket  (P^2 U) (G_{\rho \sigma})^2 \rrbracket}_{W^2 \phi^2 D^2} i\epsilon^{IJK}(D^\mu\phi^\dagger\sigma^I D^\nu\phi) W_{\mu\rho}^J W_\nu^{K\rho} \nonumber \\
		& + c^{(2), \llbracket  (P^2 U) (G_{\rho \sigma})^2 \rrbracket}_{B^2 \phi^2 D^2} (D_\mu\phi^\dagger \phi+\phi^\dagger D_\mu\phi) D_\nu B^{\mu\rho} B^\nu_{\,\,\rho} \nonumber \\
		& + c^{(1), \llbracket  (P^2 U) (G_{\rho \sigma})^2 \rrbracket}_{W B \phi^2 D^2} (D^\mu\phi^\dagger\sigma^I D_\mu\phi) B_{\nu\rho}W^{I\nu\rho} + {\rm EOM}.
	\end{align}
	The underlined terms in Eqn.~\ref{eomphi} get reduced  by the EOM for the scalar field to other classes. 
	\begin{align}
		c^{(2), \llbracket  (P^2 U) (G_{\rho \sigma})^2 \rrbracket}_{W^2 \phi^2 D^2} & = -g^2 (2 \lambda_{\mathcal{H}_2,1} + \lambda_{\mathcal{H}_2,2})   ~,&c^{(4), \llbracket  (P^2 U) (G_{\rho \sigma})^2 \rrbracket}_{W^2 \phi^2 D^2} &= -2 g^2 \lambda_{\mathcal{H}_2,2}~,  \nonumber\\
		c^{(2), \llbracket  (P^2 U) (G_{\rho \sigma})^2 \rrbracket}_{B^2 \phi^2 D^2}  & = -4 g'^2 (2 \lambda_{\mathcal{H}_2,1}+ \lambda_{\mathcal{H}_2,2})  ~,&c^{(1), \llbracket  (P^2 U) (G_{\rho \sigma})^2 \rrbracket}_{W B \phi^2 D^2} &= -8 gg' \lambda_{\mathcal{H}_2,2}~.
	\end{align}
	%%%%%%%%%%%%%%%%%%%%%%%%%%%%%%%%%%%%%%%%%%%%%%%%%%%%%%%%%%%%%%%%%%%%%%%%%%%%%%%%%%%%%%%%%
	\item \underline{Covariant operator: $(P_\mu P_\nu U) G_{\rho \mu} G_{\rho \nu} $} \\
	\begin{align}
		\tr[(P_\mu P_\nu U) G_{\rho \mu} G_{\rho \nu} ]  &\supset c^{(4), \llbracket (P_\mu P_\nu U) G_{\rho \mu} G_{\rho \nu} \rrbracket}_{W^2 \phi^2 D^2} \mathcal{O}^{(4)}_{W^2 \phi^2 D^2} + c^{(11), \llbracket (P_\mu P_\nu U) G_{\rho \mu} G_{\rho \nu} \rrbracket}_{W^2 \phi^2 D^2} \mathcal{O}^{(11)}_{W^2 \phi^2 D^2} \nonumber \\
		& + c^{(14), \llbracket (P_\mu P_\nu U) G_{\rho \mu} G_{\rho \nu}  \rrbracket}_{W^2 \phi^2 D^2} \mathcal{O}^{(14)}_{W^2 \phi^2 D^2} + c^{(4), \llbracket (P^2 U) (G_{\rho \sigma})^2 \rrbracket}_{B^2 \phi^2 D^2} \mathcal{O}^{(4)}_{B^2 \phi^2 D^2} \nonumber \\ 
		&  + c^{(8), \llbracket  (P^2 U) (G_{\rho \sigma})^2 \rrbracket}_{B^2 \phi^2 D^2} \mathcal{O}^{(8)}_{B^2 \phi^2 D^2}  +  c^{(8), \llbracket  (P^2 U) (G_{\rho \sigma})^2 \rrbracket}_{W B \phi^2 D^2} \mathcal{O}^{(8)}_{W B \phi^2 D^2} \nonumber \\
		& + c^{(10), \llbracket  (P^2 U) (G_{\rho \sigma})^2 \rrbracket}_{W B \phi^2 D^2} \mathcal{O}^{(10)}_{W B \phi^2 D^2} +c^{(11), \llbracket  (P^2 U) (G_{\rho \sigma})^2 \rrbracket}_{W B \phi^2 D^2} \mathcal{O}^{(11)}_{W B \phi^2 D^2} \nonumber\\
		& +c^{(13), \llbracket  (P^2 U) (G_{\rho \sigma})^2 \rrbracket}_{W B \phi^2 D^2} \mathcal{O}^{(13)}_{W B \phi^2 D^2} .
	\end{align}
	\begin{align}
		c^{(4), \llbracket (P_\mu P_\nu U) G_{\rho \mu} G_{\rho \nu} \rrbracket}_{W^2 \phi^2 D^2} & = - g^2 \lambda_{\mathcal{H}_2,2}   ~,&
		c^{(11), \llbracket (P_\mu P_\nu U) G_{\rho \mu} G_{\rho \nu} \rrbracket}_{W^2 \phi^2 D^2}& = \frac{g^2}{2} (2 \lambda_{\mathcal{H}_2,1} + \lambda_{\mathcal{H}_2,2})~,  \nonumber\\
		c^{(14), \llbracket (P_\mu P_\nu U) G_{\rho \mu} G_{\rho \nu} \rrbracket}_{W^2 \phi^2 D^2}& = \frac{g^2}{2} (2 \lambda_{\mathcal{H}_2,1} + \lambda_{\mathcal{H}_2,2}) ~,&
		c^{(4), \llbracket (P_\mu P_\nu U) G_{\rho \mu} G_{\rho \nu} \rrbracket}_{B^2 \phi^2 D^2}&= g'^2 (2 \lambda_{\mathcal{H}_2,1} + \lambda_{\mathcal{H}_2,2}) \nonumber~, \\
		c^{(8), \llbracket (P_\mu P_\nu U) G_{\rho \mu} G_{\rho \nu} \rrbracket}_{B^2 \phi^2 D^2}&= g'^2 (2 \lambda_{\mathcal{H}_2,1} + \lambda_{\mathcal{H}_2,2}) ~,& 
		c^{(8), \llbracket (P_\mu P_\nu U) G_{\rho \mu} G_{\rho \nu} \rrbracket}_{W B \phi^2 D^2}& = -2 g g' \lambda_{\mathcal{H}_2,2} ~,\nonumber\\
		c^{(10), \llbracket (P_\mu P_\nu U) G_{\rho \mu} G_{\rho \nu} \rrbracket}_{W B \phi^2 D^2}& = -2 g g' \lambda_{\mathcal{H}_2,2} ~,& 
		c^{(11), \llbracket (P_\mu P_\nu U) G_{\rho \mu} G_{\rho \nu} \rrbracket}_{W B \phi^2 D^2}& = 2 g g' \lambda_{\mathcal{H}_2,2} ~,\nonumber \\
		c^{(13), \llbracket (P_\mu P_\nu U) G_{\rho \mu} G_{\rho \nu} \rrbracket}_{W B \phi^2 D^2}& = -2 g g' \lambda_{\mathcal{H}_2,2}~.&&
	\end{align}
	%%%%%%%%%%%%%%%%%%%%%%%%%%%%%%%%%%%%%%%%%%%%%%%%%%%%%%%%%%%%%%%%%%%%%%%%%%%%%%%%%%%%%%%%%
	\item \underline{Covariant operator: $ U G_{\mu \nu} G_{\nu \rho} G_{\rho \mu}$} \\
	\begin{align}
		\tr[U G_{\mu \nu} G_{\nu \rho} G_{\rho \mu}] & \supset c^{(1), \llbracket U G_{\mu \nu} G_{\nu \rho} G_{\rho \mu} \rrbracket}_{W^3 \phi^2} \mathcal{O}^{(1)}_{W^3 \phi^2} .
	\end{align}
	\begin{align}
		c^{(1), \llbracket U G_{\mu \nu} G_{\nu \rho} G_{\rho \mu} \rrbracket}_{W^3 \phi^2} & = -g^3 \lambda_{\mathcal{H}_2,1}.
	\end{align}
	%%%%%%%%%%%%%%%%%%%%%%%%%%%%%%%%%%%%%%%%%%%%%%%%%%%%%%%%%%%%%%%%%%%%%%%%%%%%%%%%%%%%%%%%%
\end{enumerate}
%%%%%%%%%%%%%%%%%%%%%%%%%%%%%%%%%%%%%%%%%%%%%%%%%%%%%%%%%%%%%%%%%%%%%%%%%%%%%%%%%%%%%%%%%

\subsubsection*{\bf Dimension-eight and dimension-six operators from $\mathcal{O}(M^{-2})$ term:}
	\begin{enumerate}
		\item \underline{Covariant operator: $U^3$} 
		\begin{align}
			\tr[U^3] \supset c_{\phi^8}^{\llbracket U^3 \rrbracket} \mathcal{O}_{\phi^8} .
		\end{align}
		The matching relation at dimension-eight is:
		\begin{gather}
			c_{\phi^8}^{\llbracket U^3 \rrbracket} =  \frac{-48 \eta_H \eta_{\mathcal{H}_2}}{M_{\mathcal{H}_2}^2} \Big( \lambda_{\mathcal{H}_2,1}^2  +  \lambda_{\mathcal{H}_2,1} \lambda_{\mathcal{H}_2,2} +  \lambda_{\mathcal{H}_2,1} \lambda_{\mathcal{H}_2,3} +  \lambda_{\mathcal{H}_2,2} \lambda_{\mathcal{H}_2,3} + \frac{1}{2}\lambda_{\mathcal{H}_2,2}^2 + 4 \lambda_{\mathcal{H}_2,3}^2 \Big) .
		\end{gather}
		\begin{eqnarray}
			\text{Tr} [U^3] & \supset  c_6^{\llbracket U^3 \rrbracket} \mathcal{O}_6 = c_6 (\phi^\dag\phi)^3,
		\end{eqnarray}  
		with the matching relation at dimension-six as:
		\begin{align}
			c_6^{\llbracket U^3 \rrbracket} =& 4 \lambda_{\mathcal{H}_2,1}^3 + 6 \lambda_{\mathcal{H}_2,1}^2 \lambda_{\mathcal{H}_2,2} + 6 \lambda_{\mathcal{H}_2,1} \lambda_{\mathcal{H}_2,2}^2 + 2 \lambda_{\mathcal{H}_2,2}^2 + 24 \lambda_{\mathcal{H}_2,1} \lambda_{\mathcal{H}_2,3}^2 \nonumber\\
            & + 24 \lambda_{\mathcal{H}_2,3}^2 \lambda_{\mathcal{H}_2,2}.
		\end{align}  
		%%%%%%%%%%%%%%%%%%%%%%%%%%%%%%%%%%%%%%%%%%%%%%%%%%%%%%%%%%%%%%%%%%%%%%%%%%%%%%%%%%%%%%%%%
		\item \underline{Covariant operator: $(P_\mu U)^2$}
		\begin{align}
			\tr[(P_\mu U)^2] & \supset c^{(1),\llbracket (P_\mu U)^2 \rrbracket}_{\phi^6} \mathcal{O}^{(1)}_{\phi^6} + c^{(3),\llbracket (P_\mu U)^2 \rrbracket}_{\phi^6} \mathcal{O}^{(3)}_{\phi^6}. 
		\end{align}
		The matching relations at dimension-eight are: 
		\begin{align}
			c^{(1),\llbracket (P_\mu U)^2 \rrbracket}_{\phi^6} & = \frac{2 \eta_H \eta_{\mathcal{H}_2}}{M_{\mathcal{H}_2}^2} \Big( 6 \lambda_{\mathcal{H}_2,1} + 4 \lambda_{\mathcal{H}_2,2} + 2 \lambda_{\mathcal{H}_2,3} \Big) , \nonumber \\
			c^{(3),\llbracket (P_\mu U)^2 \rrbracket}_{\phi^6} & = \frac{2 \eta_H \eta_{\mathcal{H}_2}}{M_{\mathcal{H}_2}^2} \Big( 6 \lambda_{\mathcal{H}_2,1} + 4 \lambda_{\mathcal{H}_2,2} - \lambda_{\mathcal{H}_2,3} \Big) .
		\end{align}
		\begin{align}
			\tr[(P_\mu U)^2]  & \supset c_{H}^{\llbracket (P_\mu U)^2 \rrbracket} \mathcal{O}_{H} + c_{T} \mathcal{O}_{T} + c_{R} \mathcal{O}_{R}
		\end{align}  
		with the matching relation at dimension-six as:
		\begin{align}
			c_{H}^{\llbracket (P_\mu U)^2 \rrbracket} & = - \left(4 \lambda_{\mathcal{H}_2,1}^2 + 4 \lambda_{\mathcal{H}_2,1}  \lambda_{\mathcal{H}_2,2} + \lambda_{\mathcal{H}_2,2}^2 \right)  ~, &
			c_{T}^{\llbracket (P_\mu U)^2 \rrbracket} & =  - \lambda_{\mathcal{H}_2,2}^2 ~,  &
			c_{R}^{\llbracket (P_\mu U)^2 \rrbracket} & =  - 2 \lambda_{\mathcal{H}_2,2} ^2.
		\end{align}  
		%%%%%%%%%%%%%%%%%%%%%%%%%%%%%%%%%%%%%%%%%%%%%%%%%%%%%%%%%%%%%%%%%%%%%%%%%%%%%%%%%%%%%%%%%
		\item \underline{Covariant operator: $U (G_{\mu \nu})^2$}
		\begin{align}
			\tr[U (G_{\mu \nu})^2] & \supset c^{(1),\llbracket U (G_{\mu \nu})^2 \rrbracket}_{W^2 \phi^4} \mathcal{O}^{(1)}_{W^2 \phi^4} + c^{(1),\llbracket U (G_{\mu \nu})^2 \rrbracket}_{B^2 \phi^4} \mathcal{O}^{(1)}_{B^2 \phi^4} + c^{(1),\llbracket U (G_{\mu \nu})^2 \rrbracket}_{W B  \phi^4} \mathcal{O}^{(1)}_{W B  \phi^4}.
		\end{align}
		The matching relations at dimension-eight are: 
		\begin{align}
			c^{(1),\llbracket U (G_{\mu \nu})^2 \rrbracket}_{W^2 \phi^4} & =  \frac{3 \eta_{H} \eta_{\mathcal{H}_2}}{M_{\mathcal{H}_2}^2} g^2 ~, & c^{(1),\llbracket U (G_{\mu \nu})^2 \rrbracket}_{B^2  \phi^4} & = \frac{12 \eta_{H} \eta_{\mathcal{H}_2}}{M_{\mathcal{H}_2}^2} g'^2    ~,\nonumber\\
			c^{(1),\llbracket U (G_{\mu \nu})^2 \rrbracket}_{W B \phi^4} & =   \frac{8 \eta_{H} \eta_{\mathcal{H}_2}}{M_{\mathcal{H}_2}^2} g g'~.
		\end{align}
		\begin{align}
			\tr[U (G_{\mu \nu})^2] &  \supset c_{BB}^{\llbracket U (G_{\mu \nu})^2 \rrbracket} \mathcal{O}_{BB} + c_{WW} \mathcal{O}_{WW} + c_{WB}^{\llbracket U (G_{\mu \nu})^2 \rrbracket} \mathcal{O}_{WB}
		\end{align}  
		with the matching relation at dimension-six as:
		\begin{align}
			c_{BB}^{\llbracket U (G_{\mu \nu})^2 \rrbracket} & =  -\frac{g'^2}{2} \left( 2 \lambda_{\mathcal{H}_2,1} + \lambda_{\mathcal{H}_2,2} \right)  ~, &
			c_{WW}^{\llbracket U (G_{\mu \nu})^2 \rrbracket} = -\frac{g^2}{2} \left(2 \lambda_{\mathcal{H}_2,1} + \lambda_{\mathcal{H}_2,2} \right)~,\nonumber \\
			c_{WB}^{\llbracket U (G_{\mu \nu})^2 \rrbracket} & = - g g' \lambda_{\mathcal{H}_2,2}~.&
		\end{align}  
		%%%%%%%%%%%%%%%%%%%%%%%%%%%%%%%%%%%%%%%%%%%%%%%%%%%%%%%%%%%%%%%%%%%%%%%%%%%%%%%%%%%%%%%%%
	\end{enumerate}
\subsubsection*{\bf Dimension-eight and dimension-six operators from $\mathcal{O}(M^{0})$ term:}
	\begin{enumerate}
		\item \underline{Covariant operator: $U^2$} 
		\begin{align}
			\tr[U^2] \supset c_{\phi^8}^{\llbracket U^{2} \rrbracket} \mathcal{O}_{\phi^8} + c_{\phi^6}^{(3),\llbracket U^{2} \rrbracket} \mathcal{O}_{\phi^6}^{(3)}.
		\end{align}
		The matching relation at dimension-eight is:
		\begin{align}
			c_{\phi^8}^{\llbracket U^{2} \rrbracket} & =  \frac{4 \eta_{\mathcal{H}_2}}{M_{\mathcal{H}_2}^4}\Big( \eta_H(12 \eta_{\mathcal{H}_2} \eta_H +6 \lambda_{\mathcal{H}_2,1}^2 + 10 \lambda_{\mathcal{H}_2,1} \lambda_{\mathcal{H}_2,2} + 4 \lambda_{\mathcal{H}_2,2}^2) \nonumber \\
            &+ \frac{4 \lambda_{\mathcal{H}_2}}{M_{\mathcal{H}_2}^4} \Big(2 \eta_H^2 \lambda_{\mathcal{H}_2,3} + \eta_{\mathcal{H}_2}^2 (3 \lambda_{\mathcal{H}_2,1} + 2  \lambda_{\mathcal{H}_2,2}) \Big)~,\nonumber \\
            c_{\phi^6}^{(3), \llbracket U^{2} \rrbracket} & = -\frac{12 \eta_{H} \eta_{\mathcal{H}_2}}{M_{\mathcal{H}_2}^4} \Big( 3 \lambda_{\mathcal{H}_2,1} + 2\lambda_{\mathcal{H}_2,2} \Big)
		\end{align}
		\begin{eqnarray}
			\text{Tr} \, [U^2] & \supset  c_6^{\llbracket U^{2} \rrbracket} \mathcal{O}_6 = c_6^{\llbracket U^{2} \rrbracket} (\phi^\dag\phi)^3
		\end{eqnarray}  
		with the matching relation at dimension-six as:
		\begin{eqnarray}
			c_6^{\llbracket U^{2} \rrbracket} =  - \dfrac{4 \eta_{\mathcal{H}_2} \eta_H}{M_{\mathcal{H}_2}^2} \Big( 6 \lambda_{\mathcal{H}_2,1} + 4 \lambda_{\mathcal{H}_2,2} -2 \lambda_{\mathcal{H}_2,3}  \Big).
		\end{eqnarray}  		
		%%%%%%%%%%%%%%%%%%%%%%%%%%%%%%%%%%%%%%%%%%%%%%%%%%%%%%%%%%%%%%%%%%%%%%%%%%%%%%%%%%%%%%%%%
	\end{enumerate}
\subsubsection*{\bf Dimension-eight and dimension-six operators from $\mathcal{O}(M^{2})$ term:}
	\begin{enumerate}
		\item \underline{Covariant operator: $U$} \\
		\begin{align}
			\tr[U]   &\supset c_{\phi^8}^{\llbracket U \rrbracket} \mathcal{O}_{\phi^8} + c^{(3),{\llbracket U \rrbracket}}_{\phi^6} \mathcal{O}^{(3)}_{\phi^6} + c^{(3),{\llbracket U \rrbracket}}_{\phi^4} \mathcal{O}^{(3)}_{\phi^4} + c^{(4),{\llbracket U \rrbracket}}_{\phi^4} \mathcal{O}^{(4)}_{\phi^4}+ c^{(8),{\llbracket U \rrbracket}}_{\phi^6} \mathcal{O}^{(8)}_{\phi^6} \nonumber \\
			& + c^{(10),{\llbracket U \rrbracket}}_{\phi^4} \mathcal{O}^{(10)}_{\phi^4} + c^{(11),{\llbracket U \rrbracket}}_{\phi^4} \mathcal{O}^{(11)}_{\phi^4} .
		\end{align}
		The matching relations at dimension-eight are :
		\begin{align}
			c_{\phi^8}^{\llbracket U \rrbracket}&=  -\frac{16 \eta_H^2 \eta_{\mathcal{H}_2}}{M_{\mathcal{H}_2}^6} - \frac{8 \eta_H \eta_{\mathcal{H}_2} \lambda_{\mathcal{H}_2,1} \lambda_{\mathcal{H}_2,2}}{M_{\mathcal{H}_2}^6} - \frac{8 \eta_H \eta_{\mathcal{H}_2} \lambda_{\mathcal{H}_2,2}^2}{M_{\mathcal{H}_2}^6} + \frac{12 \eta_H \lambda_{\mathcal{H}_2,1} \lambda_H}{M_{\mathcal{H}_2}^6} + \frac{12 \eta_H \lambda_{\mathcal{H}_2,2} \lambda_H}{M_{\mathcal{H}_2}^6}~, \nonumber \\
			c^{(3),{\llbracket U \rrbracket}}_{\phi^6} &= - \frac{26 \eta_H \eta_{\mathcal{H}_2} \lambda_{\mathcal{H}_2,1}}{M_{\mathcal{H}_2}^6} - \frac{12 \eta_H \eta_{\mathcal{H}_2} \lambda_{\mathcal{H}_2,2}^2}{M_{\mathcal{H}_2}^6} - \frac{18 \eta_H^2 \lambda_H}{M_{\mathcal{H}_2}^6} ~, \nonumber 
		\end{align}
		\begin{align}
			c^{(3), \llbracket U \rrbracket}_{\phi^4} & =  \frac{24 \eta_H \eta_{\mathcal{H}_2}}{M_{\mathcal{H}_2}^6} ~, & c^{(4), \llbracket U \rrbracket}_{\phi^4} & =  \frac{24 \eta_H \eta_{\mathcal{H}_2}}{M_{\mathcal{H}_2}^6} ~,\nonumber\\
			c^{(8), \llbracket U \rrbracket}_{\phi^4} & =\frac{9 \eta_H \eta_{\mathcal{H}_2}}{M_{\mathcal{H}_2}^6} ~, & c^{(10),{\llbracket U \rrbracket}}_{\phi^4} & = \frac{6 \eta_H \eta_{\mathcal{H}_2}}{M_{\mathcal{H}_2}^6}~,\nonumber \\
			c^{(11),{\llbracket U \rrbracket}}_{\phi^4} & =\frac{18 \eta_H \eta_{\mathcal{H}_2}}{M_{\mathcal{H}_2}^6}.
		\end{align}
		\begin{align}
			\text{Tr} \, [U] &\supset c_6^{\llbracket U \rrbracket} \mathcal{O}_6 + c_H^{\llbracket U \rrbracket} \mathcal{O}_H + c_R^{\llbracket U \rrbracket} \mathcal{O}_R ,
		\end{align}  
		with the matching relations at dimension-six as:
		\begin{align}
			c_6^{\llbracket U \rrbracket} & = \dfrac{12 \eta_H \eta_{\mathcal{H}_2}}{M_{\mathcal{H}_2}^4} \Big(\, \lambda_{\mathcal{H}_2,1}+  \lambda_{\mathcal{H}_2,2}\Big) + \frac{3 \lambda_{\mathcal{H}_2} \eta_H^2}{M_{\mathcal{H}_2}^4}~,&  c_H^{\llbracket U \rrbracket} & =  - \dfrac{12 \eta_H \eta_{\mathcal{H}_2}}{M_{\mathcal{H}_2}^4} ~, &%\nonumber \\
			c_R^{\llbracket U \rrbracket}& =  - \dfrac{12 \eta_H \eta_{\mathcal{H}_2}}{M_{\mathcal{H}_2}^4}~.
		\end{align}  
		%%%%%%%%%%%%%%%%%%%%%%%%%%%%%%%%%%%%%%%%%%%%%%%%%%%%%%%%%%%%%%%%%%%%%%%%%%%%%%%%%%%%%%%%%
	\end{enumerate}
The results discussed so far are summarized in Tables~\ref{tab:matching_doublet_complex_scalar_1} and~\ref{tab:matching_doublet_complex_scalar_2}.

\subsubsection{Incorporating the effect of fermion Interactions: Complex Higgs Doublet}
\label{fermi_2hdm}
We further consider the interactions between the heavy complex doublet and the SM fermions, see Eqn.~\ref{2hdm_yukawa_1}. This modifies the classical solution of the heavy field via the functional $B$, and leads to the emergence of new operators involving SM fermions that we discuss below. 
\subsubsection*{\bf Dimension-eight operators from $\mathcal{O}(M^{-2})$ term:}
	\begin{enumerate}
		\item \underline{Covariant operator: $U^3$} \\
		\begin{align}
			\tr[U^3] \supset  c_{le\phi^5}^{\llbracket U^3 \rrbracket} \mathcal{Q}_{le\phi^5} ,
		\end{align}
		where, $\{p,\,r\}$ are the flavour indices. The matching relation is:
		\begin{gather}
			c_{le\phi^5}^{\llbracket U^3 \rrbracket} =  \frac{- 24 \eta_{\mathcal{H}_2} Y_{\mathcal{H}_2}^{(e)}}{M_{\mathcal{H}_2}^2} \Big( \lambda_{\mathcal{H}_2,1}^2  +  \lambda_{\mathcal{H}_2,1} \lambda_{\mathcal{H}_2,2} +  \lambda_{\mathcal{H}_2,1} \lambda_{\mathcal{H}_2,3} +  \lambda_{\mathcal{H}_2,2} \lambda_{\mathcal{H}_2,3} + \frac{1}{2}\lambda_{\mathcal{H}_2,2}^2 + 4 \lambda_{\mathcal{H}_2,3}^2 \Big) .
		\end{gather}
		%%%%%%%%%%%%%%%%%%%%%%%%%%%%%%%%%%%%%%%%%%%%%%%%%%%%%%%%%%%%%%%%%%%%%%%%%%%%%%%%%%%%%%%%%
		\item \underline{Covariant operator: $(P_\mu U)^2$}\\
		\begin{align}
			\tr[(P_\mu U)^2] &\supset c^{(1),\llbracket (P_\mu U)^2 \rrbracket}_{le \phi^3 D^2} \mathcal{Q}^{(1)}_{le \phi^3 D^2} + c^{(5),\llbracket (P_\mu U)^2 \rrbracket}_{le \phi^3 D^2} \mathcal{Q}^{(5)}_{le\phi^3 D^2} + \rm{EOM}\nonumber \\
			&= c^{(1),\llbracket (P_\mu U)^2 \rrbracket}_{le \phi^3 D^2} (D_\mu \phi^\dagger D^\mu \phi) (\bar{l}_p e_r \phi) + c^{(5),\llbracket (P_\mu U)^2 \rrbracket}_{le \phi^3 D^2} (\phi^\dagger D_\mu \phi) (\bar{l}_p e_r D^\mu \phi) \nonumber \\
			& \hspace{0.75cm} - \frac{2 (3 \lambda_{\mathcal{H}_2,1} + 2 \lambda_{\mathcal{H}_2,2}) \eta_{\mathcal{H}_2} Y_{\mathcal{H}_2}^{(e)}}{M_{\mathcal{H}_2}^2}((2 m^2 \phi^\dagger \phi - 2 \lambda_{\rm SM} (\phi^\dagger \phi)^2- Y_{e}^{(SM)\dagger} (\bar{e}_r l \phi^\dagger) \nonumber \\
			& \hspace{0.75cm}+ Y_{u}^{(SM)\dagger} \epsilon_{jk} \bar{q}^k u \phi^\dagger  - Y_{d}^{(SM)\dagger} (\bar{d}_p q_r \phi^\dagger))(\bar{l}_p e_r \phi) + \text{h.c.}) .
		\end{align}
		The matching relations are: 
		\begin{align}
			c^{(1),\llbracket (P_\mu U)^2 \rrbracket}_{le \phi^3 D^2} &=  \frac{1}{M_{\mathcal{H}_2}^2} \Big( (6 \lambda_{\mathcal{H}_2,1} + 4 \lambda_{\mathcal{H}_2,2}) \eta_{\mathcal{H}_2} Y_{\mathcal{H}_2}^{(e)} \Big) , \nonumber \\
			c^{(5),\llbracket (P_\mu U)^2 \rrbracket}_{le \phi^3 D^2} &=  \frac{1}{M_{\mathcal{H}_2}^2} \Big( (6 \lambda_{\mathcal{H}_2,1} + 4 \lambda_{\mathcal{H}_2,2}) \eta_{\mathcal{H}_2} Y_{\mathcal{H}_2}^{(e)} \Big)  .
		\end{align}
		%%%%%%%%%%%%%%%%%%%%%%%%%%%%%%%%%%%%%%%%%%%%%%%%%%%%%%%%%%%%%%%%%%%%%%%%%%%%%%%%%%%%%%%%%
		\item \underline{Covariant operator: $U (G_{\mu \nu})^2$} \\
		\begin{align}
			\tr[U (G_{\mu \nu})^2] &\supset  c^{(1),\llbracket U (G_{\mu \nu})^2 \rrbracket}_{le W^2 \phi} \mathcal{Q}^{(1)}_{le W^2 \phi} + c^{(1),\llbracket U (G_{\mu \nu})^2 \rrbracket}_{le B^2 \phi} \mathcal{Q}^{(1)}_{le B^2 \phi} + c^{(1),\llbracket U (G_{\mu \nu})^2 \rrbracket}_{le W B \phi} \mathcal{Q}^{(1)}_{le W B \phi} .
		\end{align}
		The matching relations are: 
		\begin{align}          
			c^{(1),\llbracket U (G_{\mu \nu})^2 \rrbracket}_{le W^2 \phi} &=  \frac{3}{2 M_{\mathcal{H}_2}^2} \eta_{\mathcal{H}_2} g^2 Y_{\mathcal{H}_2}^{(e)}~, & c^{(1),\llbracket U (G_{\mu \nu})^2 \rrbracket}_{le B^2 \phi} &=  \frac{6}{M_{\mathcal{H}_2}^2} \eta_{\mathcal{H}_2} g'^2 Y_{\mathcal{H}_2}^{(e)} ~   ,\nonumber\\
			c^{(1),\llbracket U (G_{\mu \nu})^2 \rrbracket}_{le W B \phi} &=  \frac{4}{M_{\mathcal{H}_2}^2} \eta_{\mathcal{H}_2} g g'  Y_{\mathcal{H}_2}^{(e)}~.
		\end{align}
		%%%%%%%%%%%%%%%%%%%%%%%%%%%%%%%%%%%%%%%%%%%%%%%%%%%%%%%%%%%%%%%%%%%%%%%%%%%%%%%%%%%%%%%%%
	\end{enumerate}
    
	No dimension-six fermionic operators emerge from $\mathcal{O}(M^{-2})$ term of Eqn.~\ref{eq:finite}.
   
\subsubsection*{\bf{Dimension-eight and dimension-six operators from $\mathcal{O}(M^{0})$ term:}}
	\begin{enumerate}[leftmargin=*]
		\item \underline{Covariant operator: $U^2$} \\
		\begin{align}
			\tr[U^2] & \supset c_{l^2 e^2 \phi^2}^{(1),\llbracket U^{2} \rrbracket}\mathcal{Q}_{l^2 e^2 \phi^2}^{(1)} +c_{l e q u \phi^2}^{(1),\llbracket U^{2} \rrbracket} \mathcal{Q}_{l e q u \phi^2}^{(1)} + c_{l e q d \phi^2}^{(3),\llbracket U^{2} \rrbracket} \mathcal{Q}_{l e q d \phi^2}^{(3)} \nonumber \\
			& \hspace{1cm} +c_{le\phi^3 D^2}^{(5),\llbracket U^{2} \rrbracket} \mathcal{Q}_{le\phi^3 D^2}^{(5)} + c_{le\phi^5}^{\llbracket U^{2} \rrbracket} \mathcal{Q}_{le\phi^5} + c_{e \phi}^{\llbracket U^{2} \rrbracket} \mathcal{O}_{e \phi}.
		\end{align}
		The matching relations at dimension-eight are:
		\begin{align}
			c_{l^2 e^2 \phi^2}^{(1),\llbracket U^{2} \rrbracket} &=  \frac{1}{M_{\mathcal{H}_2}^4} \Big( 40 |Y_{\mathcal{H}_2}^{(e)}|^2  \eta_{\mathcal{H}_2}^2 +  12 |Y_{\mathcal{H}_2}^{(e)}|^2 \eta_{\mathcal{H}_2} (\lambda_{\mathcal{H}_2,1} + \lambda_{\mathcal{H}_2,2}) + (4  \lambda_{\mathcal{H}_2,1} + \lambda_{\mathcal{H}_2,2} ) |Y_{\mathcal{H}_2}^{(e)}|^2 \lambda_{\mathcal{H}_2} \Big), \nonumber \\
			c_{l e q u \phi^2}^{(1),\llbracket U^{2} \rrbracket} &=  \frac{1}{M_{\mathcal{H}_2}^4} \Big( -6 Y_{\mathcal{H}_2}^{(e)} Y_{\rm SM}^{(u)} \eta_{\mathcal{H}_2} \lambda_{\mathcal{H}_2,1} - 6  Y_{\mathcal{H}_2}^{(e)}  Y_{\rm SM}^{(u)} \eta_{\mathcal{H}_2} \lambda_{\mathcal{H}_2,2}  \Big), \nonumber \\
			c_{l e q d \phi^2}^{(3),\llbracket U^{2} \rrbracket} &=  \frac{1}{M_{\mathcal{H}_2}^4} \Big(6 Y_{\mathcal{H}_2}^{(e)} Y_{\rm SM}^{(d)} \eta_{\mathcal{H}_2} \lambda_{\mathcal{H}_2,1} + 6  Y_{\mathcal{H}_2}^{(e)}  Y_{\rm SM}^{(d)} \eta_{\mathcal{H}_2} \lambda_{\mathcal{H}_2,2} \Big), \nonumber \\
			c_{le\phi^3 D^2}^{(5),\llbracket U^{2} \rrbracket} &=  \frac{1}{M_{\mathcal{H}_2}^4} \Big(6 Y_{\mathcal{H}_2}^{(e)} \eta_{\mathcal{H}_2} \lambda_{\mathcal{H}_2,1}^3 + 18  Y_{\mathcal{H}_2}^{(e)} \eta_{\mathcal{H}_2} \lambda_{\mathcal{H}_2,1}^2 \lambda_{\mathcal{H}_2,2} + 18  Y_{\mathcal{H}_2}^{(e)} \eta_{\mathcal{H}_2} \lambda_{\mathcal{H}_2,1} \lambda_{\mathcal{H}_2,2}^2  \nonumber \\
			& \hspace{1.cm}+ 3 Y_{\mathcal{H}_2}^{(e)} \eta_{\mathcal{H}_2} \lambda_{\mathcal{H}_2,1} \lambda_{\mathcal{H}_2} + Y_{\mathcal{H}_2}^{(e)} \eta_{\mathcal{H}_2} \lambda_{\mathcal{H}_2,2} \lambda_{\mathcal{H}_2} + 18 Y_{\mathcal{H}_2}^{(e)} \eta_{\mathcal{H}_2} \lambda_{\mathcal{H}_2,1} \lambda_{\rm SM} \nonumber \\ &\hspace{1.cm}+ 18 Y_{\mathcal{H}_2}^{(e)} \eta_{\mathcal{H}_2} \lambda_{\mathcal{H}_2,2} \lambda_{\rm SM} + 6 Y_{\mathcal{H}_2}^{(e)} \eta_{\mathcal{H}_2} \lambda_{\mathcal{H}_2,2}^3 \Big), \nonumber \\
			c_{le\phi^5}^{\llbracket U^{2} \rrbracket} &=  \frac{1}{M_{\mathcal{H}_2}^4} \Big(-12 Y_{\mathcal{H}_2}^{(e)} \eta_{\mathcal{H}_2} \lambda_{\mathcal{H}_2,1} - 12  Y_{\mathcal{H}_2}^{(e)} \eta_{\mathcal{H}_2} \lambda_{\mathcal{H}_2,2} \Big), \nonumber \\
			c_{e \phi}^{\llbracket U^{2} \rrbracket} &=  \frac{1}{M_{\mathcal{H}_2}^4} \Big( -18 m_H^2 Y_{\mathcal{H}_2}^{(e)} \eta_{\mathcal{H}_2} \lambda_{\mathcal{H}_2,1} - 18 m_H^2 Y_{\mathcal{H}_2}^{(e)} \eta_{\mathcal{H}_2} \lambda_{\mathcal{H}_2,2} \Big) .
		\end{align}
		\begin{eqnarray}
			\text{Tr} \, [U^2] & \supset  c_{e \phi}^{\llbracket U^{2} \rrbracket} \mathcal{O}_{e \phi} = c_{e \phi}^{\llbracket U^{2} \rrbracket} (\phi^\dag\phi)(\bar{l}_p e_r \phi),
		\end{eqnarray}  
		with the matching relation at dimension-six as:
		\begin{eqnarray}
			c_{e \phi}^{\llbracket U^{2} \rrbracket} =   -\dfrac{2}{M_{\mathcal{H}_2}^2} \Big(\, 6 \lambda_{\mathcal{H}_2,1} \eta_{\mathcal{H}_2} Y_{\mathcal{H}_2}^{(e)} + 4 \lambda_{\mathcal{H}_2,2} \eta_{\mathcal{H}_2} Y_{\mathcal{H}_2}^{(e)} - 2 \lambda_{\mathcal{H}_2,3} \eta_{\mathcal{H}_2} Y_{\mathcal{H}_2}^{(e)}  \, \Big).
		\end{eqnarray}  
		%%%%%%%%%%%%%%%%%%%%%%%%%%%%%%%%%%%%%%%%%%%%%%%%%%%%%%%%%%%%%%%%%%%%%%%%%%%%%%%%%%%%%%%%%
	\end{enumerate}

\subsubsection*{\bf{Dimension-eight and dimension-six operators from $\mathcal{O}(M^{2})$ term:}}
	\begin{enumerate}
		\item \underline{Covariant operator: $U$} \\
		\begin{align}
			\tr[U] &\supset  c_{l^2 e^2 \phi^2}^{(1), \llbracket U \rrbracket} \mathcal{Q}_{l^2 e^2 \phi^2}^{(1)}  + c^{(3),{\llbracket U \rrbracket}}_{l^2 e^2 \phi^2} \mathcal{Q}_{l^2 e^2 \phi^2}^{(3)} + c^{(1),{\llbracket U \rrbracket}}_{l e q u \phi^2} \mathcal{Q}_{l e q u \phi^2}^{(1)} + c^{(3),{\llbracket U \rrbracket}}_{l e q d \phi^2} \mathcal{Q}_{l e q d \phi^2}^{(3)}  \nonumber \\
			& \hspace{0.5cm}+ c^{{\llbracket U \rrbracket}}_{l^2 e^2 D^2} \mathcal{Q}_{l^2 e^2 D^2} + c^{{\llbracket U \rrbracket}}_{l e \phi^5} \mathcal{Q}_{l e \phi^5} + c^{{\llbracket U \rrbracket}}_{e \phi} \mathcal{O}_{e \phi} +c_{le\phi^3 D^2}^{(5),\llbracket U^{2} \rrbracket} \mathcal{Q}_{le\phi^3 D^2}^{(5)}.
		\end{align}

        %%%%%%%%%%%%%%%%%%%%%%%%%%%%%%%%%%%%%%%
\begin{table}[!t]
		\centering %\small
		\renewcommand{\arraystretch}{2.1}
        \setlength{\tabcolsep}{1pt}
        \resizebox{0.95\textwidth}{!}{
		\begin{tabular}{|c|c|c|}
			\hline
			Green's basis& Green's basis  & Relation of Green's basis operator    \vspace{-3mm}\\
			operator& coefficient & in terms of Murphy basis operators\\
			\hline
			$\mathcal{O}^{(3)}_{\phi^6}$& $c^{(3)}_{\phi^6}$ & $-4 \lambda_{\rm SM} \mathcal{O}_{\phi^8} + 2 \lambda_{\rm SM} v^2 \mathcal{O}_6 - Y_{\rm SM}^{pq} Q_{\psi^2 \phi^5}+ {\text{h.c.}}$\\
			\hline		
			$\mathcal{O}^{(4)}_{\phi^4}$& $c^{(4)}_{\phi^4}$ & $-4 \lambda_{\rm SM} \mathcal{O}_{\phi^6}^{(1)}  + 2 \lambda_{\rm SM} v^2 (\phi^\dagger \phi) (D_\mu \phi^\dagger)(D^\mu \phi)- Y_{\rm SM}^{pq} Q_{\psi^2 \phi^3 D^2}+ {\text{h.c.}}$\\
			\hline
			$\mathcal{O}^{(8)}_{\phi^4}$&  $c^{(8)}_{\phi^4}$&$2 \lambda_{\rm SM} \Big( v^4 (\phi^\dagger \phi)^2 + 4 \mathcal{O}_{\phi^8} - 4 v^2 \mathcal{O}_6\Big) - 2 \lambda_{\rm SM} (\phi^\dagger \phi) J_\phi \phi + 4 \lambda_{\rm SM} Y_{\rm SM}^{pq} Q_{\psi^2 \phi^5} + Y_{\rm SM}^{{pq}^2} Q_{\psi^4 \phi^2}^{(2)} + {\text{h.c.}}$ \\
			\hline
			$\mathcal{O}^{(10)}_{\phi^4}$& $c^{(10)}_{\phi^4}$ & $\lambda_{\rm SM} \Big( v^4 (\phi^\dagger \phi)^2 + 4 \mathcal{O}_{\phi^8} - 4 v^2 \mathcal{O}_6\Big) -  \lambda_{\rm SM} (\phi^\dagger \phi) J_\phi \phi + 2 \lambda_{\rm SM} Y_{\rm SM}^{pq} Q_{\psi^2 \phi^5} + Y_{\rm SM}^{{pq}^2} Q_{\psi^4 \phi^2}^{(1)} + {\text{h.c.}}$\\
			\hline
			$\mathcal{O}^{(11)}_{\phi^4}$& $c^{(11)}_{\phi^4}$ &  $\lambda_{\rm SM} \Big( v^4 (\phi^\dagger \phi)^2 + 4 \mathcal{O}_{\phi^8} - 4 v^2 \mathcal{O}_6\Big) -  \lambda_{\rm SM} (\phi^\dagger \phi) J_\phi \phi + 2 \lambda_{\rm SM} Y_{\rm SM}^{pq} Q_{\psi^2 \phi^5} + Y_{\rm SM}^{{pq}^2} Q_{\psi^4 \phi^2}^{(3)}$ \\&&$+ \frac{1}{2} Y_{\rm SM}^{{pq}^2} Q_{\psi^4 \phi^2}^{(1)}+ {\text{h.c.}}$\\
			\hline
			$\mathcal{O}_{W\phi^4 D^2}^{(6)}$& $c_{W\phi^4 D^2}^{(6)}$ & $\frac{g}{2} \left( \lambda_{SM}[v^2 \mathcal{O}_6 -2 \mathcal{O}_{\phi^8}] + 5 \mathcal{O}_{\phi^6}^{(1)} \right) +\frac{1}{2} Y_{\rm SM}^{pq}\left(Q^{(2)}_{\psi^2 \phi^4D} -i Q^{(4)}_{\psi^2 \phi^4D}+ {\text{h.c.}}\right)
- \frac{g}{4} Y_{\rm SM}^{pq} Q^{(1)}_{\psi^2 \phi^5} $\\
			\hline
			$\mathcal{O}_{W\phi^4 D^2}^{(7)}$& $c_{W\phi^4 D^2}^{(7)}$ &  $- 4 g \Big( Q^{(1)}_{\phi^6} - Q^{(2)}_{\phi^6} \Big) - 4 Y_{\rm SM}^{pq} \left(Q^{(4)}_{\psi^2 \phi^4D} - iQ^{(3)}_{\psi^2 \phi^4D} + {\text{h.c.}}\right) + i 4 \mathcal{O}_{W \phi^4 D^2}^{(3)}$\\
			\hline		
			$\mathcal{O}_{B\phi^4 D^2}^{(3)}$& $c_{B\phi^4 D^2}^{(3)}$ & $i\frac{g'}{2} \left( \lambda_{SM}[v^2 \mathcal{O}_6 -2 \mathcal{O}_{\phi^8}] + 3 \mathcal{O}_{\phi^6}^{(1)} + 2 \mathcal{O}_{\phi^6}^{(2)} \right) +i Y_{\rm SM}^{pq}  Q^{(1)}_{\psi^2 \phi^4D}
  -i\frac{g'}{4} Y_{\rm SM}^{pq} Q^{(1)}_{\psi^2 \phi^5} + {\text{h.c.}} $\\
			\hline 		
			$\mathcal{O}^{(13)}_{W^2 \phi^2 D^2}$& $c^{(13)}_{W^2 \phi^2 D^2}$&$\frac{g^2}{4} \left( \lambda_{SM}[v^2 \mathcal{O}_6 -2 \mathcal{O}_{\phi^8}] + 5 \mathcal{O}_{\phi^6}^{(1)} \right)
  + \frac{g}{2} Y_{\rm SM}^{pq} \left(Q^{(2)}_{\psi^2 \phi^4D} + Q^{(4)}_{\psi^2 \phi^4D} \right)
  -\frac{g^2}{8} Y_{\rm SM}^{pq} Q^{(1)}_{\psi^2\phi^5}$\\&&$
  + Y_{\rm SM}^{{pq}^2}  Q^{(5)}_{\psi^4\phi^2} + {\text{h.c.}} $\\
			\hline
			$\mathcal{O}^{(14)}_{W^2 \phi^2 D^2}$& $c^{(14)}_{W^2 \phi^2 D^2}$& $- \lambda_{\rm SM} v^2 \mathcal{O}_{\phi W} + 2 \lambda_{\rm SM} \mathcal{O}_{W^2 \phi^4}^{(1)} + Y_{\rm SM}^{pq} Q_{\psi^2 W^2 \phi}^{(1)}+{\text{h.c.}}$\\
			\hline
	$\mathcal{O}^{(8)}_{B^2 \phi^2 D^2} $& $c^{(8)}_{B^2 \phi^2 D^2} $&$-\frac{{g'}^2}{4} \left( \lambda_{SM}[v^2 \mathcal{O}_6 -2 \mathcal{O}_{\phi^8}] + 3 \mathcal{O}_{\phi^6}^{(1)} + 2 \mathcal{O}_{\phi^6}^{(2)} \right)
+i g'
  Q^{(1)}_{B \phi^4D^2} +\frac{g'}{4}\left( g' Q^{(1)}_{B^2 \phi^4}+ g Q^{(1)}_{WB \phi^4}\right)$ \\
&& $- \frac{1}{2} Y_{\rm SM}^{pq}  \left(Q^{(1)}_{\psi^2B \phi^2D} + iQ^{(3)}_{\psi^2B \phi^2D} \right)
-\frac{g'}{2} Y_{\rm SM}^{pq} Q^{(1)}_{\psi^2 \phi^4 D}
+\frac{{g'}^2}{8} Y_{\rm SM}^{pq}  Q^{(1)}_{\psi^2 \phi^5} + {\text{h.c.}} $\\
			\hline
			$\mathcal{O}^{(11)}_{W^2 \phi^2 D^2}$& $c^{(11)}_{W^2 \phi^2 D^2}$  & $
-\frac{{g}^2}{4} \left( \lambda_{SM}[v^2 \mathcal{O}_6 -2 \mathcal{O}_{\phi^8}] + 5 \mathcal{O}_{\phi^6}^{(1)} \right) + g\left(
  i \mathcal{O}^{(1)}_{W \phi^4D^2} - Q^{(3)}_{W \phi^4D^2}\right)$
    \\& &
    $ +\frac{g}{4}\left(
    g'\mathcal{O}^{(1)}_{WB \phi^4}+ 2 g \mathcal{O}^{(1)}_{W^2 \phi^4}\right) - \frac{1}{2} Y_{\rm SM}^{pq} \left(Q^{(5)}_{\psi^2W \phi^2D} + iQ^{(7)}_{\psi^2W \phi^2D} \right)
    $\\
&&$
-\frac{g}{8} Y_{\rm SM}^{pq} \left( Q^{(2)}_{\psi^2 \phi^4 D}+Q^{(4)}_{\psi^2 \phi^4 D}
-i Q^{(3)}_{\psi^2 \phi^4 D} \right)+\frac{g^2}{16} Y_{\rm SM}^{pq} \left(Q^{(1)}_{\psi^2\phi^5}+Q^{\dagger (1)}_{\psi^2 \phi^5}
\right) + {\text{h.c.}}$
   \\ \hline
			$\mathcal{O}^{(10)}_{W B \phi^2 D^2}$& $c^{(10)}_{W B \phi^2 D^2}$& $\frac{gg'}{4} \left( \lambda_{SM}[v^2 \mathcal{O}_6 -2 \mathcal{O}_{\phi^8}] + 3 \mathcal{O}_{\phi^6}^{(1)} + 2 \mathcal{O}_{\phi^6}^{(2)} \right) -\frac{gg'}{8} Y_{\rm SM}^{pq} \left(Q^{(1)}_{\psi^2 \phi^5}+Q^{\dagger (1)}_{\psi^2 \phi^5}\right)
  +\frac{g}{2} Y_{\rm SM}^{pq}  Q^{(1)}_{\psi^2 \phi^4D}$ \\&&
 $ +\frac{g'}{4} Y_{\rm SM}^{pq}  \left(Q^{(2)}_{\psi^2 \phi^4D} - Q^{(4)}_{\psi^2 \phi^4D}\right)
  + Y_{\rm SM}^{{pq}^2}  Q^{(7)}_{\psi^4 \phi^2}$\\\hline
		\end{tabular}}
		\caption{\small  Translation of redundant dimension-eight \emph{Green's basis} coefficients, emerging in the complex triplet model and complex doublet model into the  other non-redundant dimension-eight operators and dimension-eight fermionic operators of \emph{Murphy}-\emph{basis} form. Here, $Y_{\rm SM}^{pq}$ denotes the SM Yukawa coupling, $\{p,\,q\} \in (1,2,3)$ are the flavour indices.}
		\label{Tab:nonredundant_basis_1}
\end{table}
%%%%%%%%%%%%%%%%%%%%%%%%%%%%%%%%%%%%%%%

%%%%%%%%%%%%%%%%%%%%%%%%%%%%%%%%%%%%%%%
\begin{table}[!t]
		\centering %\small
		\renewcommand{\arraystretch}{1.8}
        \setlength{\tabcolsep}{1pt}
        \resizebox{0.95\textwidth}{!}{
		\begin{tabular}{|c|c|c|}
			\hline
			Green's basis& Green's basis  & Relation of Green's basis operator    \vspace{-3mm}\\
			operator& coefficient & in terms of Murphy basis operators\\
			\hline
			$\mathcal{O}^{(11)}_{W B \phi^2 D^2}$& $c^{(11)}_{W B \phi^2 D^2}$&$-2 g' \left(  i
Q^{(2)}_{W \phi^4D^2} +   Q^{(4)}_{W \phi^4D^2}\right)-\frac{gg'}{4}
\left( Q^{(2)}_{W^2 \phi^4}-  Q^{(4)}_{W^2 \phi^4}\right) - 2 Y_{\rm SM}^{pq} \left(Q^{(2)}_{\psi^2W \phi^2D} - iQ^{(4)}_{\psi^2W \phi^2D} \right) $\\&&
$+\frac{g g'}{2} \left( \lambda_{SM}[v^2 \mathcal{O}_6 -2 \mathcal{O}_{\phi^8}] + 3 \mathcal{O}_{\phi^6}^{(1)} + 2 \mathcal{O}_{\phi^6}^{(2)} \right)- 2 \mathcal{O}_{WB\phi^2 D^2}^{(4)} -i 6 g Q^{(1)}_{B \phi^4D^2}  $
\\&&$- \frac{g}{2}\left(
g Q^{(1)}_{WB \phi^4} +g' Q^{(1)}_{B^2 \phi^4}\right) +g Q^{(1)}_{\psi^2 \phi^4D} 
-\frac{gg'}{4} \left(Q^{(1)}_{\psi^2 \phi^5}+{\text h.c.}\right) $
\nonumber\\
&& $
 - gg'  \left( \lambda_{SM}[v^2 \mathcal{O}_6 -2 \mathcal{O}_{\phi^8}] + 3 \mathcal{O}_{\phi^6}^{(1)} + 2 \mathcal{O}_{\phi^6}^{(2)} \right) - 4 Q^{(7)}_{\psi^4 \phi^2} $
\\&&
$ + \frac{gg'}{2} Y_{\rm SM}^{pq} \left(Q^{(1)}_{\psi^2 \phi^5}+Q^{\dagger (1)}_{\psi^2 \phi^5}\right)
-2 g Y_{\rm SM}^{pq} Q^{(1)}_{\psi^2 \phi^4D} 
 - g' Y_{\rm SM}^{pq} \left(Q^{(2)}_{\psi^2 \phi^4D} - Q^{(4)}_{\psi^2 \phi^4D}\right)$\\
			\hline
			$\mathcal{O}^{(13)}_{W B \phi^2 D^2}$& $c^{(13)}_{W B \phi^2 D^2}$&$+\frac{g g'}{4}  \left( \lambda_{SM}[v^2 \mathcal{O}_6 -2 \mathcal{O}_{\phi^8}] + 3 \mathcal{O}_{\phi^6}^{(1)} + 2 \mathcal{O}_{\phi^6}^{(2)} \right)
-i 3 g Q^{(1)}_{B \phi^4D^2}- \frac{g}{4}\left(
g Q^{(1)}_{WB \phi^4} +g' Q^{(1)}_{B^2 \phi^4}\right)  $\\
&& $+\frac{g}{2} Y_{\rm SM}^{pq} Q^{(1)}_{\psi^2 \phi^4D} - Y_{\rm SM}^{pq} \left(Q^{(5)}_{\psi^2B \phi^2D} + iQ^{(7)}_{\psi^2B \phi^2D}\right)
-\frac{gg'}{8} Y_{\rm SM}^{pq} \left(Q^{(1)}_{\psi^2 \phi^5}+{\text h.c.}\right)$
\\&&
$- \frac{gg'}{4}  \left( \lambda_{SM}[v^2 \mathcal{O}_6 -2 \mathcal{O}_{\phi^8}] + 3 \mathcal{O}_{\phi^6}^{(1)} + 2 \mathcal{O}_{\phi^6}^{(2)} \right) + \frac{gg'}{8} Y_{\rm SM}^{pq} \left(Q^{(1)}_{\psi^2 \phi^5}+Q^{\dagger (1)}_{\psi^2 \phi^5}\right)$ \\ &&$
-\frac{g}{2} Y_{\rm SM}^{pq} Q^{(1)}_{\psi^2 \phi^4D} 
 - \frac{g'}{4} Y_{\rm SM}^{pq} \left(Q^{(2)}_{\psi^2 \phi^4D} - Q^{(4)}_{\psi^2\phi^4D}\right)
- Y_{\rm SM}^{{pq}^2} Q^{(7)}_{\psi^4 \phi^2}$\\
			\hline
			$\mathcal{O}_{B^2 D^4} $&  $c_{B^2 D^4}$ & 
  $\frac{g^2{g'}^2}{4} \lambda_{\rm SM} \Big( \left( v^2 \mathcal{O}_6 -2 \mathcal{O}_{\phi^8}\right) + 2 \mathcal{O}_{\phi^6}^{(1)} + 2 \mathcal{O}_{\phi^6}^{(2)} \Big) +
  {g'}^2\left(Q^{(1)}_{\phi^4}-Q^{(2)}_{\phi^4}\right)
  $ \\ & &
$ +\frac{{g'}^2}{8}\left({g'}^2 Q^{(1)}_{B^2\phi^4}-g^2 Q^{(1)}_{W^2\phi^4}
\right) +i {g'}^2\left(g' Q^{(1)}_{B\phi^4D^2}-g Q^{(1)}_{W\phi^4D^2}\right) -\frac{g'}{2} Y_{\rm SM}^{pq} Q^{(1)}_{\psi^2 \phi^2 D^3}+ Q^{(2)}_{\psi^4 D^2}
  $ \\& & $
  -\frac{{g'}^2 g}{4} Y_{\rm SM}^{pq} Q^{(2)}_{\psi^2 \phi^4D}
  -\frac{g^2 {g'}^2}{8} Y_{\rm SM}^{pq} Q^{(1)}_{\psi^2\phi^5} -\frac{3 {g'}^2}{4} Y_{\rm SM}^{{pq}^2}
  \left(- Q^{(1)}_{\psi^4 \phi^2}
    +Q^{(2)}_{\psi^4 \phi^2} 
    -2Q^{(3)}_{\psi^4 \phi^2}\right)$\\
			\hline
			$\mathcal{O}_{W^2 D^4} $ & $c_{W^2 D^4}$ & $\frac{g^4}{4}\lambda_{\rm SM} \Big( \left( v^2 \mathcal{O}_6 -2 \mathcal{O}_{\phi^8}\right) + 5 \mathcal{O}_{\phi^6}^{(1)}  \Big) +\frac{g^2 {g'}^2}{2}\lambda_{\rm SM} \Big( \left( v^2 \mathcal{O}_6 -2 \mathcal{O}_{\phi^8}\right) + 3 \mathcal{O}_{\phi^6}^{(1)} + 2 \mathcal{O}_{\phi^6}^{(2)} \Big)
   $\\
    &&$ -g^2\left(Q^{(1)}_{\phi^4}+Q^{(2)}_{\phi^4}-2Q^{(3)}_{\phi^4}\right) -\frac{{g}^2}{8}\left(3{g'}^2 Q^{(1)}_{B^2 \phi^4}+g^2 Q^{(1)}_{W^2 \phi^4}
  +4 g g' Q^{(1)}_{WB \phi^4}\right)
 $ \\ &
& $ -  i g^2 \left(3 g' Q^{(1)}_{B \phi^4D^2}+g Q^{(1)}_{W \phi^4D^2}\right)
+\frac{g^2}{4} Y_{\rm SM}^{pq} \left(g Q^{(2)}_{\psi^2 \phi^4 D}+g Q^{(4)}_{\psi^2 \phi^4 D}+4g' Q^{(1)}_{\psi^2 \phi^4 D}
  \right) + {\text{h.c.}}$
\\
& & $-\frac{g}{2} Y_{\rm SM}^{pq} Q^{(2)}_{\psi^2 \phi^2 D^3}-g^2 Y_{\rm SM}^{pq} Q^{(11)}_{\psi^2 W \phi^2 D}-\frac{g^2(g^2+2{g'}^2)}{8} Y_{\rm SM}^{pq}  Q^{(1)}_{\psi^2 \phi^5} $\\&&
  $-\frac{3 {g}^2}{4} Y_{\rm SM}^{{pq}^2}
  \left(- 3 Q^{(1)}_{\psi^4 \phi^2} +Q^{(2)}_{\psi^4 \phi^2}+Q^{\dagger (2)}_{\psi^4 \phi^2} + 2Q^{(3)}_{\psi^4 \phi^2}\right)
    +  Y_{\rm SM}^{{pq}^2} Q^{(3)}_{\psi^4 D^2} + {\text{h.c.}}$ \\
			\hline
		\end{tabular}}
		\caption{\small  Translation of redundant dimension-eight \emph{Green's basis} coefficients, emerging in the complex triplet model and complex doublet model into the  other non-redundant dimension-eight operators and dimension-eight fermionic operators of \emph{Murphy}-\emph{basis} form. Here, $Y_{\rm SM}^{pq}$ denotes the SM Yukawa coupling, $\{p,\,q\} \in (1,2,3)$ are the flavour indices.}
		\label{Tab:nonredundant_basis_2}
\end{table}

		The matching relations at dimension-eight are:
		\begin{align}
			c_{l^2 e^2 \phi^2}^{(1), \llbracket U \rrbracket} &=  -\frac{1}{M_{\mathcal{H}_2}^6} \Big( 6 |Y_{\mathcl{H}_2}^{(e)}|^2 \eta_{\mathcal{H}_2}^2 + Y_{\mathcl{H}_2}^{(e)} Y_{\rm SM}^{(e)} \eta_{\mathcal{H}_2} (\lambda_{\mathcal{H}_2,1} - \lambda_{\mathcal{H}_2,2} ) \nonumber \\
			& \hspace{1.cm}- 3 \lambda_{\mathcal{H}_2} |Y_{\mathcl{H}_2}^{(e)}|^2 (\lambda_{\mathcal{H}_2,1} + \lambda_{\mathcal{H}_2,2}) + 3 Y_{\mathcl{H}_2}^{(e)} Y_{\rm SM}^{(e)}   \eta_{\mathcal{H}_2}  \lambda_{\mathcal{H}_2} \nonumber \\
            & \hspace{1.cm}+3 Y_{\mathcl{H}_2}^{(e)} Y_{\rm SM}^{(e)} \eta_{\mathcl{H}_2} (2 \lambda_{\mathcal{H}_2,1} + \lambda_{\mathcal{H}_2,2}) \Big)~,\nonumber\\
			c^{(3),{\llbracket U \rrbracket}}_{l^2 e^2 \phi^2} &=-\frac{1}{M_{\mathcal{H}_2}^6} \Big(6 Y_{\mathcl{H}_2}^{(e)}{}^2 \eta_{\mathcal{H}_2} + Y_{\mathcl{H}_2}^{(e)} Y_{\rm SM}^{(e)} \eta_{\mathcal{H}_2} (\lambda_{\mathcal{H}_2,1} - \lambda_{\mathcal{H}_2,2} ) \Big) ~, \nonumber \\
			c^{(1),{\llbracket U \rrbracket}}_{l e q u \phi^2} &=  \frac{1}{M_{\mathcal{H}_2}^6} \Big(-7 Y_{\mathcl{H}_2}^{(e)} Y_{\rm SM}^{(u)} \eta_{\mathcl{H}_2} \lambda_{\mathcal{H}_2,1}  - 2 Y_{\mathcl{H}_2}^{(e)} Y_{\rm SM}^{(u)} \eta_{\mathcl{H}_2} \lambda_{\mathcal{H}_2,2}  - 3 Y_{\mathcl{H}_2}^{(e)} Y_{\rm SM}^{(u)}   \eta_{\mathcal{H}_2}  \lambda_{\mathcal{H}_2}\Big) ~, \nonumber \\
			c^{(3),{\llbracket U \rrbracket}}_{l e q d \phi^2} &=  \frac{1}{M_{\mathcal{H}_2}^6} \Big(-Y_{\mathcl{H}_2}^{(e)} Y_{\rm SM}^{(d)} \eta_{\mathcl{H}_2} (\lambda_{\mathcal{H}_2,1} + \lambda_{\mathcal{H}_2,2}) - 3 Y_{\mathcl{H}_2}^{(e)} Y_{\rm SM}^{(d) \ast} \eta_{\mathcl{H}_2} (2 \lambda_{\mathcal{H}_2,1} + \lambda_{\mathcal{H}_2,2}) \nonumber \\  &\hspace{1.cm} + 3 Y_{\mathcl{H}_2}^{(e)} Y_{\rm SM}^{(d)}   \eta_{\mathcal{H}_2}  \lambda_{\mathcal{H}_2}\Big)~,\nonumber\\
			c^{{\llbracket U \rrbracket}}_{l^2 e^2 D^2} &=  \frac{3}{M_{\mathcal{H}_2}^6}  \lambda_{H_2} |Y_{\mathcl{H}_2}^{(e)}|^2 ~, ~
			c_{le\phi^3 D^2}^{(5),\llbracket U \rrbracket} =  \frac{1}{M_{\mathcal{H}_2}^6} \Big( -6  Y_{\mathcl{H}_2}^{(e)} \eta_{\mathcl{H}_2}^{(e)}  \lambda_{\mathcl{H}_2} \Big)~, \nonumber \\
			c^{{\llbracket U \rrbracket}}_{l e \phi^5} & =  \frac{1}{M_{\mathcal{H}_2}^6} \Big(-3 Y_{\mathcl{H}_2}^{(e)} \eta_{\mathcl{H}_2} \lambda_{\mathcal{H}_2,1}^2 - 6 Y_{\mathcl{H}_2}^{(e)} \eta_{\mathcl{H}_2} \lambda_{\mathcal{H}_2,1} \lambda_{\mathcal{H}_2,2} -3 Y_{\mathcl{H}_2}^{(e)} \eta_{\mathcl{H}_2} \lambda_{\mathcal{H}_2,2}^2 \nonumber \\
			&  \hspace{1.cm}- 3 Y_{\mathcl{H}_2}^{(e)} \eta_{\mathcl{H}_2} (2 \lambda_{\mathcal{H}_2,1} + \lambda_{\mathcal{H}_2,2} ) \lambda_{\rm SM} - 2 Y_{\mathcl{H}_2}^{(e)} \eta_{\mathcl{H}_2} (\lambda_{\mathcal{H}_2,1} + \lambda_{\mathcal{H}_2,2}) \lambda_{\rm SM}
			 \nonumber \\
			& \hspace{1.cm}+ 6 Y_{\mathcl{H}_2}^{(e)} \lambda_{\mathcl{H}_2} \eta_{\mathcal{H}_2}  \lambda_{\mathcal{H}_2,1} + 6 Y_{\mathcl{H}_2}^{(e)} \lambda_{\mathcl{H}_2} \eta_{\mathcal{H}_2}  \lambda_{\mathcal{H}_2,2} + 9 Y_{\mathcl{H}_2}^{(e)} \eta_{\mathcl{H}_2} \lambda_{\mathcl{H}_2} \lambda_{\rm SM} \Big)~,\nonumber \\
			c^{{\llbracket U \rrbracket}}_{e \phi}&= \frac{1}{M_{\mathcal{H}_2}^6} \Big( m_H^2 Y_{\mathcal{H}_2}^{(e)} \eta_{\mathcal{H}_2} (8 \lambda_{\mathcal{H}_2,1} + \lambda_{\mathcal{H}_2,2}) - 9 m_H^2 Y_{\mathcal{H}_2}^{(e)} \eta_{\mathcal{H}_2} \lambda_{\mathcal{H}_2} \Big)~.
		\end{align}\begin{align}
			\text{Tr} \, [U] &\supset c_{e\phi}^{\llbracket U \rrbracket} \mathcal{Q}_{e\phi} + c_{le}^{\llbracket U \rrbracket} \mathcal{Q}_{le} + c_{ledq}^{\llbracket U \rrbracket} \mathcal{Q}_{ledq} +c_{lequ}^{(1),\llbracket U \rrbracket} \mathcal{Q}_{lequ}^{(1)}  + \eta_{\mathcal{H}_2} m_H^2 Y_{\mathcal{H}_2}^{(e)} \bar{l}_p e_r \phi +\rm{h.c},
		\end{align}  
		with the matching relations at dimension-six as:
		\begin{align}
			c_{e\phi}^{\llbracket U \rrbracket} & = \dfrac{1}{M_{\mathcal{H}_2}^4} \Big(-3 \eta_{\mathcal{H}_2} \lambda_{\rm SM} Y_{\mathcal{H}_2}^{(e)} + 3 \eta_{\mathcal{H}_2} Y_{\mathcal{H}_2}^{(e)} (\lambda_{\mathcal{H}_2,1}+ \lambda_{\mathcal{H}_2,2}) \Big),&&\nonumber \\
			c_{le}^{\llbracket U \rrbracket} & =  \dfrac{1}{M_{\mathcal{H}_2}^4} \Big(-3 \eta_{\mathcal{H}_2} Y_{\rm SM}^{(e)}  Y_{\mathcal{H}_2}^{(e)} - \frac{3}{2} \lambda_{\mathcal{H}_2} Y_{\mathcal{H}_2}^{(e)} Y_{\mathcal{H}_2}^{(e)\ast} \Big), &&\nonumber \\
			c_{ledq}^{\llbracket U \rrbracket}& =   - \dfrac{1}{M_{\mathcal{H}_2}^4} \Big(-3 \eta_{\mathcal{H}_2} Y_{\rm SM}^{(d) \ast}  Y_{\mathcal{H}_2}^{(e)} \Big), ~~~~c_{lequ}^{(1),\llbracket U \rrbracket} ~=~\dfrac{1}{M_{\mathcal{H}_2}^4} \Big(-3 \eta_{\mathcal{H}_2} Y_{\rm SM}^{(u)}  Y_{\mathcal{H}_2}^{(e)} \Big).&
		\end{align}
		%%%%%%%%%%%%%%%%%%%%%%%%%%%%%%%%%%%%%%%%%%%%%%%%%%%%%%%%%%%%%%%%%%%%%%%%%%%%%%%%%%%%%%
	\end{enumerate}
	We summarize all the operators and their associated WCs in  Table~\ref{tab:matching_doublet_complex_scalar_3}.
%%%%%%%%%%%%%%%%%%%%%%%%%%%%%%%%%%%%%%%%%%%%%%%%%%%%%%%%%%%%%%%%%%%%%%%%

%===============================================
\section{Parametrising dimension-8 terms after removing redundancies}
\label{eom_fdrefinition}
%===============================================
We provide effective operators and related WCs on a Green's basis, which contains redundant structures. These operators can be recast on a non-redundant operator basis following the conventions depicted in \cite{Murphy:2020rsh} after employing the equation of motions, Fierz and Bianchi identities, and related information discussed in \cite{Banerjee:2022thk, Corbett:2024yoy}. 
The operator relations among the Green and Murphy bases are given in Tables~\ref{Tab:nonredundant_basis_1}, \ref{Tab:nonredundant_basis_2} that can be used to find the non-redundant dimension-eight effective operators.

%%%%%%%%%%%%%%%%%%%%%%%%%%%%%%%%%%%%%%%%%%%%%%%%%%%%%
%  bibliography
%%%%%%%%%%%%%%%%%%%%%%%%%%%%%%%%%%%%%%%%%%%%%%%%%%%%%

\bibliographystyle{JHEP}
\bibliography{references}

\providecommand{\href}[2]{#2}\begingroup\raggedright\begin{thebibliography}{10}

\bibitem{Weinberg:1978kz}
S.~Weinberg, {\it {Phenomenological Lagrangians}},  {\em Physica A} {\bf 96}
  (1979), no.~1-2 327--340.

\bibitem{Georgi:1991ch}
H.~Georgi, {\it {On-shell effective field theory}},  {\em Nucl. Phys. B} {\bf
  361} (1991) 339--350.

\bibitem{BUCHMULLER1986621}
W.~Buchmüller and D.~Wyler, {\it Effective lagrangian analysis of new
  interactions and flavour conservation},  {\em Nuclear Physics B} {\bf 268}
  (1986), no.~3 621 -- 653.

\bibitem{Grzadkowski:2010es}
B.~Grzadkowski, M.~Iskrzynski, M.~Misiak, and J.~Rosiek, {\it {Dimension-Six
  Terms in the Standard Model Lagrangian}},  {\em JHEP} {\bf 10} (2010) 085,
  [\href{http://arxiv.org/abs/1008.4884}{{\tt arXiv:1008.4884}}].

\bibitem{Dedes:2019uzs}
A.~Dedes, M.~Paraskevas, J.~Rosiek, K.~Suxho, and L.~Trifyllis, {\it {SmeftFR
  \textendash{} Feynman rules generator for the Standard Model Effective Field
  Theory}},  {\em Comput. Phys. Commun.} {\bf 247} (2020) 106931,
  [\href{http://arxiv.org/abs/1904.03204}{{\tt arXiv:1904.03204}}].

\bibitem{Giudice:2007fh}
G.~F. Giudice, C.~Grojean, A.~Pomarol, and R.~Rattazzi, {\it {The
  Strongly-Interacting Light Higgs}},  {\em JHEP} {\bf 06} (2007) 045,
  [\href{http://arxiv.org/abs/hep-ph/0703164}{{\tt hep-ph/0703164}}].

\bibitem{Englert:2015hrx}
C.~Englert, R.~Kogler, H.~Schulz, and M.~Spannowsky, {\it {Higgs coupling
  measurements at the LHC}},  {\em Eur. Phys. J. C} {\bf 76} (2016), no.~7 393,
  [\href{http://arxiv.org/abs/1511.05170}{{\tt arXiv:1511.05170}}].

\bibitem{Ellis:2018gqa}
J.~Ellis, C.~W. Murphy, V.~Sanz, and T.~You, {\it {Updated Global SMEFT Fit to
  Higgs, Diboson and Electroweak Data}},  {\em JHEP} {\bf 06} (2018) 146,
  [\href{http://arxiv.org/abs/1803.03252}{{\tt arXiv:1803.03252}}].

\bibitem{Banerjee:2019twi}
S.~Banerjee, R.~S. Gupta, J.~Y. Reiness, S.~Seth, and M.~Spannowsky, {\it
  {Towards the ultimate differential SMEFT analysis}},  {\em JHEP} {\bf 09}
  (2020) 170, [\href{http://arxiv.org/abs/1912.07628}{{\tt arXiv:1912.07628}}].

\bibitem{Banerjee:2018bio}
S.~Banerjee, C.~Englert, R.~S. Gupta, and M.~Spannowsky, {\it {Probing
  Electroweak Precision Physics via boosted Higgs-strahlung at the LHC}},  {\em
  Phys. Rev. D} {\bf 98} (2018), no.~9 095012,
  [\href{http://arxiv.org/abs/1807.01796}{{\tt arXiv:1807.01796}}].

\bibitem{Pomarol:2013zra}
A.~Pomarol and F.~Riva, {\it {Towards the Ultimate SM Fit to Close in on Higgs
  Physics}},  {\em JHEP} {\bf 01} (2014) 151,
  [\href{http://arxiv.org/abs/1308.2803}{{\tt arXiv:1308.2803}}].

\bibitem{Butter:2016cvz}
A.~Butter, O.~J.~P. \'Eboli, J.~Gonzalez-Fraile, M.~C. Gonzalez-Garcia,
  T.~Plehn, and M.~Rauch, {\it {The Gauge-Higgs Legacy of the LHC Run I}},
  {\em JHEP} {\bf 07} (2016) 152, [\href{http://arxiv.org/abs/1604.03105}{{\tt
  arXiv:1604.03105}}].

\bibitem{Berthier:2015gja}
L.~Berthier and M.~Trott, {\it {Consistent constraints on the Standard Model
  Effective Field Theory}},  {\em JHEP} {\bf 02} (2016) 069,
  [\href{http://arxiv.org/abs/1508.05060}{{\tt arXiv:1508.05060}}].

\bibitem{Berthier:2015oma}
L.~Berthier and M.~Trott, {\it {Towards consistent Electroweak Precision Data
  constraints in the SMEFT}},  {\em JHEP} {\bf 05} (2015) 024,
  [\href{http://arxiv.org/abs/1502.02570}{{\tt arXiv:1502.02570}}].

\bibitem{deBlas:2019rxi}
J.~de~Blas et~al., {\it {Higgs Boson Studies at Future Particle Colliders}},
  {\em JHEP} {\bf 01} (2020) 139, [\href{http://arxiv.org/abs/1905.03764}{{\tt
  arXiv:1905.03764}}].

\bibitem{Buckley:2015lku}
A.~Buckley, C.~Englert, J.~Ferrando, D.~J. Miller, L.~Moore, M.~Russell, and
  C.~D. White, {\it {Constraining top quark effective theory in the LHC Run II
  era}},  {\em JHEP} {\bf 04} (2016) 015,
  [\href{http://arxiv.org/abs/1512.03360}{{\tt arXiv:1512.03360}}].

\bibitem{Brivio:2019ius}
I.~Brivio, S.~Bruggisser, F.~Maltoni, R.~Moutafis, T.~Plehn, E.~Vryonidou,
  S.~Westhoff, and C.~Zhang, {\it {O new physics, where art thou? A global
  search in the top sector}},  {\em JHEP} {\bf 02} (2020) 131,
  [\href{http://arxiv.org/abs/1910.03606}{{\tt arXiv:1910.03606}}].

\bibitem{Ellis:2021jhep}
J.~Ellis, M.~Madigan, K.~Mimasu, V.~Sanz, and T.~You, {\it {Top, Higgs, Diboson
  and Electroweak Fit to the Standard Model Effective Field Theory}},  {\em
  JHEP} {\bf 04} (2021) 279, [\href{http://arxiv.org/abs/2012.02779}{{\tt
  arXiv:2012.02779}}].

\bibitem{Durieux:2019rbz}
G.~Durieux, A.~Irles, V.~Miralles, A.~Pe\~nuelas, R.~P\"oschl, M.~Perell\'o,
  and M.~Vos, {\it {The electro-weak couplings of the top and bottom quarks
  \textemdash{} Global fit and future prospects}},  {\em JHEP} {\bf 12} (2019)
  98, [\href{http://arxiv.org/abs/1907.10619}{{\tt arXiv:1907.10619}}].
  [Erratum: JHEP 01, 195 (2021)].

\bibitem{Biswas:2021qaf}
T.~Biswas, A.~Datta, and B.~Mukhopadhyaya, {\it {Following the trail of new
  physics via the vector boson fusion Higgs boson signal at the Large Hadron
  Collider}},  {\em Phys. Rev. D} {\bf 105} (2022), no.~5 055028,
  [\href{http://arxiv.org/abs/2107.05503}{{\tt arXiv:2107.05503}}].

\bibitem{Biswas:2022fsr}
T.~Biswas and A.~Datta, {\it {Exploring Higgs-photon production at the LHC}},
  {\em JHEP} {\bf 05} (2023) 104, [\href{http://arxiv.org/abs/2208.08432}{{\tt
  arXiv:2208.08432}}].

\bibitem{Ethier:2021bye}
{\bf SMEFiT} Collaboration, J.~J. Ethier, G.~Magni, F.~Maltoni, L.~Mantani,
  E.~R. Nocera, J.~Rojo, E.~Slade, E.~Vryonidou, and C.~Zhang, {\it {Combined
  SMEFT interpretation of Higgs, diboson, and top quark data from the LHC}},
  {\em JHEP} {\bf 11} (2021) 089, [\href{http://arxiv.org/abs/2105.00006}{{\tt
  arXiv:2105.00006}}].

\bibitem{Hays:2018zze}
C.~Hays, A.~Martin, V.~Sanz, and J.~Setford, {\it {On the impact of
  dimension-eight SMEFT operators on Higgs measurements}},  {\em JHEP} {\bf 02}
  (2019) 123, [\href{http://arxiv.org/abs/1808.00442}{{\tt arXiv:1808.00442}}].

\bibitem{Degrande:2013kka}
C.~Degrande, {\it {A basis of dimension-eight operators for anomalous neutral
  triple gauge boson interactions}},  {\em JHEP} {\bf 02} (2014) 101,
  [\href{http://arxiv.org/abs/1308.6323}{{\tt arXiv:1308.6323}}].

\bibitem{Dawson:2021xei}
S.~Dawson, S.~Homiller, and M.~Sullivan, {\it {Impact of dimension-eight SMEFT
  contributions: A case study}},  {\em Phys. Rev. D} {\bf 104} (2021), no.~11
  115013, [\href{http://arxiv.org/abs/2110.06929}{{\tt arXiv:2110.06929}}].

\bibitem{Dedes:2023zws}
A.~Dedes, J.~Rosiek, M.~Ryczkowski, K.~Suxho, and L.~Trifyllis, {\it {SmeftFR
  v3 \textendash{} Feynman rules generator for the Standard Model Effective
  Field Theory}},  {\em Comput. Phys. Commun.} {\bf 294} (2024) 108943,
  [\href{http://arxiv.org/abs/2302.01353}{{\tt arXiv:2302.01353}}].

\bibitem{Bakshi:2024wzz}
S.~D. Bakshi, M.~Chala, A.~D\'\i{}az-Carmona, Z.~Ren, and F.~Vilches, {\it
  {Renormalization of the SMEFT to dimension eight: Fermionic interactions I}},
   \href{http://arxiv.org/abs/2409.15408}{{\tt arXiv:2409.15408}}.

\bibitem{DasBakshi:2022mwk}
S.~Das~Bakshi, M.~Chala, A.~D\'\i{}az-Carmona, and G.~Guedes, {\it {Towards the
  renormalisation of the Standard Model effective field theory to dimension
  eight: bosonic interactions II}},  {\em Eur. Phys. J. Plus} {\bf 137} (2022),
  no.~8 973, [\href{http://arxiv.org/abs/2205.03301}{{\tt arXiv:2205.03301}}].

\bibitem{Li:2023pfw}
H.-L. Li, Y.-H. Ni, M.-L. Xiao, and J.-H. Yu, {\it {Complete UV resonances of
  the dimension-8 SMEFT operators}},  {\em JHEP} {\bf 05} (2024) 238,
  [\href{http://arxiv.org/abs/2309.15933}{{\tt arXiv:2309.15933}}].

\bibitem{Degrande:2023iob}
C.~Degrande and H.-L. Li, {\it {Impact of dimension-8 SMEFT operators on
  diboson productions}},  {\em JHEP} {\bf 06} (2023) 149,
  [\href{http://arxiv.org/abs/2303.10493}{{\tt arXiv:2303.10493}}].

\bibitem{Chen:2023bhu}
Q.~Chen, K.~Mimasu, T.~A. Wu, G.-D. Zhang, and S.-Y. Zhou, {\it {Capping the
  positivity cone: dimension-8 Higgs operators in the SMEFT}},  {\em JHEP} {\bf
  03} (2024) 180, [\href{http://arxiv.org/abs/2309.15922}{{\tt
  arXiv:2309.15922}}].

\bibitem{Ellis:2023zim}
J.~Ellis, K.~Mimasu, and F.~Zampedri, {\it {Dimension-8 SMEFT analysis of
  minimal scalar field extensions of the Standard Model}},  {\em JHEP} {\bf 10}
  (2023) 051, [\href{http://arxiv.org/abs/2304.06663}{{\tt arXiv:2304.06663}}].

\bibitem{Dawson:2024ozw}
S.~Dawson, M.~Forslund, and M.~Schnubel, {\it {SMEFT matching to Z' models at
  dimension eight}},  {\em Phys. Rev. D} {\bf 110} (2024), no.~1 015002,
  [\href{http://arxiv.org/abs/2404.01375}{{\tt arXiv:2404.01375}}].

\bibitem{Adhikary:2025pbb}
N.~Adhikary, J.~Das, and D.~Dey, {\it {Two-loop dimension Six Effective Action:
  Integrating Out Heavy Scalar}},  \href{http://arxiv.org/abs/2501.01313}{{\tt
  arXiv:2501.01313}}.

\bibitem{Li:2020gnx}
H.-L. Li, Z.~Ren, J.~Shu, M.-L. Xiao, J.-H. Yu, and Y.-H. Zheng, {\it {Complete
  Set of Dimension-8 Operators in the Standard Model Effective Field Theory}},
  \href{http://arxiv.org/abs/2005.00008}{{\tt arXiv:2005.00008}}.

\bibitem{Murphy:2020rsh}
C.~W. Murphy, {\it {Dimension-8 Operators in the Standard Model Effective Field
  Theory}},  \href{http://arxiv.org/abs/2005.00059}{{\tt arXiv:2005.00059}}.

\bibitem{Brivio:2017vri}
I.~Brivio and M.~Trott, {\it {The Standard Model as an Effective Field
  Theory}},  {\em Phys. Rept.} {\bf 793} (2019) 1--98,
  [\href{http://arxiv.org/abs/1706.08945}{{\tt arXiv:1706.08945}}].

\bibitem{Isidori:2023pyp}
G.~Isidori, F.~Wilsch, and D.~Wyler, {\it {The standard model effective field
  theory at work}},  {\em Rev. Mod. Phys.} {\bf 96} (2024), no.~1 015006,
  [\href{http://arxiv.org/abs/2303.16922}{{\tt arXiv:2303.16922}}].

\bibitem{Anisha:2021hgc}
Anisha, S.~Das~Bakshi, S.~Banerjee, A.~Biek\"otter, J.~Chakrabortty,
  S.~Kumar~Patra, and M.~Spannowsky, {\it {Effective limits on single scalar
  extensions in the light of recent LHC data}},
  \href{http://arxiv.org/abs/2111.05876}{{\tt arXiv:2111.05876}}.

\bibitem{Gorbahn:2015gxa}
M.~Gorbahn, J.~M. No, and V.~Sanz, {\it {Benchmarks for Higgs Effective Theory:
  Extended Higgs Sectors}},  {\em JHEP} {\bf 10} (2015) 036,
  [\href{http://arxiv.org/abs/1502.07352}{{\tt arXiv:1502.07352}}].

\bibitem{delAguila:2016zcb}
F.~del Aguila, Z.~Kunszt, and J.~Santiago, {\it {One-loop effective lagrangians
  after matching}},  {\em Eur. Phys. J. C} {\bf 76} (2016), no.~5 244,
  [\href{http://arxiv.org/abs/1602.00126}{{\tt arXiv:1602.00126}}].

\bibitem{Branco:2011iw}
G.~C. Branco, P.~M. Ferreira, L.~Lavoura, M.~N. Rebelo, M.~Sher, and J.~P.
  Silva, {\it {Theory and phenomenology of two-Higgs-doublet models}},  {\em
  Phys. Rept.} {\bf 516} (2012) 1--102,
  [\href{http://arxiv.org/abs/1106.0034}{{\tt arXiv:1106.0034}}].

\bibitem{Gunion:1989we}
J.~F. Gunion, H.~E. Haber, G.~L. Kane, and S.~Dawson, {\em {The Higgs Hunter's
  Guide}}, vol.~80.
\newblock 2000.

\bibitem{delAguila:2008ks}
F.~del Aguila, J.~A. Aguilar-Saavedra, J.~de~Blas, and M.~Perez-Victoria, {\it
  {Electroweak constraints on see-saw messengers and their implications for
  LHC}},  in {\em {43rd Rencontres de Moriond on Electroweak Interactions and
  Unified Theories}}, pp.~45--52, 6, 2008.
\newblock \href{http://arxiv.org/abs/0806.1023}{{\tt arXiv:0806.1023}}.

\bibitem{Akeroyd:2005gt}
A.~G. Akeroyd and M.~Aoki, {\it {Single and pair production of doubly charged
  Higgs bosons at hadron colliders}},  {\em Phys. Rev. D} {\bf 72} (2005)
  035011, [\href{http://arxiv.org/abs/hep-ph/0506176}{{\tt hep-ph/0506176}}].

\bibitem{Akeroyd:2009hb}
A.~G. Akeroyd and C.-W. Chiang, {\it {Doubly charged Higgs bosons and
  three-lepton signatures in the Higgs Triplet Model}},  {\em Phys. Rev. D}
  {\bf 80} (2009) 113010, [\href{http://arxiv.org/abs/0909.4419}{{\tt
  arXiv:0909.4419}}].

\bibitem{Han:2015sca}
Z.-L. Han, R.~Ding, and Y.~Liao, {\it {LHC phenomenology of the type II seesaw
  mechanism: Observability of neutral scalars in the nondegenerate case}},
  {\em Phys. Rev. D} {\bf 92} (2015), no.~3 033014,
  [\href{http://arxiv.org/abs/1506.08996}{{\tt arXiv:1506.08996}}].

\bibitem{Akeroyd:2009nu}
A.~G. Akeroyd, M.~Aoki, and H.~Sugiyama, {\it {Lepton Flavour Violating Decays
  $\tau \to \bar l ll$ and $\mu \to e \gamma$ in the Higgs Triplet Model}},
  {\em Phys. Rev. D} {\bf 79} (2009) 113010,
  [\href{http://arxiv.org/abs/0904.3640}{{\tt arXiv:0904.3640}}].

\bibitem{Dey:2018uvu}
A.~Dey, J.~Lahiri, and B.~Mukhopadhyaya, {\it {Extended scalar sectors,
  effective operators and observed data}},  {\em JHEP} {\bf 11} (2018) 127,
  [\href{http://arxiv.org/abs/1808.04869}{{\tt arXiv:1808.04869}}].

\bibitem{Cao:2009as}
J.~Cao, P.~Wan, L.~Wu, and J.~M. Yang, {\it {Lepton-Specific Two-Higgs Doublet
  Model: Experimental Constraints and Implication on Higgs Phenomenology}},
  {\em Phys. Rev. D} {\bf 80} (2009) 071701,
  [\href{http://arxiv.org/abs/0909.5148}{{\tt arXiv:0909.5148}}].

\bibitem{Bhattacharyya:2015nca}
G.~Bhattacharyya and D.~Das, {\it {Scalar sector of two-Higgs-doublet models: A
  minireview}},  {\em Pramana} {\bf 87} (2016), no.~3 40,
  [\href{http://arxiv.org/abs/1507.06424}{{\tt arXiv:1507.06424}}].

\bibitem{Bambhaniya:2013yca}
G.~Bambhaniya, J.~Chakrabortty, S.~Goswami, and P.~Konar, {\it {Generation of
  neutrino mass from new physics at TeV scale and multilepton signatures at the
  LHC}},  {\em Phys. Rev. D} {\bf 88} (2013), no.~7 075006,
  [\href{http://arxiv.org/abs/1305.2795}{{\tt arXiv:1305.2795}}].

\bibitem{Cai:2017mow}
Y.~Cai, T.~Han, T.~Li, and R.~Ruiz, {\it {Lepton Number Violation: Seesaw
  Models and Their Collider Tests}},  {\em Front. in Phys.} {\bf 6} (2018) 40,
  [\href{http://arxiv.org/abs/1711.02180}{{\tt arXiv:1711.02180}}].

\bibitem{BhupalDev:2018tox}
P.~S. Bhupal~Dev and Y.~Zhang, {\it {Displaced vertex signatures of doubly
  charged scalars in the type-II seesaw and its left-right extensions}},  {\em
  JHEP} {\bf 10} (2018) 199, [\href{http://arxiv.org/abs/1808.00943}{{\tt
  arXiv:1808.00943}}].

\bibitem{Fuks:2019clu}
B.~Fuks, M.~Nemev\v{s}ek, and R.~Ruiz, {\it {Doubly Charged Higgs Boson
  Production at Hadron Colliders}},  {\em Phys. Rev. D} {\bf 101} (2020), no.~7
  075022, [\href{http://arxiv.org/abs/1912.08975}{{\tt arXiv:1912.08975}}].

\bibitem{Antusch:2018svb}
S.~Antusch, O.~Fischer, A.~Hammad, and C.~Scherb, {\it {Low scale type II
  seesaw: Present constraints and prospects for displaced vertex searches}},
  {\em JHEP} {\bf 02} (2019) 157, [\href{http://arxiv.org/abs/1811.03476}{{\tt
  arXiv:1811.03476}}].

\bibitem{delAguila:2013mia}
F.~del \'Aguila and M.~Chala, {\it {LHC bounds on Lepton Number Violation
  mediated by doubly and singly-charged scalars}},  {\em JHEP} {\bf 03} (2014)
  027, [\href{http://arxiv.org/abs/1311.1510}{{\tt arXiv:1311.1510}}].

\bibitem{CMS:2022cbe}
{\bf CMS} Collaboration, {\it {Prospects for a Search for Doubly Charged Higgs
  Bosons at the HL-LHC}}, .

\bibitem{Henning:2014wua}
B.~Henning, X.~Lu, and H.~Murayama, {\it {How to use the Standard Model
  effective field theory}},  {\em JHEP} {\bf 01} (2016) 023,
  [\href{http://arxiv.org/abs/1412.1837}{{\tt arXiv:1412.1837}}].

\bibitem{Bakshi:2018ics}
S.~Das~Bakshi, J.~Chakrabortty, and S.~K. Patra, {\it {CoDEx: Wilson
  coefficient calculator connecting SMEFT to UV theory}},  {\em Eur. Phys. J.
  C} {\bf 79} (2019), no.~1 21, [\href{http://arxiv.org/abs/1808.04403}{{\tt
  arXiv:1808.04403}}].

\bibitem{Henning:2016lyp}
B.~Henning, X.~Lu, and H.~Murayama, {\it {One-loop Matching and Running with
  Covariant Derivative Expansion}},  {\em JHEP} {\bf 01} (2018) 123,
  [\href{http://arxiv.org/abs/1604.01019}{{\tt arXiv:1604.01019}}].

\bibitem{vonGersdorff:2022kwj}
G.~von Gersdorff and K.~Santos, {\it {New covariant Feynman rules for effective
  field theories}},  {\em JHEP} {\bf 04} (2023) 025,
  [\href{http://arxiv.org/abs/2212.07451}{{\tt arXiv:2212.07451}}].

\bibitem{Fuentes-Martin:2022jrf}
J.~Fuentes-Mart\'\i{}n, M.~K\"onig, J.~Pag\`es, A.~E. Thomsen, and F.~Wilsch,
  {\it {A Proof of Concept for Matchete: An Automated Tool for Matching
  Effective Theories}},  \href{http://arxiv.org/abs/2212.04510}{{\tt
  arXiv:2212.04510}}.

\bibitem{Cohen:2020qvb}
T.~Cohen, X.~Lu, and Z.~Zhang, {\it {STrEAMlining EFT Matching}},  {\em SciPost
  Phys.} {\bf 10} (2021), no.~5 098,
  [\href{http://arxiv.org/abs/2012.07851}{{\tt arXiv:2012.07851}}].

\bibitem{Carmona:2021xtq}
A.~Carmona, A.~Lazopoulos, P.~Olgoso, and J.~Santiago, {\it {Matchmakereft:
  automated tree-level and one-loop matching}},  {\em SciPost Phys.} {\bf 12}
  (2022), no.~6 198, [\href{http://arxiv.org/abs/2112.10787}{{\tt
  arXiv:2112.10787}}].

\bibitem{Criado:2017khh}
J.~C. Criado, {\it {MatchingTools: a Python library for symbolic effective
  field theory calculations}},  {\em Comput. Phys. Commun.} {\bf 227} (2018)
  42--50, [\href{http://arxiv.org/abs/1710.06445}{{\tt arXiv:1710.06445}}].

\bibitem{Celis:2017hod}
A.~Celis, J.~Fuentes-Martin, A.~Vicente, and J.~Virto, {\it {DsixTools: The
  Standard Model Effective Field Theory Toolkit}},  {\em Eur. Phys. J.} {\bf
  C77} (2017), no.~6 405, [\href{http://arxiv.org/abs/1704.04504}{{\tt
  arXiv:1704.04504}}].

\bibitem{Haisch:2020ahr}
U.~Haisch, M.~Ruhdorfer, E.~Salvioni, E.~Venturini, and A.~Weiler, {\it
  {Singlet night in Feynman-ville: one-loop matching of a real scalar}},  {\em
  JHEP} {\bf 04} (2020) 164, [\href{http://arxiv.org/abs/2003.05936}{{\tt
  arXiv:2003.05936}}]. [Erratum: JHEP 07, 066 (2020)].

\bibitem{Jiang:2018pbd}
M.~Jiang, N.~Craig, Y.-Y. Li, and D.~Sutherland, {\it {Complete one-loop
  matching for a singlet scalar in the Standard Model EFT}},  {\em JHEP} {\bf
  02} (2019) 031, [\href{http://arxiv.org/abs/1811.08878}{{\tt
  arXiv:1811.08878}}]. [Erratum: JHEP 01, 135 (2021)].

\bibitem{Gherardi:2020det}
V.~Gherardi, D.~Marzocca, and E.~Venturini, {\it {Matching scalar leptoquarks
  to the SMEFT at one loop}},  {\em JHEP} {\bf 07} (2020) 225,
  [\href{http://arxiv.org/abs/2003.12525}{{\tt arXiv:2003.12525}}]. [Erratum:
  JHEP 01, 006 (2021)].

\bibitem{Du:2022vso}
Y.~Du, X.-X. Li, and J.-H. Yu, {\it {Neutrino seesaw models at one-loop
  matching: discrimination by effective operators}},  {\em JHEP} {\bf 09}
  (2022) 207, [\href{http://arxiv.org/abs/2201.04646}{{\tt arXiv:2201.04646}}].

\bibitem{Banerjee:2023iiv}
U.~Banerjee, J.~Chakrabortty, S.~U. Rahaman, and K.~Ramkumar, {\it {One-loop
  effective action up to dimension eight: integrating out heavy scalar(s)}},
  {\em Eur. Phys. J. Plus} {\bf 139} (2024), no.~2 159,
  [\href{http://arxiv.org/abs/2306.09103}{{\tt arXiv:2306.09103}}].

\bibitem{Chakrabortty:2023yke}
J.~Chakrabortty, S.~U. Rahaman, and K.~Ramkumar, {\it {One-loop effective
  action up to dimension eight: Integrating out heavy fermion(s)}},  {\em Nucl.
  Phys. B} {\bf 1000} (2024) 116488,
  [\href{http://arxiv.org/abs/2308.03849}{{\tt arXiv:2308.03849}}].

\bibitem{Barrie:2022cub}
N.~D. Barrie, C.~Han, and H.~Murayama, {\it {Type II Seesaw leptogenesis}},
  {\em JHEP} {\bf 05} (2022) 160, [\href{http://arxiv.org/abs/2204.08202}{{\tt
  arXiv:2204.08202}}].

\bibitem{Chala:2021cgt}
M.~Chala, A.~D\'\i{}az-Carmona, and G.~Guedes, {\it {A Green\textquoteright{}s
  basis for the bosonic SMEFT to dimension 8}},  {\em JHEP} {\bf 05} (2022)
  138, [\href{http://arxiv.org/abs/2112.12724}{{\tt arXiv:2112.12724}}].

\bibitem{DasBakshi:2018vni}
S.~Das~Bakshi, J.~Chakrabortty, and S.~K. Patra, {\it {CoDEx: Wilson
  coefficient calculator connecting SMEFT to UV theory}},  {\em Eur. Phys. J.
  C} {\bf 79} (2019), no.~1 21, [\href{http://arxiv.org/abs/1808.04403}{{\tt
  arXiv:1808.04403}}].

\bibitem{Li:2018jns}
T.~Li, {\it {Type II Seesaw and tau lepton at the HL-LHC, HE-LHC and FCC-hh}},
  {\em JHEP} {\bf 09} (2018) 079, [\href{http://arxiv.org/abs/1802.00945}{{\tt
  arXiv:1802.00945}}].

\bibitem{Bolton:2022lrg}
P.~D. Bolton and S.~T. Petcov, {\it {Measurements of $\mu \to 3e$ decay with
  polarised muons as a probe of new physics}},  {\em Phys. Lett. B} {\bf 833}
  (2022) 137296, [\href{http://arxiv.org/abs/2204.03468}{{\tt
  arXiv:2204.03468}}].

\bibitem{Primulando:2019evb}
R.~Primulando, J.~Julio, and P.~Uttayarat, {\it {Scalar phenomenology in
  type-II seesaw model}},  {\em JHEP} {\bf 08} (2019) 024,
  [\href{http://arxiv.org/abs/1903.02493}{{\tt arXiv:1903.02493}}].

\bibitem{Peskin:1991sw}
M.~E. Peskin and T.~Takeuchi, {\it {Estimation of oblique electroweak
  corrections}},  {\em Phys. Rev. D} {\bf 46} (1992) 381--409.

\bibitem{Peskin:1990zt}
M.~E. Peskin and T.~Takeuchi, {\it {A New constraint on a strongly interacting
  Higgs sector}},  {\em Phys. Rev. Lett.} {\bf 65} (1990) 964--967.

\bibitem{Bagnaschi:2022whn}
E.~Bagnaschi, J.~Ellis, M.~Madigan, K.~Mimasu, V.~Sanz, and T.~You, {\it {SMEFT
  Analysis of $m_{W}$}},  \href{http://arxiv.org/abs/2204.05260}{{\tt
  arXiv:2204.05260}}.

\bibitem{Asadi:2022xiy}
P.~Asadi, C.~Cesarotti, K.~Fraser, S.~Homiller, and A.~Parikh, {\it {Oblique
  Lessons from the $W$ Mass Measurement at CDF II}},
  \href{http://arxiv.org/abs/2204.05283}{{\tt arXiv:2204.05283}}.

\bibitem{Corbett:2021eux}
T.~Corbett, A.~Helset, A.~Martin, and M.~Trott, {\it {EWPD in the SMEFT to
  dimension eight}},  {\em JHEP} {\bf 06} (2021) 076,
  [\href{http://arxiv.org/abs/2102.02819}{{\tt arXiv:2102.02819}}].

\bibitem{Belvedere:2024wzg}
A.~Belvedere, C.~Englert, R.~Kogler, and M.~Spannowsky, {\it {Dispelling the
  $\sqrt{\mathcal {L}} $ myth for the High-Luminosity LHC}},  {\em Eur. Phys.
  J. C} {\bf 84} (2024), no.~7 715,
  [\href{http://arxiv.org/abs/2402.07985}{{\tt arXiv:2402.07985}}].

\bibitem{Corbett:2023qtg}
T.~Corbett, J.~Desai, O.~J.~P. \'Eboli, M.~C. Gonzalez-Garcia, M.~Martines, and
  P.~Reimitz, {\it {Impact of dimension-eight SMEFT operators in the
  electroweak precision observables and triple gauge couplings analysis in
  universal SMEFT}},  {\em Phys. Rev. D} {\bf 107} (2023), no.~11 115013,
  [\href{http://arxiv.org/abs/2304.03305}{{\tt arXiv:2304.03305}}].

\bibitem{Hamoudou:2022tdn}
S.~Hamoudou, J.~Kumar, and D.~London, {\it {Dimension-8 SMEFT matching
  conditions for the low-energy effective field theory}},  {\em JHEP} {\bf 03}
  (2023) 157, [\href{http://arxiv.org/abs/2207.08856}{{\tt arXiv:2207.08856}}].

\bibitem{Dawson:2022cmu}
S.~Dawson, D.~Fontes, S.~Homiller, and M.~Sullivan, {\it {Role of
  dimension-eight operators in an EFT for the 2HDM}},  {\em Phys. Rev. D} {\bf
  106} (2022), no.~5 055012, [\href{http://arxiv.org/abs/2205.01561}{{\tt
  arXiv:2205.01561}}].

\bibitem{DasBakshi:2024krs}
S.~Das~Bakshi, S.~Dawson, D.~Fontes, and S.~Homiller, {\it {Relevance of
  one-loop SMEFT matching in the 2HDM}},  {\em Phys. Rev. D} {\bf 109} (2024),
  no.~7 075022, [\href{http://arxiv.org/abs/2401.12279}{{\tt
  arXiv:2401.12279}}].

\bibitem{Baak:2014ora}
{\bf Gfitter Group} Collaboration, M.~Baak, J.~C\'uth, J.~Haller, A.~Hoecker,
  R.~Kogler, K.~M\"onig, M.~Schott, and J.~Stelzer, {\it {The global
  electroweak fit at NNLO and prospects for the LHC and ILC}},  {\em Eur. Phys.
  J. C} {\bf 74} (2014) 3046, [\href{http://arxiv.org/abs/1407.3792}{{\tt
  arXiv:1407.3792}}].

\bibitem{Banerjee:2022thk}
U.~Banerjee, J.~Chakrabortty, C.~Englert, S.~U. Rahaman, and M.~Spannowsky,
  {\it {Integrating out heavy scalars with modified equations of motion:
  Matching computation of dimension-eight SMEFT coefficients}},  {\em Phys.
  Rev. D} {\bf 107} (2023), no.~5 055007,
  [\href{http://arxiv.org/abs/2210.14761}{{\tt arXiv:2210.14761}}].

\bibitem{Corbett:2024yoy}
T.~Corbett, J.~Desai, O.~J.~P. Eboli, and M.~C. Gonzalez-Garcia, {\it
  {Dimension-eight operator basis for universal standard model effective field
  theory}},  {\em Phys. Rev. D} {\bf 110} (2024), no.~3 033003,
  [\href{http://arxiv.org/abs/2404.03720}{{\tt arXiv:2404.03720}}].

\end{thebibliography}\endgroup

\end{document}